\documentclass[fleqn,usenatbib]{mnras}

\usepackage{mathptmx}
\usepackage{txfonts}

\usepackage[T1]{fontenc}
\usepackage{ae,aecompl}

\usepackage{graphicx}	

\newcommand{\gaia}{\textit{Gaia}}

\newcommand{\kmps}{\mbox{km\,s$^{-1}$}}

\title[\gaia\ white dwarfs within 40\,pc I]{\gaia\ white dwarfs within 40\,pc I:  spectroscopic observations of new candidates}

\author[Tremblay et al]{P.-E. Tremblay,$^{1}$\thanks{E-mail: P-E.Tremblay@warwick.ac.uk} M.~A.~Hollands,$^{1}$ N.~P.~Gentile Fusillo,$^{1,2}$ J.~McCleery,$^{1}$
\newauthor P.~Izquierdo,$^{3,4}$ B.~T.~G\"ansicke,$^{1}$ E.~Cukanovaite,$^{1}$ D.~Koester,$^{5}$ W. R. Brown,$^{6}$
\newauthor
S.~Charpinet,$^{7}$ T. Cunningham,$^{1}$ J.~Farihi,$^{8}$ N.~Giammichele,$^{7}$ V.~van~Grootel,$^{9}$ 
\newauthor
J.~J.~Hermes,$^{10}$ M.~J.~Hoskin,$^{1}$, S. Jordan,$^{11}$ 
S.~O.~Kepler,$^{12}$ S.~J.~Kleinman,$^{13}$ 
\newauthor
C.~J.~Manser,$^{1}$ T.~R.~Marsh,$^{1}$ D.~de Martino,$^{14}$ A.~Nitta,$^{13}$ S.~G.~Parsons,$^{15}$ 
\newauthor
I.~Pelisoli,$^{16}$ R.~Raddi,$^{17}$ A. Rebassa-Mansergas,$^{17,18}$ J.-J.~Ren,$^{19}$
\newauthor
M.~R.~Schreiber,$^{20,21}$ R. Silvotti,$^{22}$ O. Toloza,$^{1}$ S. Toonen,$^{23}$ and  S.~Torres$^{17,18}$
\\
(affiliations can be found after the references)
}

\date{Accepted XXX. Received YYY; in original form ZZZ}

\pubyear{2020}
\begin{document}
\label{firstpage}
\pagerange{\pageref{firstpage}--\pageref{lastpage}}
\maketitle

\begin{abstract}
We present a spectroscopic survey of 230 white dwarf candidates within 40\,pc of the Sun from the William Herschel Telescope and Gran Telescopio Canarias. All candidates were selected from \textit{Gaia} Data Release 2 (DR2) and in almost all cases had no prior spectroscopic classifications. We find a total of 191 confirmed white dwarfs and 39 main-sequence star contaminants. The majority of stellar remnants in the sample are relatively cool ($\langle T_{\rm eff} \rangle$ = 6200\,K), showing either hydrogen Balmer lines or a featureless spectrum, corresponding to 89 DA and 76 DC white dwarfs, respectively. We also recover two DBA white dwarfs and 9--10 magnetic remnants. We find two carbon-bearing DQ stars and 14 new metal-rich white dwarfs. This includes the possible detection of the first ultra-cool white dwarf with metal lines. We describe three DZ stars for which we find at least four different metal species, including one which is strongly Fe- and Ni-rich, indicative of the accretion of a planetesimal with core-Earth composition. We find one extremely massive (1.31 $\pm$ 0.01 M$_{\odot}$) DA white dwarf showing weak Balmer lines, possibly indicating stellar magnetism. Another white dwarf shows strong Balmer line emission but no infrared excess, suggesting a low-mass sub-stellar companion. High spectroscopic completeness ($>$99\%) has now been reached for \textit{Gaia} DR2 sources within 40\,pc sample, in the northern hemisphere ($\delta >$ 0\,deg) and located on the white dwarf cooling track in the Hertzsprung-Russell diagram. A statistical study of the full northern sample is presented in a companion paper.

\end{abstract}



\section{Introduction}
\label{intro}

The nearest stars to the Sun and Solar System have a remarkable historical importance in astronomy and continue to represent some of the best known examples of spectral types and exoplanetary hosts. Past efforts have generally been focused on the relatively small volume complete sample of $\approx$ 315 systems including main-sequence stars, brown dwarfs, white dwarfs and planets within 10\,pc of the Sun \citep{henry2018}. White dwarfs make up about 6\% of stellar objects in that sample, but up to 95\% of local stars and their planetary systems are destined to that ultimate fate. These abundant stellar remnants serve several distinctive purposes in modern astrophysics such as tracing the local and Galactic stellar formation history \citep{winget87,rowell13,tremblay14,fantin19} and calibrating stellar ages \citep{Rebassa2016,Fouesneau2019}. The significance of these results is enhanced from unbiased volume-limited samples upon which statistical studies can be performed \citep{holberg02,giammichele12,limoges15,holberg16,subasavage17,hollands18Gaia}.

Some of the closest white dwarfs are prototypes for several wide ranging applications. The brightest and closest stellar remnant Sirius B has led to important input on stellar evolution and white dwarf structure \citep{bond17}, fundamental physics through gravitational redshift measurements \citep{joyce18} as well as providing insight on the local binary population \citep{holberg13,toonen17}. GRW~+70~8247, at 12.9\,pc, was the first white dwarf to show circularly polarized light from its strong magnetic field \citep{kemp70,bagnulo19}. The first degenerate star with metal lines in its spectrum, now recognised as the signature of a planetary system, was detected in the closest single white dwarf Van Maanen 2 \citep{vanMaanen17}. Seventy years later, G29-38, a ZZ Ceti pulsator at 17.5\,pc, became the prototype for white dwarfs with dusty debris discs \citep{zuckerman87}, leading to a very active research area around evolved planetary systems \citep{veras16} and a unique window into the bulk composition of other rocky worlds \citep{zuckerman07}. With increasing distances, further rare examples of stellar or planetary evolution are being discovered, such as double white dwarf merger remnants \citep{kawka20,hollands20}, extremely low mass white dwarfs that will most likely merge \citep{brown16,kawka20b}, or stellar remnants with transiting planetesimals and planets on close-in orbits \citep{vanderburg16,manser19,gaensicke19}. Hot white dwarfs might even serve to study the composition of gas giant planet atmospheres \citep{schreiber19}. A proper characterisation of the occurrence of these systems requires the definition of larger volume-complete samples.

The \textit{Gaia} spacecraft has shed a new light on the nearby Milky Way stellar population from its extremely accurate astrometry and photometry \citep{gaia-dr2}. The Hertzsprung-Russell (HR) diagram of the local stellar population within 100\,pc constructed from \textit{Gaia} DR2 \citep{gaiaHR} presents, for the first time, a near volume-complete census inclusive of all low-luminosity stars and white dwarfs. While the increase in size is exemplary, a full understanding of the local white dwarf population is still a major challenge. \textit{Gaia} DR2 has significant technical limitations even within 100\,pc, with a number of problematic sources that have been filtered from most early science results \citep{gaiaHR,jimenez2018,gentile19}. We nevertheless estimate the completeness of the \textit{Gaia} DR2 white dwarf selection to be $\approx$ 96 per cent within 20--40\,pc \citep{hollands18Gaia,gentile19,mccleery2020}. Furthermore, only $\approx$ 10\% of \textit{Gaia} white dwarfs within 100\,pc have published spectroscopy \citep{gentile19,tremblay2019Nature}. Effective temperatures ($T_{\rm eff}$), surface gravities and masses can in principle be derived from \textit{Gaia} DR2 astrometry and photometry with a high precision of 1--2\% for nearby white dwarfs where reddening is small or negligible \citep{hollands18Gaia,tremblay19,bergeron19}. This technique is based on relatively well constrained white dwarf evolution models but the atmospheric composition must be known to transform colours into an atmospheric temperature. Erroneously assuming pure-hydrogen or pure-helium atmospheres can lead to 5--20\% systematic errors in mass \citep{bergeron19} and a strong bias in the evaluation of total ages. Upcoming multi-object medium resolution spectroscopic follow-ups such as WEAVE, 4MOST, DESI and SDSS-V \citep{WEAVE,4MOST,DESI,SDSSV}, and to a lesser degree upcoming \textit{Gaia} low-resolution spectrophotometry \citep{carrasco14}, will be essential to improve the accuracy of local white dwarf parameters and detect subtypes corresponding to metal pollution, magnetic fields or binarity. Nevertheless, it will be many years before these surveys cover large enough areas of the sky to significantly overlap with the $\approx$ 100\,pc \textit{Gaia} defined volume-limited white dwarf sample.

In this work we describe our dedicated spectroscopic follow-up and analysis of new {\it Gaia} white dwarf candidates within 40\,pc with the William Herschel Telescope (WHT) and Gran Telescopio Canarias (GTC). This directly follows upon the previous survey of \citet{limoges15} who relied on reduced proper motion for white dwarf identification.  We discuss the nature of 230 {\it Gaia} white dwarf candidates across all sky, among which a handful were recently confirmed separately in the literature \citep[see, e.g.,][]{scholz18,landstreet2019,landstreet2020} or had earlier ambiguous classifications. Among all observed targets, 155 are located in the northern hemisphere ($\delta > 0$\,deg). Combining this work and existing spectroscopic confirmations from the literature, there are 521 spectroscopically confirmed white dwarfs found within \textit{Gaia} DR2 in the northern 40\,pc hemisphere and only three unobserved high-probability white dwarf candidates \citep{gentile19}, indicating a very high spectroscopic completeness. A detailed statistical analysis of the full northern 40\,pc white dwarf sample, including a list of spectral types and references, is presented in a companion paper \citep[Paper II;][]{mccleery2020}. Here we also report on spectroscopic data of an additional 75 \textit{Gaia} white dwarf candidates in the southern hemisphere, which leaves approximately 200 unobserved high-probability white dwarf candidates in that region of the sky. The southern 40\,pc sample will be part of a future statistical analysis once spectroscopic completeness has reached a higher level.

\section{Observations}

\subsection{Catalogue photometry and astrometry}

\citet{gentile19} used spectroscopically confirmed white dwarfs from the Sloan Digital Sky Survey (SDSS) to map the regions of the \textit{Gaia} DR2 HR diagram encompassed by these stellar remnants. Based on the fraction of SDSS white dwarfs and contaminants, they calculated a probability of being a white dwarf ($P_{\rm WD}$) for all \textit{Gaia} sources that passed the initial selection. The authors recommend using $P_{\rm WD} > 0.75$ as a balance between completeness and contamination, a cut which recovers 96 per cent of the spectroscopically confirmed SDSS white dwarfs in the catalogue at all distances and up to a \textit{Gaia} magnitude limit of $G \approx 20$.

We selected white dwarf candidates from the catalogue of \citet{gentile19} with $\varpi > 25$ mas $\pm 1\sigma$ as the only requirement, resulting in 1233 sources, among which 184 are low probability candidates ($P_{\rm WD} < 0.75$). The following step was to perform a detailed cross-match of the literature to identify previous spectroscopic classifications. We have found 410 and 256 white dwarfs with a published spectral type in the northern and southern hemispheres, respectively, that we generally did not attempt to re-observe unless the spectral type was ambiguous or the spectrum not published (see Paper II for spectral types and references). The highest priority was given to high probability candidates without a spectral type in the northern hemisphere. We nevertheless kept low probability candidates in our target list, especially those with high proper motions that might reveal to be peculiar white dwarfs. In addition to 155 sources selected in the northern hemisphere, we also had the opportunity to observe an additional 75 {\it Gaia} white dwarf candidates in the southern hemisphere. We use the WD\,J naming convention introduced by \citet{gentile19} throughout all tables and figures, although we employ short names in the text for readability. 

\subsection{Spectroscopy}

We observed a total of 230 white dwarf candidates at the Observatorio del Roque de los Muchachos, both with the Intermediate-dispersion Spectrograph and Imaging System (ISIS) on the WHT and the Optical System for Imaging and low-Resolution Integrated Spectroscopy (OSIRIS) on the GTC. Table~\ref{tab:log} sumarises the different observations. Time was obtained through the International Time Programme ITP08 (PI Tremblay) and individual allocations (PIs Izquierdo, Marsh, Manser and G\"ansicke, see Table~\ref{tab:log}). 

\begin{table*}
	\centering
        \caption{Observing log}
        \label{tab:log}
        \begin{tabular}{lllllllll}
                \hline
                Dates & Telescope/ & Programme ID & Grating & No. of objects \\
                -- & Instrument & -- & -- & in this work \\
                \hline
                \hline
                2016--2017 & FAST & -- & \citet{FAST} & 3 \\
                -- & LAMOST & -- & \citet{LAMOST} & 2 \\
                2018 July -- December & GTC/OSIRIS & ITP08 & R1000B, R2500R & 26 \\
                2018 August 6--7 & WHT/ISIS & C117 & R600B, R600R & 31\\
                2018 August 9--23 & WHT/ISIS & ITP08 & R600B, R600R & 67\\
                2018 August 28 -- September 4 & WHT/ISIS & P8/N13 & R600B, R600R & 3\\
                2018 October 13--15 & WHT/ISIS & ITP08 & R600B, R600R & 22\\
                2019 February 9--10 & WHT/ISIS & P9 & R600B, R600R & 29\\
                2019 February 21--26 & WHT/ISIS & ITP08 & R600B, R600R & 23\\
                2019 April 13--14 & WHT/ISIS & ITP08 & R600B, R600R & 6\\
                2019 July 3--5 & WHT/ISIS & C82 & R600B, R600R & 16 \\
                2019 Aug 1--3 & WHT/ISIS & P23 & R600B, R600R & 2 \\
                \hline
        \end{tabular}\\
        Notes: Selected observations from FAST and LAMOST are included to ensure a coverage as complete as possible of \textit{Gaia} DR2 white dwarfs in the northern hemisphere (see Paper II).
\end{table*}

ISIS allows simultaneous observations using blue (R600B grating, R $\approx$ 2000) and red (R600R grating, R $\approx$ 3900) optimised CCDs via a dichroic beam-splitter. In our initial setup we employed central wavelengths of 4540\,\AA\  and 6562\,\AA\ for the blue and red arms, respectively, ensuring coverage of all Balmer lines and Ca H+K lines (wavelength ranges $\approx$ 3730--5350, 5730--7290\,\AA). Objects with metal lines were typically re-observed with central wavelengths of 3930\,\AA\  and 8200\,\AA\ for the blue and red arms, respectively, to ensure maximal wavelength coverage from the limit of atmospheric transmission to 9000\,\AA. Slit width varied between 1$^{\prime\prime}$ and 1.5$^{\prime\prime}$ depending on the observing conditions and we employed a binning of $2\times2$, resulting in an average resolution of $\approx$ 2~\AA. As much as possible we tried to adjust exposure times to ensure signal-to-noise ratio (S/N) greater than 30 at H$\alpha$ but there is a correlation between S/N and apparent magnitude. Cool and featureless DC white dwarfs ($T_{\rm eff}$ < 4800\,K) have on average slightly lower S/N ratios. 

Amongst the faintest sources in our target list, 26 candidates were observed with OSIRIS. Slit width was 1$^{\prime\prime}$ and we employed the standard binning of $2\times2$. For objects with the reddest colours corresponding to \textit{Gaia} $T_{\rm eff} \lesssim$ 4500\,K, we favoured low-resolution identification spectra using the R500B  grating (R $\approx$ 540, wavelength range 3600--7200\,\AA) which minimised overheads. For a few warmer objects we favoured the R1000R grating (R $\approx$ 1100, wavelength range 5100--10\,000\,\AA) in order to possibly detect H$\alpha$. One DZ white dwarf was serendipitously discovered using this second setup (WD\,J1922+0233). It implies other metal lines are likely to be present in the unobserved blue portion of the spectrum, which would allow a detailed chemical abundance analysis. 

The spectra were de-biased and flat-fielded 
using the standard \textsc{starklink}\footnote{The \textsc{starklink} Software Group homepage website is \url{http://starlink.jach.hawaii.edu/starlink}} packages \textsc{kappa}, \textsc{figaro} and \textsc{convert}. We carried out optimal extraction of spectra using the package \textsc{pamela}\footnote{\textsc{pamela} was written by T.~R.~Marsh and can be found in the \textsc{starklink} distribution Hawaiki and later releases.} \citep{marsh89}. The extracted 1D spectra were wavelength calibrated  and flux calibrated using \textsc{molly} \citep{marsh19}.

We also rely on external spectroscopic observations for five additional \textit{Gaia} white dwarfs in the northern hemisphere for which we could not find a spectral type in the literature. Those were included to ensure a coverage as complete as possible of white dwarfs in the northern hemisphere for the statistical analysis in the companion Paper II. Three spectra (WD\,J02215+0445, WD\,J0925+6120 and WDJ\,0319+4230) were acquired at the 1.5m Fred Lawrence Whipple Observatory telescope with the FAST spectrograph \citep{FAST} using the 600 l/mm grating and the 1.5$^{\prime\prime}$ slit, which provides 3550-5500 \AA\ wavelength coverage at 1.7 \AA\ spectral resolution. Two spectra (WD\,J0657+0550 and WD\,J0723+1617) were acquired from the LAMOST survey \citep{LAMOST}, which provides 3800-8800\,\AA\ wavelength coverage at 3\,\AA\ spectral resolution.

\section{Atmosphere and Evolution Models}
\label{sec:models}

For all observed targets we have used spectroscopic and photometric data to determine spectral types  by human inspection. We have also derived atmospheric parameters and chemical abundances using the different methods described in this section. Effective temperatures and $\log g$ values can be derived for most white dwarfs using photometric fits \citep{koester79,bergeron01}. For DA and DB stars, an independent method is to determine these parameters from spectroscopic line fits \citep{bergeron92,beauchamp99}. For these spectral types we refrain from performing simultaneous fits of the photometric and spectroscopic data, which would not improve accuracy because of the known systematic offset between both techniques \citep{tremblay19}. Instead in Section~\ref{results} we compare the results of both methods in order to better understand possible model atmosphere systematics and photometric colour calibration issues.

All objects in the northern hemisphere are also included in the sample discussed in Paper II, where for homogeneity only photometric parameters are employed. In comparison here we also report on our detailed spectroscopic analysis and describe the properties of 75 additional white dwarf candidates in the southern hemisphere.

\subsection{Photometric parameters}

Photometric temperatures, $\log g$ and masses are determined from \textit{Gaia} DR2 photometry and astrometry using the same method as that outlined in \citet{gentile19}. In brief we rely on our own grids of pure-H, pure-He and mixed H/He 1D model atmospheres \citep{tremblay11,cukanovaite19} and the mass-radius relation of \citet{fontaine01} for thick (H-atmospheres) and thin (He-atmospheres) hydrogen layers. The only differences are that we have neglected reddening and adopted grid of model fluxes that correspond to the newly identified atmospheric composition. We also rely on Pan-STARRS photometry when available, resulting in two different sets of atmospheric parameters using the same model atmospheres and \textit{Gaia} parallaxes.  Paper II demonstrates that \textit{Gaia} DR2 and Pan-STARRS are generally in good agreement except for crowded regions of the sky or white dwarfs with close, bright companions.

\citet{bergeron19} have demonstrated that pure-He atmospheres result in spuriously high masses and are unable to accurately reproduce the so-called bifurcation in the \textit{Gaia} DR2 HR diagram corresponding to 7000\,K $< T_{\rm eff} <$ 11\,000\,K \citep{gaiaHR,gentile19}. As a consequence, here and in Paper II we now rely on mixed atmospheres with [H/He] = $-5$ in number, which is well below spectroscopic detection limits \citep{rolland18}, for all photometric fits of DC white dwarfs above 7000\,K. For DC stars within 5000\,K $< T_{\rm eff} <$ 7000\,K we use pure-helium atmospheres if we can rule out the presence of the predicted H$\alpha$ line from pure-H model atmospheres. We note that pure-H and pure-He solutions are similar to within a few per cent in this temperature range. Below these temperatures, we assume a pure-hydrogen atmosphere for all DC white dwarfs because it is difficult to constrain the atmospheric composition with optical, near-IR and mid-IR photometry alone \citep{gentile2020}. 

In all cases we quote the small intrinsic fitting errors and refer to Section~\ref{results} for a discussion on extrinsic errors from photometric calibration. Paper II relies on the same set of photometric parameters as described above in their statistical analysis.

For metal-rich DZ, DZA and DAZ white dwarfs, and similarly for carbon-bearing DQ objects, we use the atmosphere code of \citet{koester10} to derive atmospheric parameters and chemical abundances from iterative fits of the photometry, astrometry and spectroscopy \citep{hollands18DZ,coutu19}. This method allows for the non-negligible feedback of metal lines on predicted photometric colours, which determine $T_{\rm eff}$ and $\log g$ values. With these quantities fixed, spectroscopy largely determines the chemical composition. The procedure is iterated until convergence. For a handful of these objects we postpone a dedicated analysis to future papers (Hollands et al.\ and G\"ansicke et al., in preparation).

\subsection{Spectroscopic parameters}

We derive $T_{\rm eff}$ and $\log g$ values from spectroscopic fits of non-magnetic DA white dwarfs with \textit{Gaia} temperatures above 6000\,K. We rely on the model atmospheres of \citet{tremblay11} with 3D corrections from \citet{tremblay13c}. The Balmer line fitting procedure is the same as that reported in \citet{gianninas11} and \citet{tremblay11}. In brief, we first normalise the individual Balmer lines to a continuum set to unity, defined by fitting a model spectrum to the observations where we include a polynomial with of the order of ten free parameters to account for residual errors in the flux calibration. We then perform a $\chi^2$ minimisation between the observed and predicted line profiles, convolved with a Gaussian instrumental profile with a resolution of 2\,\AA\ (FWHM). In all cases we only quote the intrinsic fitting errors. External errors from the flux calibration and fitting procedure are estimated at 0.038 dex in $\log g$ and
1.2\% in $T_{\rm eff}$ \citep{liebert05}.

For DA white dwarfs with temperatures below 6000\,K, the lines are too weak for a meaningful determination of the spectroscopic parameters. Therefore we simply compare observed H$\alpha$ line profiles with predictions using photometric atmospheric parameters to flag any outliers and systematic model effects.

For two DBA white dwarfs we rely on the 3D model atmospheres of \citet{cukanovaite18,cukanovaite19} to constrain the spectroscopic atmospheric parameters and hydrogen abundances with a fitting procedure similar to that of \citep{bergeron11}.

The spectra of DC and magnetic white dwarfs are not fitted. White dwarf candidates that were found to be main-sequence stars are not analysed further as it is outside of the scope of this paper, although we attempt to explain the reason why they contaminated the initial selection of \citet{gentile19} in the \textit{Gaia} DR2 HR diagram. 

\begin{figure}
    \centering
	\includegraphics[viewport= 1 230 600 600,scale=0.40]{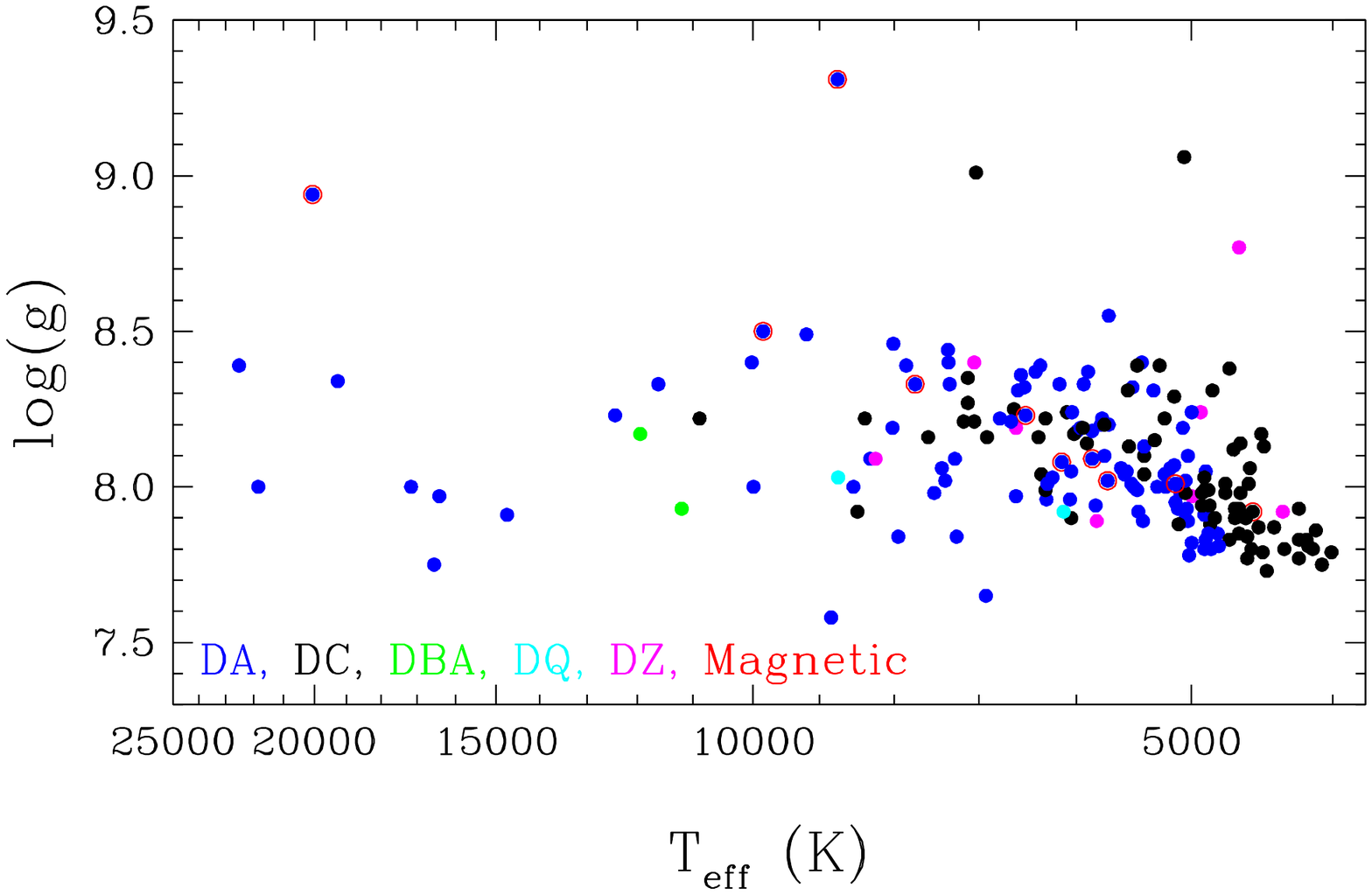}
	\caption{$\log g$ versus effective temperature distribution for 191 confirmed white dwarfs within 40\,pc based on Pan-STARRS photometric parameters. For 19 objects with missing or unreliable Pan-STARRS photometry, we relied on \textit{Gaia} parameters instead. The ultra-cool DZ WD J1922+0233 does not have reliable atmospheric parameters and is omitted from the figure. Spectral types are colour coded for DA (blue), DC (black), DBA (green), DQ (cyan) and DZ (magenta) white dwarfs. Magnetic stellar remnants have red contours.}
        \label{fig:distribution}
\end{figure}

\section{Results}
\label{results}

From the 230 objects observed, all 191 spectroscopically confirmed white dwarfs have spectral types in Table~\ref{tab:final_all}. These objects also have at least one set of either \textit{Gaia} or Pan-STARRS photometric atmospheric parameters (using either pure-H, pure-He or mixed H/He model atmospheres), and in the case of warmer DA and DBA white dwarfs, an independent set of atmospheric parameters from the analysis of the optical spectrum. The resulting Pan-STARRS photometric distribution of $\log g$ as a function of $T_{\rm eff}$ is shown in Fig.~\ref{fig:distribution}. We flag with an asterisk in the WD\,J name those objects where the parallax value is below 25 mas but for which 40\,pc membership is still possible within 1$\sigma$. Many of these objects may therefore be excluded from the sample by the upcoming third data release of \textit{Gaia} (DR3). 

\setcounter{table}{1}

\begin{table*}
	\centering
	    \scriptsize
        \caption{Spectral types and parameters of the white dwarf sample}
        \label{tab:final_all}
        \begin{tabular}{lllllllll}
                \hline
                WD\,J name  & SpT & $T_{\rm eff}$ [K] & $\log g$ & $T_{\rm eff}$ [K] & $\log g$ & $T_{\rm eff}$ & $\log g$ & Note\\
                &  & 3D Spectro &  		3D Spectro &  Gaia &  Gaia &  Pan-STARRS &  Pan-STARRS & \\
                \hline
                \hline
                001324.45+543757.64 			& DC   	& --		& --		& --		& -- 		& 4070 (20)	& 7.75 (0.02)	& \\
                001333.22$-$021319.42 		& DC   	& --		& --		& 4440 (90)	& 7.80 (0.07)	& 4590 (30)	& 7.90	(0.03)	& \\
                002116.21+253134.45			& DA	& 9680 (50) 	& 8.62 (0.04) 	& 9190	(170) 	& 8.49 (0.04)	& -- 		& -- 		& \\
                002450.37+683446.85 			& DC   	& --		& --		& 5520 (70)	& 8.20 (0.04)	& 5390	(20)	& 8.10 (0.02)	& \\
                002702.93+055433.40 			& DC   	& --		& --		& 5050 (70)	& 8.28 (0.06)	& 5260 (20)	& 8.39 (0.02)	& \\
                003047.74+034657.93			& DC   	& --		& --		 & 6390 (30)	& 8.07 (0.02)	& 6340	(40)	& 8.04	(0.04)	& \\
                005503.58+101005.56			& DA	& 6160 (100) 	& 7.60 (0.24) 	& 6190 (30) 	& 8.02 (0.02) 	& 6230 (10)	& 8.03 (0.01) 	&  \\
                010338.56$-$052251.96 		& DAH:    & -- 		& --		& 8960 (180)	& 9.34 (0.03)	& 8750 (70)	& 9.31 (0.01)	&  \\
                010416.07$-$035025.39		& DA	& -- 		& -- 		& 5280 (70)	& 8.29 (0.05)	& 5310 (20)	& 8.31 (0.02)	& \\
                012923.99+510846.97 			& DA	& 22550 (90) 	& 8.01 (0.01)	& 21850 (270)	& 8.00 (0.02)	& -- 		& -- 		& (a)\\
                013055.01+441423.29 			& DZA	& -- 		& -- 		& 4990 (40)	& 7.99 (0.04)	& 5000 (20)	& 7.97 (0.02)	& \\
                013705.08$-$020738.75 		& DA	& 7570 (40) 	& 8.35 (0.06)	& 7260 (80)	& 8.31 (0.03)	& 7330 (40)	& 8.33 (0.02)	& \\
                014258.08+073045.39 			& DA	& -- 		& -- 		& 5500 (70)	& 8.01 (0.05)	& 5560 (20)	& 8.04 (0.02)	& \\
                015825.83+253051.31 			& DC   	& --		& --		& -- 		& -- 		& 4220 (20)	& 7.77 (0.03) 	&\\
                020210.60+160203.31 			& DZ 	& -- 		& -- 		& 4760 (70)	& 8.13 (0.05)	& 4930 (20)	& 8.24 (0.02)	& \\
                020809.31+372939.12 			& DA	& -- 		& -- 		& 5560	(50)	& 8.50 (0.03)	& 5700 (25)	& 8.55 (0.02)	& \\
                021839.49+501351.28 			& DA	& -- 		& -- 		& 4840 (30)	& 7.76 (0.02)	& 4900 (20)	& 7.80 (0.01)	& \\
                022111.49+533330.39 			& DC   	& --		& --		& 6630 (50)	& 8.15 (0.03)	& 6050	(40)	& 7.90	(0.01)	& \\
                022157.89+044517.91 			& DA	& 7520 (100) 	& 8.24 (0.15)	& 7150 (100)	& 8.06 (0.04)	& 7270 (20)	& 8.09 (0.01)	& \\
                022704.24+591502.04 			& DA	& 7330 (40) 	& 7.75 (0.06)	& 7290	(50)	& 7.85 (0.02)	& 7250 (40)	& 7.84 (0.02)	& \\
                022724.62+180724.03 			& DA	& 8800 (20) 	& 8.03 (0.03)	& 8440	(90)	& 7.97 (0.05)	& 8530 (60)	& 8.00 (0.04)	& \\
                023117.04+285939.88 			& DA	& 7200 (20) 	& 7.97 (0.03)	& 6980	(30)	& 7.67 (0.02)	& 6920 (10)	& 7.65 (0.02)	& (b)\\
                025007.11+081753.42 			& DC   	& --		& --		& 4330 (90)	& 7.65 (0.08)	& 4550	(30)	& 7.80	(0.02)	& \\
                025328.32+375959.38 			& DA	& 6550 (60) 	& 7.70 (0.12)	& 6480	(40)	& 7.92 (0.03)	& 6600 (40)	& 7.97 (0.02)	& \\
                030350.56+060748.75 			& DXP & -- 		& --		& 20050 (7000)	& 8.94 (0.30)	& -- 		& -- 		& (c) \\
                030850.43+512822.32 			& DA	& -- 		& --		& 5090	(40)	& 7.92 (0.04)	& 5210 (20)	& 8.00 (0.02)	& \\
                031124.57$-$085324.98 		& DA	& -- 		& --		& 4920 (60)	& 7.79 (0.05)	& 5060 (20)	& 7.89 (0.03)	& \\
                031138.80$-$055117.55 		& DC   	& --		& --		& -- 		& -- 		& 4220 (10)	& 7.83	(0.01) 	&	\\
                031907.61+423045.45 			& DC   	& --		& --		& 11020 (80)	& 8.23 (0.02)	& 10880 (60)	& 8.22 (0.01)	& \\
                032631.46+155714.79 			& DA	& -- 		& --		& 5680	(70)	& 8.18 (0.05)	& 5760 (20)	& 8.22 (0.02)	& \\
                034501.53$-$034849.73 		& DC   	& --		& --		& -- 		& -- 		& 4390 (20)	& 7.87 (0.02) 	& (d)\\
                034501.70$-$034844.85 		& DA	& -- 		& --		& 4960 (50)	& 7.85 (0.05)	& 5030 (10)	& 7.89 (0.02)	& (d)\\
                034511.83+194026.08 			& DA	& 12780 (70) 	& 8.17 (0.02)	& 12450 (100)	& 8.23 (0.01)	& 12430 (80)	& 8.23 (0.01)	& \\
                035556.50+452510.26 			& DA	& -- 		& --		& 5060 (30)	& 7.91 (0.03)	& 5130 (20)	& 7.95 (0.02)	& \\
                035826.49+215726.16 			& DAZ 	& -- 		& -- 		&  6660 (30)	& 8.17 (0.02)	& 6770 (30)	& 8.22 (0.01)	& \\
                040242.39+152742.47 			& DC   	& --		& --		& 6850 (30)	& 8.14 (0.02)	& 6910 (30)	& 8.16 (0.02)	&  \\
                041246.85+754942.26 			& DA(e)	& -- 		& --		& 8510	(90)	& 8.25 (0.02)	& 8380 (50)	& 8.22 (0.01)	& \\
                042313.75+574526.76 			& DC   	& --		& --		& 7110 (40)	& 8.18 (0.02)	& 7170	(40)	& 8.21 (0.01)	& \\
                042731.73$-$070802.80 		& DC   	& --		& --		& 6850 (40)	& 8.15 (0.02)	& 7050 (40)	& 8.21 (0.02)	&  \\
               $*$ 052400.25$-$060402.72 		& DC 	& -- 		& -- 		& 6620 (110)	& 8.25 (0.05)	& -- 		& -- 		&  \\
                052436.27$-$053510.52 		& DA	& 17510 (80)	& 8.02 (0.02)	& 17300 (130)	& 8.03 (0.01)	& 16410 (150)	& 7.97 (0.01)	& \\
                052913.45+430025.89 			& DQ	& -- 		& -- 		& 8880	(80)	& 8.05 (0.02)	& 8740 (60)	& 8.03 (0.02)	& \\
                053026.01+393917.04 			& DA	& -- 		& --		& 5450 (50)	& 7.92 (0.04)	& 5440 (20)	& 7.92 (0.02)	& \\
                053714.90+675950.51     & DAH & -- & --	& 7750 (40)	& 8.33 (0.01)	& 7740 (30)	& 8.33 (0.01)		& (e) \\
                053916.45+435234.70 			& DC   	& --		& --		& 5910 (50)	& 8.14 (0.03)	& 5900	(20)	& 8.14 (0.02)	& \\
                054839.48+132551.76 			& DC   	& --		& --		& -- 		& -- 		& 4110 (60)	& 7.86 (0.05)	& \\
               $*$ 055231.03+164250.27 		& DBA 	& -- 		& -- 		& 11950 (90)	& 8.17 (0.02)	& -- 		& -- 		& \\
                055443.04$-$103521.34 		& DZ 	& -- 		& -- 		& 6570 (20)	& 8.17 (0.01)	& 6600 (20)	& 8.19 (0.01)	& \\
                061848.64+620425.54 			& DC   	& --		& --		& 4530 (50)	& 7.80 (0.04)	& 4670	(30)	& 7.90	(0.03)	& \\
                062006.01+420544.38 			& DA	& -- 		& --		& 6650 (70)	& 8.33 (0.03)	& 6580	(30)	& 8.31 (0.03)	& (f)\\
                063235.80+555903.12 			& DAH & --		& --		& 9970 (80)	& 8.52 (0.02)	& 9840 (40)	& 8.50 (0.01)	& \\
                063931.88+243546.15 			& DC   	& --		& --		& 8480 (110)	& 7.92 (0.04)	& -- 		& -- 		& \\
                064111.93$-$043212.31		& DC   	& --		& --		 & -- 		& -- 		& 4130 (20)	& 7.80 (0.01)	& \\
                064400.61+092605.76 			& DAH & --		& --		& 6080 (50)	& 8.05 (0.03)	& 6140 (30)	& 8.08 (0.02)	& \\
                064926.55+752124.97         & DAH & --       & --	& 6440 (40)	& 8.21 (0.02)	& 6500 (30)	& 8.23 (0.01)		& (g)\\ 
                065729.40+055047.87 			& DC   	& --		& --		& 6020 (60)	& 8.21 (0.04)	& 6090 (30)	& 8.24 (0.02)	 &	\\
                065845.23$-$011552.84 		& DA	& -- 		& --		& 4840 (90)	& 8.02 (0.08)	& 4890 (30)	& 8.05 (0.03)	& \\
                065910.86+122526.52 			& DA	& --		& --		& 4920 (60)	& 7.85 (0.05)	& 5040 (10)	& 7.93 (0.02)	& \\
                070356.98+780504.72 			& DA	& --		& --		& 5340	(20)	& 7.86	(0.02)	& 5400 (10)	& 7.89 (0.01)	& \\
                070357.43+253418.34 			& DBA	&11710 (180)	& 8.01 (0.14) 	& 11500 (100)	& 7.97 (0.02)	& 11190 (50)	& 7.93 (0.01)	& \\
                070549.32+514250.52 			& DA	& --		& --		& 4970 (90)	& 8.14 (0.08)	& 5070 (20)	& 8.19 (0.02)	& \\
                071206.15$-$042815.30 		& DA	& -- 		& --		& 5280 (60)	& 8.06 (0.04)	& 5390 (20)	& 8.13 (0.02)	& \\
                071703.10+112541.55 			& DA	& -- 		& --		& 4690 (40)	& 7.78 (0.03)	& 4800 (20)	& 7.85 (0.02)	& \\
                072205.61+280626.98 			& DA	& -- 		& -- 		& 5200 (70)	& 8.03	(0.06)	& 5220 (20)	& 8.04 (0.02)	& \\
                072300.22+161703.98 			& DA	& 11760 (80)	& 8.29 (0.02)	& 11580 (140)	& 8.31 (0.02)	& 11610 (110)	& 8.33 (0.01) 	&\\
                072434.96$-$133828.38 		& DA	& -- 		& -- 		& 4940 (50)	& 8.04 (0.05)	& 5030 (70)	& 8.10 (0.05)	& \\
                073024.53+533211.95 			& DC   	& --		& --		& 4530 (60)	& 7.86 (0.06)	& 4740 (30)	& 8.01 (0.03)	 &	\\
                075252.85$-$030707.97 		& DC   	& --		& --		& -- 		& -- 		& 4470 (40)	& 7.79 (0.03)	& \\
                                $*$ 080247.02+564640.62 		& DC 	& -- 		& -- 		& 4320 (50)	& 7.80 (0.06)	&  -- 		& -- 		& \\
                081325.13+195729.18 			& DC   	& --		& --		& --		& -- 		& 4010 (30)	& 7.79 (0.03)	& \\
                082532.35$-$072823.21 		& DA	& 15560 (110)	& 7.97 (0.02)	& 15550 (90)	& 7.97 (0.01)	& 14750 (170)	& 7.91 (0.01)	& \\
                083150.62$-$164329.97 		& DC   	& --		& --		& -- 		& -- 		& 4220 (40)	& 7.93 (0.04)	& \\
             \hline
        \end{tabular}
        \\
        Notes: (a) Also in \citet{scholz18}, (b) Double degenerate candidate, (c) Also in \citet{landstreet2020}, (d) Wide double white dwarf, (e) DAH: in \citet{limoges15}, (f) likely He-rich white dwarf, (g) DA in \citet{limoges15}, (h) D: in \citet{leggett18}, (i) FAST spectrum has poor S/N, Balmer line fit uncertain, (j) Main-sequence star in \citet{reid04}, (k) WD in \citet{greenstein84}, (l) DC in \citet{oswalt94}, (m) DA in \citet{motch98}, (n) Also in \citet{landstreet2019}, (o) DC: in \citet{limoges15}. Objects with an asterisk before their name have a parallax value outside of 40\,pc but may still be within that volume at 1$\sigma$. For all photometric fits we have used either pure-H (DA, DAZ, DAH and DC coooler than 5000\,K), pure-He (He-rich DA and DC in the range 5000\,K $< T_{\rm eff} <$ 7000\,K) or mixed [H/He] = $-5$ model atmospheres (DZ, DZH, DZA, DQ, DBA and DC warmer than 7000\,K).
\end{table*}

\setcounter{table}{1}

\begin{table*}
\centering
\scriptsize
        \caption{Spectral types and parameters of the white dwarf sample (continued)}
        \begin{tabular}{llllllllll}
                \hline
                WD\,J name  & SpT & $T_{\rm eff}$ [K] & $\log g$ & $T_{\rm eff}$ [K] & $\log g$ & $T_{\rm eff}$ & $\log g$ & Note\\
                &  & 3D Spectro &  		3D Spectro &  Gaia &  Gaia &  Pan-STARRS &  Pan-STARRS & \\
                \hline
                \hline
                084515.55+611705.79 			& DA	& -- 		& -- 		& 5470	(30)	& 8.02	(0.02)	& 5540 (20)	& 8.05 (0.02)	& (d)\\
                084516.87+611704.81 			& DAH & --		& --		& 5820 (40)	& 8.09 (0.02)	& 5850 (20)	& 8.09 (0.01)	& (d)\\
                084957.48$-$015612.38 		& DC   	& --		& --		& 4820 (20)	& 7.89 (0.02)	& 4880 (10)	& 7.93 (0.01)	& \\
                085534.72$-$083345.34 		& DC   	& --		& --		& 4500 (60)	& 7.75 (0.06)	& 4640 (10)	& 7.85 (0.02)	& \\
                085804.42+681338.66 			& DC   	& --		& --		& 4580 (60)	& 7.94 (0.06)	& 4630 (10)	& 7.98 (0.02)	& \\
                090912.98$-$224625.86 		& DC   	& --		& --		& 4510 (100)	& 7.98 (0.09)	& 4570 (40)	& 8.01 (0.04)	& \\
                091353.95+620601.69 			& DA	& -- 		& -- 		& 5650	(30)	& 8.06	(0.02)	& 5740 (20)	& 8.10 (0.01)	& \\
                092542.84+612012.85 			& DA	& 8200 (135)	& 8.65 (0.17)	& 8010 (30)	& 7.85 (0.01)	& 7950 (30)	& 7.84 (0.01)	& (i) \\
                093948.69$-$145836.69 		& DC   	& --		& --		& -- 		& -- 		& 4170 (20)	& 7.83 (0.02)	& \\
                095447.49+670208.00 			& DA	& -- 		& -- 		& 5700 (20)	& 8.20	(0.01)	& 5700 (20)	& 8.20 (0.01)	& \\
                100424.18$-$050614.92 		& DC   	& --		& --		& -- 		& -- 		& 4160 (20)	& 7.81 (0.02)	& (j)\\
               $*$  102203.66+824310.00			& DA 	& -- 		& -- 		& 5490 (70)	& 8.32 (0.04)	& -- 		& -- 		& \\
                103055.44$-$142400.53 		& DA	& 5980 (80)	& 7.89 (0.19)	& 6040 (50)	& 8.21 (0.03)	& 5990 (20)	& 8.18 (0.01)	& \\
                111913.56$-$083137.22 		& DA	& -- 		& -- 		& 5530 (30)	& 8.04 (0.02)	& 5590 (20)	& 8.06 (0.02)	& \\
                113444.64+610826.68 			& DAZ 	& 7610 (50) 	& 8.03 (0.08) 	& 7440 (40)	& 7.95 (0.01)	& 7510 (20)  	& 7.98 (0.01) 	& \\
                113840.67$-$131338.55 		& DC   	& --		& --		& 5940 (20)	& 8.15 (0.01)	& 6020 (30)	& 8.17 (0.02)	& \\
                120055.89$-$103220.61 		& DA	& 8070 (60)	& 8.40 (0.08)	& 7950 (50)	& 8.44 (0.02)	& 8010 (30)	& 8.46 (0.01)	& \\
                121701.84+684851.45 			& DA	& 6440 (110)	& 8.25 (0.20)	& 6420	(40)	& 8.38 (0.02)	& 6400 (60)	& 8.37 (0.02)	& (k) \\
                122956.02$-$070727.57 		& DA	& -- 		& -- 		& 4950 (40)	& 7.80 (0.03)	& 5000 (20)	& 7.82 (0.02)	& \\
                124828.17$-$102857.82 		& DC   	& --		& --		& 7130 (70)	& 8.27 (0.03)	&  7120 (30)	& 8.27 (0.01)	& \\
                130503.44+702243.05 			& DC   	& --		& --		& 4860 (280)	& 8.99 (0.14)	& 5060 (170)	& 9.06 (0.07)	& \\
                135509.42$-$262248.95 		& DA	& 5922 (120)	& 7.72 (0.25)	& 6080 (40)	& 8.06 (0.02)	& 6050 (30)	& 8.05 (0.01)	& \\
               $*$  140841.83$-$264948.55 		& DC 	& -- 		& -- 		& 7120 (90)	& 8.35 (0.04)	& -- 		& -- 		& \\
                144318.17$-$143715.32 		& DA	& 6570 (60)	& 7.88 (0.11)	& 6610 (50)	& 8.20 (0.02)	& 6650 (40)	& 8.21 (0.02)	& (a)\\
                144528.12+292124.29 			& DA	& -- 		& -- 		& 5350	(20)	& 7.94 (0.02)	& 5450 (20)	& 7.99 (0.02)	& \\
                151534.80+823028.99 			& DZH 	& -- 		&  -- 		& 4360 (80)	& 7.80 (0.06)	& 4540 (80)	& 7.92 (0.06)	&  \\
                160041.14$-$165430.24 		& DC   	& --		& --		& -- 		& -- 		& 4460 (80)	& 8.13 (0.06)	& \\
                160415.07$-$072658.01 		& DC   	& --		& --		& 4770 (60)	& 8.47 (0.04)	& 4710 (40)	& 8.38 (0.02)	& \\
                160420.47$-$133123.84 		& DA	& -- 		& -- 		& 4960 (40)	& 7.76 (0.04)	& 5020 (20)	& 7.78 (0.01)	& \\
                160606.17+702226.94 			& DA	& -- 		& -- 		& 6290 (40)	& 7.97 (0.02)	& 6290 (40)	& 7.96 (0.03)	& (a)\\
                160700.89$-$140423.88		& DAH & --		& --		& 5700 (30)	& 8.05 (0.02)	& 5710 (30)	& 8.02 (0.02)	& \\
                161330.58+442754.13 			& DA	& -- 		& -- 		& 5280 (140)	& 8.00 (0.10)	& -- 		& -- 		& \\
                161916.31$-$183114.19 		& DA	& -- 		& -- 		& 5340 (90)	& 8.34 (0.06)	& 5410 (20)	& 8.40 (0.02)	& \\
                162125.64$-$055219.84 		& DC   	& --		& --		& 4570 (70)	& 7.74 (0.06)	& 4710 (20)	& 7.83 (0.03)	& \\
                162818.90$-$173917.89 		& DA	& -- 		& -- 		& 4820 (40)	& 7.79 (0.03)	& 4890 (20)	& 7.83 (0.02)	& \\
                164951.45$-$215503.96 		& DA	& -- 		& -- 		& 5040 (40)	& 7.94 (0.04)	& 5090 (20)	& 7.98 (0.02)	& \\
                170438.32$-$144620.79 		& DA	& 7520 (40)	& 8.50 (0.06)	& 7440 (100)	& 8.42 (0.03)	& 7340 (30)	& 8.40 (0.01)	& \\
                170502.87$-$014502.70 			& DC   	& --		& --		& 4730 (40)	& 7.97 (0.03)	& 4740 (10)	& 7.98 (0.01)	& \\
              170552.58+260551.20 			& DA	& -- 		& -- 		& 5950 (40)	& 8.20 (0.02)	& 6040 (30)	& 8.24 (0.02)	& (l)  \\
              171620.72$-$082118.60 		& DQ	&-- 		& -- 		& 6000 (90)	& 7.86 (0.05)	& 6120 (20)	& 7.92 (0.02)	& \\
                172006.79+102227.98 			& DC   	& --		& --		& 4940 (60)	& 7.90 (0.05)	& 5050 (20)	& 7.98 (0.02)	& \\
                172945.19+143541.28 			& DC   	& --		& --		& 4730 (50)	& 8.25 (0.04)	& 4840 (20)	& 8.31 (0.02)	& \\
                173337.18+290338.04 			& DC   	& --		& --		& 6340 (40)	& 8.15 (0.02)	& 6370 (20)	& 8.16	(0.01)	& \\
                173404.42+442303.09 			& DA	& -- 		& -- 		& 5000 (60)	& 7.99 (0.05)	& 5140 (20)	& 8.07 (0.02)	& \\
                174620.41$-$123425.48 		& DA	& -- 		& -- 		& 5970 (70)	& 8.24 (0.04)	& 6160 (30)	& 8.33 (0.02)	& \\
                174935.61$-$235500.63 		& DAZ 	& 7200 (30) 	& 7.75 (0.06) 	& 7000 (50)	& 7.87 (0.02)	& 7420 (40)	& 8.06 (0.01) 	& \\
                175352.16+330622.62 			& DA	& 17480 (80)	& 8.06 (0.01)	& 17120 (110)	& 7.99 (0.01)	& 17160 (150)	& 8.00 (0.01)	& \\ 
                175919.66+392504.95 			& DC   	& --		& --		& 4540 (70)	& 7.82 (0.05)	& 4580 (20)	& 7.84 (0.02)	& \\
              $*$  180218.60+135405.46			& DAZ 	& 8570 (30)	& 8.39 (0.05)	& 8310 (110)	& 8.09 (0.04)	& -- 		& -- 		& \\	
                180919.46+295720.85 			& DA	& 23280 (70)	& 8.40 (0.01)	& 22515 (160)	& 8.39 (0.01)	& -- 		& -- 		& (a) \\
                181539.13$-$114041.83 		& DC   	& --		& --		& 4830 (70)	& 7.86 (0.06)	& 4855 (20)	& 7.88 (0.02)	& \\
                181745.57$-$133531.54 		& DA	& 6290 (220)	& 8.76 (0.36)	& 6050 (70)	& 8.38 (0.03)	& 5930 (40)	& 8.33 (0.02)	& \\
                181909.96$-$193438.00 		& DC   	& --		& --		& -- 		& -- 		& 4500 (30)	& 7.87 (0.03)	& \\
                182021.81+261936.58 			& DA	& -- 		& -- 		& 4890 (70)	& 8.19 (0.06)	& 5000 (20)	& 8.24 (0.02)	& \\
                182147.11+550906.70 			& DA	& -- 		& -- 		& 4890 (60)	& 7.86 (0.05)	& 4870 (50)	& 7.85 (0.04)	& \\
                182347.60$-$112347.38 		& DA	& -- 		& -- 		& 5560 (110)	& 8.01 (0.07)	& 5850 (30)	& 8.18 (0.02)	& \\
                182359.62+202248.81 			& DC   	& --		& --		& 4950 (40)	& 8.01 (0.04)	& 4920	(20)	& 7.98	(0.02)	& \\
                182417.72+120945.86 			& DA	& -- 		& -- 		& 5160 (70)	& 8.05 (0.05)	& 5170 (20)	& 8.06 (0.01)	& \\
                182458.45+121316.82 			& DZ 	& -- 		& -- 		& -- 		& -- 		& 4330 (31)	& 7.92 (0.04)	& \\
                182524.24+113557.34 			& DA	& -- 		& -- 		& 4850 (30)	& 7.87 (0.02)	& 4900	(10)	& 7.91 (0.01)	& \\
                182624.44+112049.58 			& DA	& -- 		& -- 		& 4860 (50)	& 7.85	(0.04)	& 4850 (10)	& 7.80 (0.01)	& \\
                182951.89$-$053623.17 		& DA	& -- 		& -- 		& 5450 (60)	& 7.97 (0.04)	& 5480 (30)	& 8.00 (0.02)	& (d)\\
                182952.07$-$053622.88 		& DC   	& --		& --		& 6490 (80)	& 8.07 (0.04)	& 6300 (30)	& 7.99 (0.03)	& (d) \\
                183158.72+465828.98 			& DA	& 7650 (20)	& 8.11 (0.03)	& 7380 (30)	& 8.02 (0.01)	& 7380 (30)	& 8.02 (0.01)	& \\
                183518.23+642117.68 			& DC   	& --		& --		& 4860 (30)	& 7.99 (0.04)	& 4870 (20)	& 7.99 (0.02)	& \\
                183352.68+321757.25			& DZA 	& -- 		& -- 		& 7540 (100)	& 7.88 (0.04)	& 8240 (50)	& 8.09 (0.02)	& \\
               $*$  184733.18+282057.54 		& DC 	& -- 		& -- 		& 4630 (70)	& 8.14 (0.07)	& -- 		& -- 		&  \\
                184741.53+122631.75 			& DA	& 10450 (30)	& 8.42 (0.02)	& 10020 (150)	& 8.40 (0.03)	& -- 		& -- 		& \\
                184907.50$-$073619.82 		& DC   	& --		& --		& 6190 (110)	& 8.15 (0.06)	& 6300 (20)	& 8.22 (0.02)	& \\
                185517.99+535923.18 			& DC   	& --		& --		& 4610 (50)	& 7.88 (0.04)	& 4670 (10)	& 7.93 (0.01)	& \\
                191246.12+024239.11 			& DZ 	& -- 		& -- 		& 6280 (40)	& 8.11 (0.02)	& 7050 (50)	& 8.40 (0.02)	& \\
                192126.76+061322.71 			& DA	& -- 		& -- 		& 5880 (20)	& 8.14	(0.01)	& 5960 (10)	& 8.19 (0.01)	&  \\
               $*$  192206.20+023313.29			& DZ  	& -- 		& -- 		& 5800 (390)	& 9.10 (0.02)	& -- 		& -- 		& \\
                192359.24+214103.62 			& DA	& 9280 (20)	& 8.06 (0.02)	& 8750 (50)	& 7.55 (0.02)	& 8840 (40)	& 7.58 (0.02)	& (b)\\
                192626.93+462015.10 			& DA	& 8170 (30)	& 8.30 (0.04)	& 8130 (50)	& 8.21 (0.02)	& 8020 (30)	& 8.19 (0.02)	& \\
                192724.75+564455.34 			& DA	& 6750 (70)	& 8.45 (0.12)	& 6530 (60)	& 8.36 (0.02)	& 6550 (20)	& 8.36 (0.01)	& \\
                192938.65+111752.41 			& DA	& 21130 (90)	& 8.00 (0.01)	& 20220 (300)	& 7.94 (0.02)	& 16550 (370)	& 7.75 (0.03)	& \\
                193019.71$-$005730.56 		& DC   	& --		& --		& 7620 (80)	& 8.17 (0.03)	& 7580 (30)	& 8.16 (0.01)	& \\
                \hline
        \end{tabular}
        \\
\end{table*}

\setcounter{table}{1}

\begin{table*}
\centering
\scriptsize
        \caption{Spectral types and parameters of the white dwarf sample (continued)}
        \begin{tabular}{llllllllll}
                \hline
                WD\,J name  & SpT & $T_{\rm eff}$ [K] & $\log g$ & $T_{\rm eff}$ [K] & $\log g$ & $T_{\rm eff}$ & $\log g$ & Note\\
                &  & 3D Spectro &  		3D Spectro &  Gaia &  Gaia &  Pan-STARRS &  Pan-STARRS & \\
                \hline
                \hline
               $*$  193500.68$-$172443.11 		& DC 	& -- 		& -- 		& 4480 (60)	& 8.17 (0.02)	& -- 		& -- 		& \\
                193618.58+263255.79 			& DA	& 25220 (90)	& 8.54 (0.01)	& 24380 (170)	& 8.53 (0.01)	& 19270 (300)	& 8.34 (0.02)	& (m) \\
                193955.83+661856.08 			& DC   	& --		& --		& 5070 (40)	& 8.14 (0.03)	& 5220 (20)	& 8.22 (0.02)	& (k)\\
                195003.62+003357.09 			& DA	& -- 		& -- 		& 5800 (40)	& 7.93 (0.03)	& 5820 (20)	& 7.94 (0.01)	& \\
                195119.15+420941.40 			& DA	& -- 		& -- 		& 5050 (50)	& 8.02 (0.05)	& 5050 (20)	& 8.02 (0.02)	& \\
                195151.76+402629.07 			& DC   	& --		& --		& 5050 (80)	& 8.22 (0.06)	& 5140 (30)	& 8.29 (0.02)	& \\
                200445.49+010929.21 			& DA	& 6770 (60)	& 8.30 (0.10)	& 6510 (70)	& 8.32 (0.03)	& -- 		& -- 		& \\
                200632.25$-$210142.90 		& DA	& --		& -- 		& 5070 (60)	& 7.90 (0.05)	& 5110 (20)	& 7.93 (0.02)	& \\
                200850.81$-$161943.62 		& DC   	& --		& --		& 5470 (70)	& 8.28 (0.04)	& 5530 (20)	& 8.31 (0.02)	& \\
                201216.01$-$221023.03 		& DC   	& --		& --		& 5500 (70)	& 8.12 (0.04)	& 5520 (10)	& 8.13 (0.01)  	&	\\
                201530.35+000111.80 			& DC   	& --		& --		& 4760 (40)	& 7.92 (0.03)	& 4900 (20)	& 8.03 (0.02)	& \\
                202157.83+545438.25 			& DC   	& --		& --		& 5940 (60)	& 8.18 (0.03)	& 5940 (30)	& 8.19 (0.02)	& \\
                203100.56$-$145041.38 		& DC   	& --		& --		& 4460 (90)	& 8.03 (0.08)	& 4560 (20)	& 8.06 (0.03)	& \\
                203102.15+393454.05 			& DC   	& --		& --		& 4530 (70)	& 7.87 (0.06)	& 4640	(20)	& 7.93 (0.03)	& \\
                203321.86+395409.76 			& DA	& -- 		& -- 		& 5930 (20)	& 8.40 (0.01)	& 5890 (20)	& 8.37 (0.01)	& \\
                204832.02+511047.25 			& DC   	& --		& --		& 4840 (80)	& 7.92 (0.07)	& 4860 (20)	& 7.94 (0.02)	& \\
                210646.77+010635.24 			& DA	& 6250 (120)	& 8.15 (0.22)	& 6030 (40)	& 7.94 (0.03)	& 6060 (20)	& 7.96 (0.02)	& \\
                210854.87$-$031202.98 		& DC   	& --		& --		& 5230 (50)	& 7.93 (0.04)	& 5390 (20)	& 8.04 (0.02)	& \\
                213343.70+241457.72 			& DA	& 6650 (80)	& 8.53 (0.14)	& 6320 (40)	& 8.40 (0.02)	& 6350 (30)	& 8.39 (0.01)	& \\
                213517.95+463318.21 			& DA	& -- 		& -- 		& 4900 (100)	& 7.92 (0.09)	& 4790 (30)	& 7.81 (0.02)	& (l)\\
                213957.13$-$124549.09 		& DA	& 7780 (40)	& 8.25 (0.06)	& 7730 (160)	& 8.36 (0.05)	& 7850 (180)	& 8.39 (0.05)	& \\
                215008.33$-$043900.36		& DC   	& --		& --		 & 5320 (80)	& 8.32 (0.06)	& 5450 (30)	& 8.39 (0.03)	& \\
                215140.11+591734.85 			& DAH & --		& --		& 5100 (10)	& 7.98	(0.01)	& 5130 (10)	& 8.01 (0.01)	& (n)\\
                215759.55+270519.13 			& DC   	& --		& --		& 4500 (60)	& 7.73 (0.05)	& 4580 (20)	& 7.77 (0.02)	& \\
                215839.14$-$023916.44		& DC   	& --		& --		 & 4780 (40)	& 7.84 (0.03)	& 4920 (10)	& 7.94 (0.02)	& (d)\\
                215847.13$-$024024.42 		& DC   	& --		& --		& 4730 (50)	& 7.83 (0.04)	& 4820 (10)	& 7.90 (0.02)	& (d)\\
                220052.62+582202.29 			& DA	& -- 		& -- 		& 5380	(50)	& 7.94 (0.04)	& 5500 (30)	& 8.01 (0.02)	& \\
                220253.65+023741.53 			& DA	& -- 		& -- 		& 5750 (50)	& 8.20 (0.04)	& 5770 (20)	& 8.20 (0.02)	& \\
                220751.81+342845.79 			& DA	& 10190 (30)	& 8.05 (0.03)	& 10070 (40)	& 8.01 (0.01)	& 9990 (20)	& 8.00 (0.01)	& \\
                221800.59+560214.92			& DC   	& --		& --		 & -- 		& -- 		& 4440 (20)	& 7.73 (0.02)	& \\
                222547.07+635727.37			& DC   	& --		& --		 & 5030 (50)	& 7.81 (0.03)	& 5103	(20)	& 7.88 (0.02)	& \\
                223059.16+225454.09 			& DC   	& --		& --		& 5720 (30)	& 8.20 (0.02)	& 5740 (20)	& 8.20 (0.01)	& \\
                223418.83+145654.42 			& DA	& 6380 (60)	& 7.96 (0.12)	& 6290 (50)	& 8.02 (0.03)	& 6280 (20)	& 8.01 (0.01)	&\\
                225257.98+392817.40 			& DC   	& --		& --		& 4830 (50)	& 7.94 (0.05)	& 4900 (20)	& 7.99 (0.02)	& \\
                225338.11+813039.98 			& DC   	& --		& --		& 5300 (80)	& 8.15 (0.06)	& -- 		& -- 		& (o) \\
                225725.27+513008.56 			& DA	& 7260 (50)	& 8.30 (0.07)	& 7210 (310)	& 8.41 (0.10)	& 7350 (50)	& 8.44 (0.02)	& \\
                230056.46+640815.95 			& DC   	& --		& --		& 4530 (70)	& 7.85 (0.06)	& 4590 (20)	& 7.90 (0.02)	& \\
                230303.62+463241.98 			& DC   	& --		& --		& 4580 (70)	& 8.05 (0.06)	& 4680	(30)	& 8.12 (0.03)	& \\
                230550.09+392232.88 			& DC   	& --		& --		& 6550 (120)	& 8.88 (0.04)	& 7030 (60)	& 9.01 (0.02)	& \\
                231726.74+183052.75 			& DZ	& -- 		& -- 		& 4600 (220)	& 8.78 (0.13)	& 4640 (40)	& 8.77 (0.04)	&  \\
               $*$  235750.73+194905.90 		& DZ 	& -- 		& -- 		& 5810 (50)	&7.89 (0.04)	& -- 		& -- 		&  \\
                \hline
        \end{tabular}
        \\
\end{table*}

All white dwarfs with traces of metals or carbon, and for which we have performed a combined fit of photometry and spectroscopy, are also listed in Tables~\ref{tab:marksfit_H} (hydrogen-dominated atmospheres with metals), \ref{tab:marksfit_He} (helium-dominated atmospheres with metals) and \ref{tab:marksfit_DQ} (DQ white dwarfs), where our best estimates of atmospheric parameters and chemical abundances are found. All 39 main-sequence star contaminants are discussed in Section \ref{MS}.

\subsection{DA white dwarfs}
\label{DAsec}

The spectra for all 89 DA white dwarfs are shown in Figs.~\ref{fig:DA1}--\ref{fig:DA4}. In most cases the Balmer lines are too weak for a meaningful fit. However for the subset of 40 objects with {\it Gaia} $T_{\rm eff} \geq$  6000\,K (excluding two He-rich DA white dwarfs), fits are presented in Figs.~\ref{fig:fitsDA1}-\ref{fig:fitsDA2}, with best fit atmospheric parameters corrected for 3D convection \mbox{\citep{tremblay13c}} identified in Table~\ref{tab:final_all}. \textit{Gaia} photometric temperatures are systematically lower by 2.4\% compared to the spectroscopic values, an offset very similar to that previously identified in \citet{tremblay19} and \citet{genest-beaulieu19a} from larger spectroscopic samples. Results with Pan-STARRS are very similar, with photometric temperatures systematically lower by 2.2\% (excluding two hot white dwarfs in crowded fields). Considering external errors on spectroscopic fits \citep{liebert05}, in the majority of individual cases the spectroscopic and photometric solutions agree within 2$\sigma$. For a number of cool objects with $T_{\rm eff} < 7\,000$\,K, the spectroscopic parameters are likely reliable but with a precision well below the photometric parameters. 

\begin{figure}
    \centering
	\includegraphics[viewport= 20 1 600 750,scale=0.22]{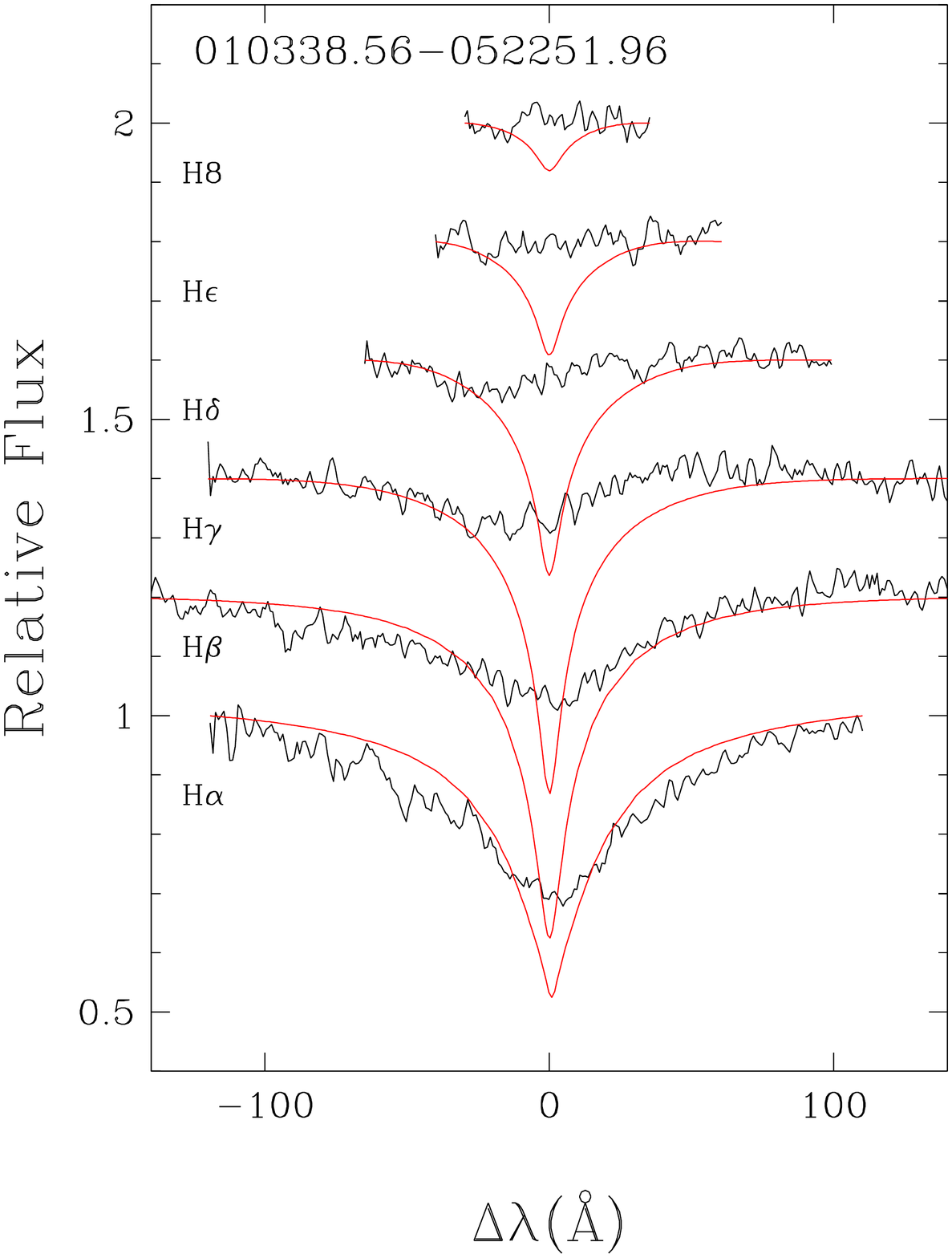}
	\hskip -0.95cm
	\includegraphics[viewport= 20 1 600 750,scale=0.22]{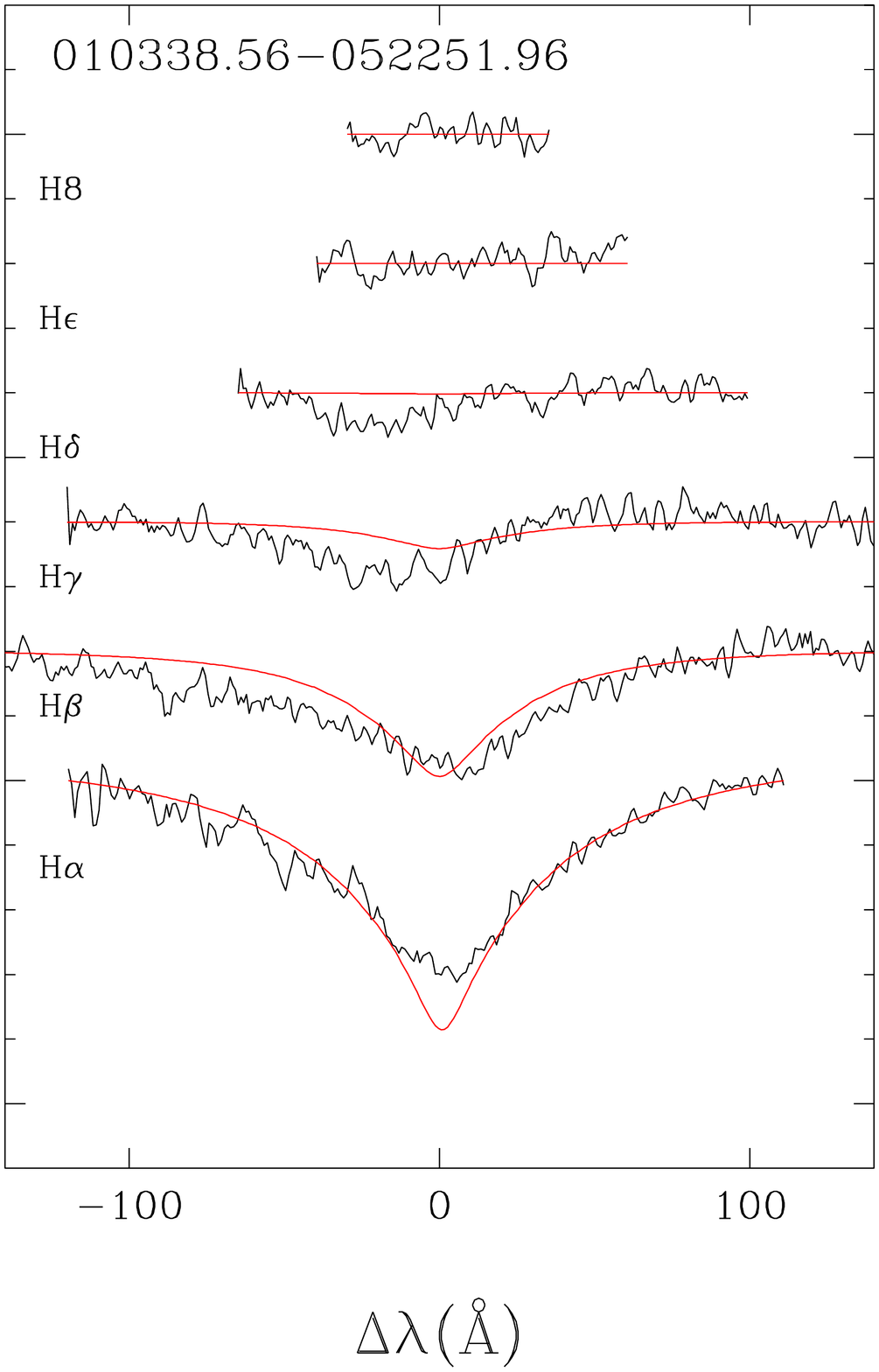}
	\caption{(Left:) Comparison of the normalised observed Balmer line profiles for WD\,J0103$-$0522 with a DA model atmosphere at the \textit{Gaia} derived photometric parameters ($T_{\rm eff}$ = 8960\,K and $\log g$ = 9.34). (Right:) Fit of the observations using He-rich model atmospheres with [H/He] = $-1.70$. The resulting best fit parameters ($T_{\rm eff}$ = 8520\,K and $\log g$ = 9.06), excluding the constraint from parallax, should be taken with caution given the relatively poor quality of the fit and we consider the photometric parameters more reliable.}
        \label{fig:WDJ0103}
\end{figure}

For all objects with {\it Gaia} $T_{\rm eff} <$ 6000\,K, in Figs.~\ref{fig:HalphaDA1}-\ref{fig:HalphaDA4} we compare instead the H$\alpha$ line profile with the prediction from the best \textit{Gaia} photometric fit identified in Table~\ref{tab:final_all}. In doing so we have identified a systematic shift where predicted H$\alpha$ equivalent widths (or line strengths) are systematically too small compared to the observed lines. In other words, \textit{Gaia} colours are systematically too red resulting in temperatures that are too low by 2.7\%. Pan-STARRS is only marginally better and predicted temperatures are still too low by 1.8\%. Neutral broadening dominates in cool DA stars \citep{tremblay10} and the predicted equivalent width of H$\alpha$ depends on $T_{\rm eff}$ and $\log g$ but the influence of non-ideal gas effects is very weak. 

The offset between observed and predicted H$\alpha$ is of very similar amplitude and in the same direction as the issue identified above between photometric temperatures and spectroscopic temperatures from Balmer line fits of warmer DA white dwarfs, where Stark broadening dominates. This raises doubts that issues with the current implementation of Stark broadening theory \citep{tremblay2009} is the source of the offset for hotter DA white dwarfs, as there would be no reason for the similar observed pattern at cool temperatures. Furthermore, there is no clear evidence of an offset between predicted and observed \textit{Gaia} or Pan-STARRS absolute magnitudes in warm DA white dwarfs \citep{tremblay19}, suggesting that the issue is with observed colours rather than a constant shift in photometric zero points. A possible explanation is that the photometric colour calibration of \textit{Gaia} and Pan-STARRS is the source of the offset \citep{maiz18} and that the hot DA white dwarf spectroscopic temperature scale remains adequate \citep{narayan2019,gentile2020}. In  Figs.~\ref{fig:HalphaDA1}-\ref{fig:HalphaDA4} we have corrected the \textit{Gaia} $T_{\rm eff}$ values by +2.7\% to demonstrate the good agreement with observed spectroscopy and to flag outliers. We now discuss peculiar DA stars in turn.

\textbf{WD\,J0103$-$0522} is one of the most massive known white dwarfs with a photometric $\log g$ = 9.34 $\pm$ 0.03, corresponding to a mass of 1.31 $\pm$ 0.01 M$_{\odot}$. Within 40\,pc, only the previously known pure-hydrogen atmosphere DA star WD\,2349$-$031 has a larger mass based on {\it Gaia} photometry \citep{gentile19}. In comparison, Balmer lines are clearly detected in WD\,J0103$-$0522 but have an unusual asymmetric profile with line centres shifted towards the blue (Fig.~\ref{fig:WDJ0103}). Each member of the Balmer series corresponding to a transition to upper level $n$ is progressively more blue-shifted compared to the line that precedes with a transition to the level $n-1$, which is inconsistent with a large radial velocity. Furthermore the lines are much broader and shallower than expected for a pure-hydrogen white dwarf at the \textit{Gaia} temperature of 8960 $\pm$ 180~K.  The object is strikingly similar to PG\,1157+004 ($T_{\rm eff} = 9425$ $\pm$ 50\,K and $\log g$ = 8.66 $\pm$ 0.01; \citealt{gentile19}) from figure 17 of \citet{limoges15}, also a member of the 40\,pc sample. These authors had flagged the star as a double degenerate candidate, but given that both PG\,1157+004 and WD\,J0103$-$0522 are very massive, this appears unlikely. Another possibility is that the atmosphere is a mixture of helium and hydrogen, and that neutral helium broadening in the dense atmosphere is responsible for disrupting the line profiles. Fig.~\ref{fig:WDJ0103} (right panel) demonstrates that a helium-dominated atmosphere with [H/He] $\approx$ $-1.70$ best reproduces the lower Balmer line equivalent widths at the {\it Gaia} temperature, but is still a rather poor fit to the line asymmetries, wavelength shifts and Balmer line decrement. 

The presence of a magnetic field is common (25--50\%) for {\it Gaia} white dwarfs with $>$1M$_{\odot}$ \mbox{\citep[see the SDSS-\textit{Gaia} catalogue of][]{gentile19}.} One possibility is that stellar magnetism is responsible for the unusual line shapes. While we do not observe Zeeman splitting, the asymmetry and shifts of the Balmer series can be explained as quadratic Zeeman effect of the $\pi$-component. The absence of $\sigma$-components can arise from a complex field geometry that leaves these them washed out: for example an offset-dipole configuration where the offset is towards the observer, resulting in a broad distribution of field strengths across the visible hemisphere of the star.

To investigate this possibility, we considered that the quadratic Zeeman shift of each atomic energy level has an $n^4$ dependence, where $n$ is the principle quantum number. This implies that each member of the Balmer series will be progressively more blue-shifted than the one that precedes it (as appears to be the case in \mbox{Fig.~\ref{fig:WDJ0103}}).

While we stop short of a full analysis of the field geometry, we show that the flux minimum of each Balmer line (H$\alpha$--H$\delta$) follows the expected pattern of wavelength shifts. For the purpose of constraining the location of the Balmer line centers, $\lambda_0$, we fit each with an asymmetric Lorentzian profile

\begin{equation}
    L(\lambda, \lambda_0) = \exp\left(-A\left[1 + \left(\frac{\lambda-\lambda_0}{\Gamma(\lambda, \lambda_0)}\right)^2\right]^{-1}\right),
\end{equation}
where $A$ is fit parameter scaling the line depth, and
\begin{equation}
    \Gamma(\lambda, \lambda_0) = a\left[1+\tanh(b(\lambda-\lambda_0)) \right],
\end{equation}

\begin{figure}
    \centering
	\includegraphics[width=\columnwidth]{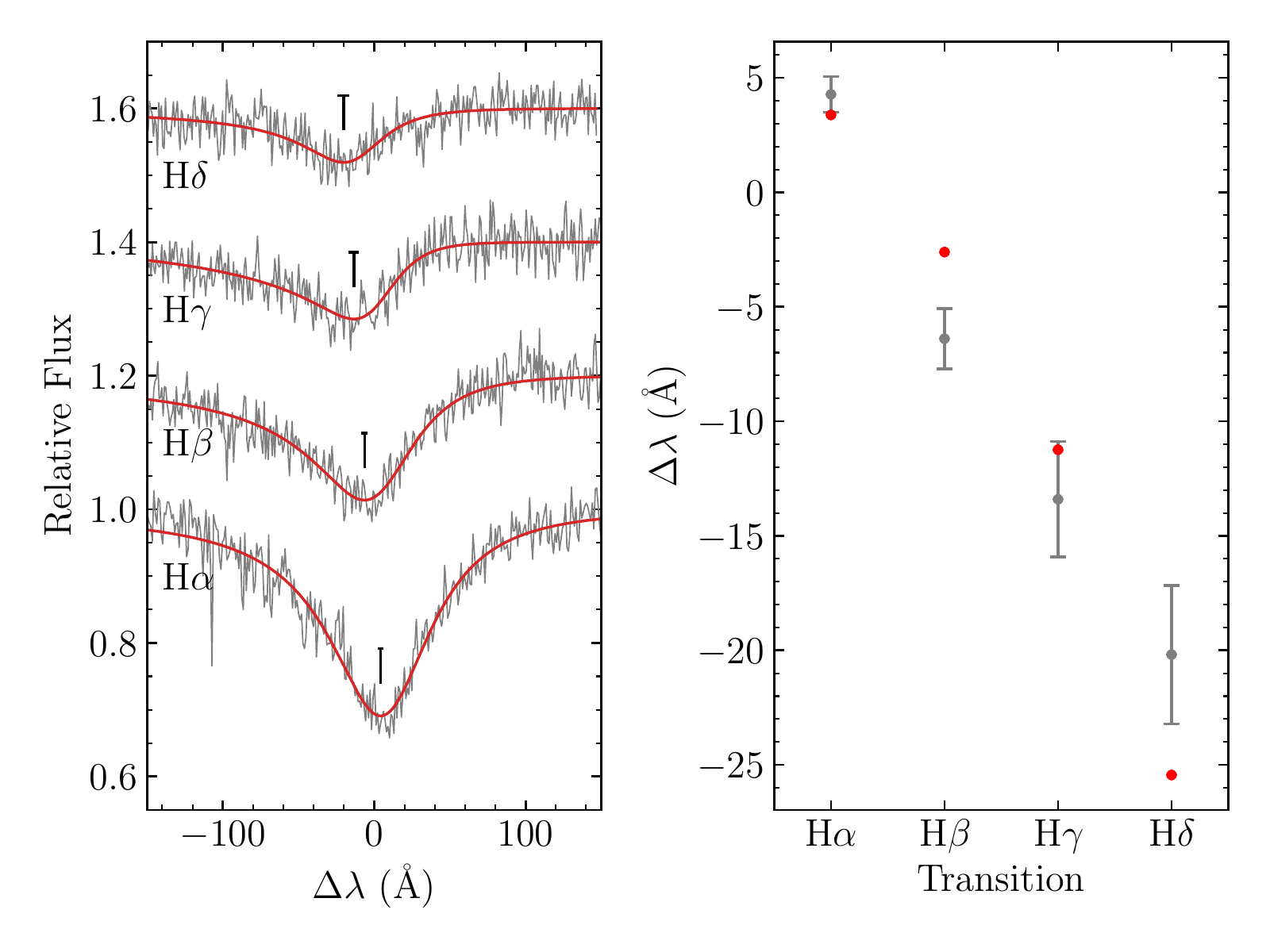}
	\caption{Left: Asymmetric-Lorentzian profile fits to H$\alpha$--H$\delta$ for WD\,J0103$-$0522.
	The black T symbols mark the position of the best fit to the profile minimum, with their widths
	indicating the 1$\sigma$ uncertainties. Right: The measured values of the profile minima,
	fitted with a redshift of $261\pm40$\,\kmps, and field strength of $4.8\pm0.4$\,MG (red points).}
    \label{fig:WDJ0103b}
\end{figure}

{\noindent}where $a$ and $b$ are fit parameters that control the profile asymmetry. Note that the use of $(1+\tanh)$ acts to keep $\Gamma(\lambda, \lambda_0)$ positive, but bounded for all values of $\lambda$. Additionally this formulation ensures that the minimum is located at $\lambda_0$. Continuum normalisation was performed
by fitting a quadratic polynomial. The resulting fits to the first four Balmer lines are shown in Fig.~\ref{fig:WDJ0103b}, left. The minima (in vacuum wavelengths) for H$\alpha$--H$\delta$ were found to be $6568.9\pm0.8$\,\AA, $4856.3\pm1.3$\,\AA, $4328.3\pm2.5$\,\AA, and $4082.7\pm3.0$\,\AA, respectively.

It is reasonable to assume that the four minima correspond to approximately the same field strength, from the distribution of fields on the visible hemisphere of the star. We therefore fitted the shifts according to those expected from the quadratic Zeeman effect with a single field strength and a redshift as free parameters. We used the shifts as given in \citet{hamada71}, for transitions of the type $2S\rightarrow nP$ and $2P \rightarrow n(S/D)$, i.e.

\begin{equation} 
    \Delta k(2S,nP) = 4.96\times10^{-3} B^2 (n^4-n^2-28),
    \label{eq:quadS}
\end{equation}
and
\begin{equation} 
    \Delta k(2Pm,nlm) = 
        2.37\times10^{-4} B^2 \left[(5n^4-17n^2)(5+m^2)-252(1+m^2)\right],
    \label{eq:quadP}
\end{equation}

{\noindent}where $\Delta k$ is the wavenumber shift in cm$^{-1}$, and $m$ is the magnetic quantum number. Since we are only interested in transitions belonging to the $\pi$-component, we have set $m_\mathrm{up}=0$ in equation~(\ref{eq:quadS}), and $m=m_\mathrm{lo}=m_\mathrm{hi}$ in equation~(\ref{eq:quadP}). The resulting fit for the $2S\rightarrow nP$ transitions are given in Fig.~\ref{fig:WDJ0103b}, right, demonstrating good agreement with the quadratic Zeeman effect, and thus providing moderate evidence that WD\,J0103$-$0522 is magnetic. The fits to the $2P \rightarrow n(S/D)$ transitions showed similarly good fits, though with some variance between the redshift and field strength parameters. We therefore averaged the parameters across the set of fits, weighted by their respective oscillator strengths ($f_{ik}$ values), which are shown individually and with the mean in Table~\ref{tab:magparams}.

\begin{table}
	\centering
        \caption{Results from fitting the line shifts of WD\,J0103$-$0522 for the different sub components
        of the H$\alpha$--H$\delta$ $\pi$-components.}
        \label{tab:magparams}
        \begin{tabular}{lccc}
                \hline
                 & m & redshift [\kmps] & $B$ [MG] \\
                \hline
                \hline
                $2S\rightarrow nP$     &      0 & $277\pm39$ & $5.35\pm0.25$ \\
                $2P\rightarrow n(S/D)$ &      0 & $271\pm39$ & $5.04\pm0.24$ \\
                $2P\rightarrow nD$     & $\pm1$ & $251\pm38$ & $4.59\pm0.22$ \\
                \hline
                mean                   &        & $261\pm40$ & $4.82\pm0.37$ \\
                
                \hline
        \end{tabular}\\
\end{table}

At first glance the redshift appears to be extremely large, however the ultra-massive nature of WD\,J0103$-$0522 indicates a gravitational redshift of $\simeq 210$\,\kmps, implying a radial velocity of $50\pm40$\,\kmps. This is consistent with the moderate tangential velocity of 24.20\,\kmps\ from \textit{Gaia} DR2. The magnetic field of $4.82\pm0.37$\,MG, can not be taken as representative of the global magnetic field. More likely, it corresponds to the weakest end of the field strength distribution on the stars visible hemisphere, with the higher field strengths causing asymmetry towards bluer wavelengths.

We note that this object has been observed at only a single epoch, and so if WD\,J0103$-$0522 is a rotator, the sigma components may become visible when viewed from a more favourable orientation, permitting detailed analysis of the field structure.

\begin{figure}
    \centering
	\includegraphics[viewport= 20 10 600 750,scale=0.27]{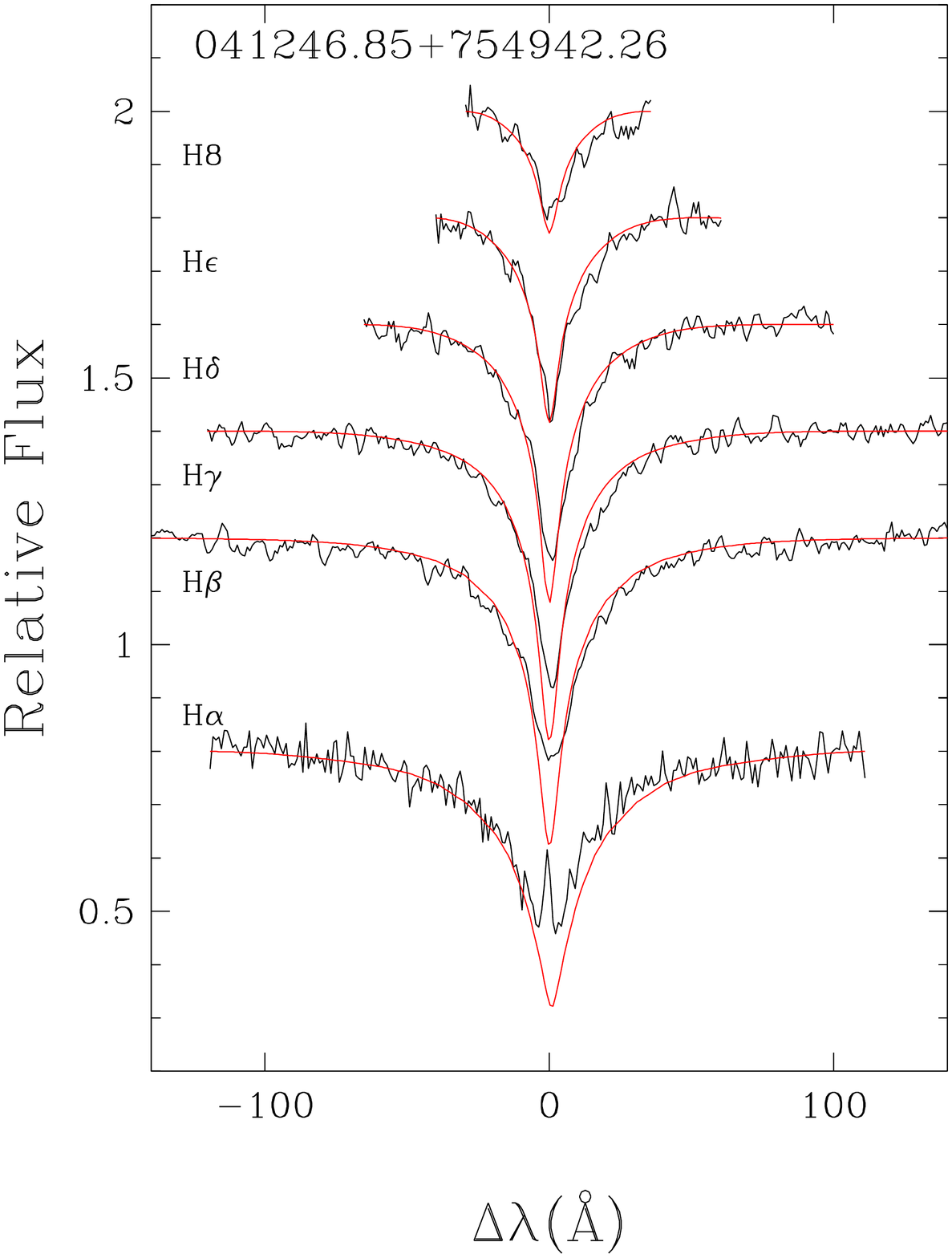}
	\caption{Comparison of the normalised observed Balmer line profiles for WD\,J0412$+$7549 with a DA model atmosphere at the \textit{Gaia} derived photometric parameters ($T_{\rm eff}$ = 8510\,K and $\log g$ = 8.25). Lines are offset vertically for clarity.}
        \label{fig:WDJ0412}
\end{figure}

\textbf{WD\,J0412+7549} shows strong Balmer line core emission (see Fig.~\ref{fig:WDJ0412}), hence best fit parameters are omitted in Table~\ref{tab:final_all}. \textit{Gaia} and Pan-STARRS agree on the photometric parameters ($T_{\rm eff} \approx$ 8500\,K and $\log g \approx$ 8.25), which are likely more accurate. The object was observed on three consecutive nights (14--16 October 2018) and those show near-identical spectra with no obvious phase difference, which makes the possibility of a close double degenerate or a close irradiated low-mass companion rather unlikely. Furthermore we made an attempt to fit simultaneously the spectroscopic and {\it Gaia} data using a composite of two white dwarf models but no specific set of atmospheric parameters could provide a good match to the observations. In particular the observed Balmer lines are fully inconsistent with two massive white dwarfs or a combination of a cool and hot white dwarfs, which would be required to explain the relatively small \textit{Gaia} absolute fluxes. Therefore we conclude that the observations show line emission that is likely to originate from or close to the white dwarf photosphere. We have verified that the object has 2MASS JHK and WISE W1 and W2 absolute fluxes consistent with a single white dwarf at the \textit{Gaia} atmospheric parameters. This rules out a stellar companion although a late T-type brown dwarf (given the system cooling age of $\approx$ 1.5 Gyr) or giant planet would still avoid detection. The white dwarf is close ($\approx$ 15") to an edge-on dusty galaxy with an estimated redshift of $z=$ 0.07 \citep{gal-cat}. The redshift of the galaxy rules out that it is associated with the Balmer line emission, although it may complicate future efforts to obtain reliable IR observations of a possible companion.

\textbf{WD\,J0620+4205} and \textbf{WD\,J1606+7022} are likely He-rich DA stars as they only show weak H$\alpha$ lines given their relatively warm photometric temperatures of 6300--6600\,K (see Figs.~\ref{fig:HalphaDA1}-\ref{fig:HalphaDA2}). Binarity (DC+DA) is unlikely because both stars have photometric surface gravities close to or above the canonical $\log g = 8.0$ value. From the equivalent width of the weak H$\alpha$ lines at fixed \textit{Gaia} atmospheric parameters, we find [H/He] $\sim$ $-2.5$ for both objects. At the effective temperature of these white dwarfs, these large hydrogen abundances can be explained by either convective mixing \citep[see figure 16 of ][]{rolland18} or prior accretion of water-rich planetary debris \citep{raddi15}.

\textbf{WD\,J0021+2531} and  \textbf{WD\,J0129+5108} have problematic (non-WD like) Pan-STARRS photometry that does not agree with \textit{Gaia} colours. \textit{Gaia} agrees with the spectroscopic fits and therefore we favour these solutions. 

\textbf{WD\,J1613+4427, WD\,J1809+2957, WD\,J1847+1226, WD\,J2004+0109} and \textbf{WD\,J2253+8130} are each less than 30$^{\prime\prime}$ from their bright G/K/M-star common proper motion companions, and the Pan-STARRS photometry is unreliable and likely contaminated.  \textbf{WD\,J1929+1117} and \textbf{WD\,J1936+2632} are in crowded fields and Pan-STARRS photometry appears redder than expected from \textit{Gaia} and spectroscopic parameters. 

\textbf{WD\,J0231+2859} and \textbf{WD\,J1923+2141} are likely DA+DA double degenerates as the photometric $\log g$ are much lower than the spectroscopic ones, despite unremarkable spectroscopic fits. Furthermore, both have photometric $\log g$ values in the range 7.50--7.70, consistent with two normal mass white dwarfs at similar temperatures.

\subsection{DAZ white dwarfs}
\label{sec:DAZ}

\begin{table*}
	\centering
        \caption{Atmospheric parameters and chemical abundances of hydrogen-dominated atmosphere metal-rich white dwarfs.}
        \label{tab:marksfit_H}
        \begin{tabular}{llllccc}
                \hline
                WD\,J name & SpT & $T_{\rm eff}$ [K] & $\log g$ & [Na/H] & [Mg/H] & [Ca/H] \\
                \hline
                \hline
                  013055.01+441423.29 & DZA & 4950 (20) & 7.94 (0.02) & $-$8.99 (0.08) & -- & $-$9.90 (0.07) \\
                  020210.60+160203.31 & DZ & 4910 (50) & 8.13 (0.03) & -- & -- & $-$9.43 (0.03) \\
                113444.64+610826.68 & DAZ & 7590 (40) & 7.96 (0.02) & -- & $-$6.79 (0.08) & $-$7.95 (0.04) \\
                  151534.80+823028.99 & DZH & 4490 (60) & 7.90 (0.05) & $-$9.03 (0.02) & -- & $-$9.72 (0.11) \\
                174935.61$-$235500.63 & DAZ & 7430 (70) & 8.03 (0.02) & -- & -- & $-$9.44 (0.15) \\
                  180218.60+135405.46 & DAZ & 8280 (70) & 8.06 (0.05) & -- & -- & $-$8.81 (0.09) \\
                \hline
        \end{tabular}
        \\
        Note: Atmospheric parameters are based on an iterative fit of photometry and spectroscopy. All quoted uncertainties represent the intrinsic fitting errors.
\end{table*}

Figure~\ref{fig:DAZ} shows the spectra of four new DAZ white dwarfs. Balmer line fits are shown in Figs.~\ref{fig:fitsDA1}-\ref{fig:fitsDA2} for three of them where the presence of metals at these cool temperatures is not expected to influence the Balmer line shapes. The very metal-rich \textbf{WD\,J0358+2157} is a notable discovery because it includes a large number of metal lines possibly from different chemical elements. Given the relatively low temperature of 6600\,K, the convection zone is large \citep{tremblay15} and diffusion timescales long \citep{koester09}. Hence the total accreted mass must be relatively large, but there is no evidence that the white dwarf has a debris disc because the 2MASS JHK and WISE W1 and W2 absolute fluxes are consistent with the \textit{Gaia} white dwarf parameters. This object will have a dedicated analysis in G\"ansicke et al.~(in preparation).

For {\bf WD\,J1134+6108}, {\bf WD\,J1749$-$2355} and {\bf WD\,J1802+1354} we have measured Ca/H ratios (as well as Mg/H in the first case) using the model atmospheres of \citet{koester10} and results are presented in Table~\ref{tab:marksfit_H}. $T_{\rm eff}$ and $\log g$ were allowed to vary for internal consistency but the atmospheric parameters were found to be similar to those otherwise derived in Table~\ref{tab:final_all}.

\subsection{Magnetic white dwarfs}

Figure~\ref{fig:Mag} shows eight magnetic white dwarfs, amongst which \textbf{WD\,J0303+0607} and \textbf{WD\,J2151+5917} were observed concurrently and recently analysed in \citet{landstreet2019,landstreet2020}. Two further objects shown in the figure are new observations of known white dwarfs initially part of the \citet{limoges15} 40\,pc sample. Both have a clear detection of Zeeman splitting at H$\alpha$. \textbf{WD\,J0537+6759} was already identified as possibly magnetic in \citet{limoges15} while \textbf{WD\,J0649+7521} is a new detection. There is also a new metal-rich magnetic white dwarf discussed separately in Section \ref{DZsec}, and the likely magnetic ultra-massive DAH: white dwarf WD\,J0103$-$0522 discussed in Section~\ref{DAsec}, for a total of 9--10 new magnetic field detections not known before \textit{Gaia} DR2.

\begin{table}
	\centering
        \caption{Field strengths for magnetic white dwarfs}
        \label{tab:Mag_st}
        \begin{tabular}{lll}
                \hline
                WD\,J name & SpT  & $\langle B \rangle$ (MG)  \\
                \hline
                \hline
                  030350.56+060748.75 & DXP & $>$100 \\
                  053714.90+675950.51 & DAH & 0.7 (0.2)\\ 
                  063235.80+555903.12 & DAH & 1.0 (0.2)  \\
                  064400.61+092605.76 & DAH & 3.2 (0.2)  \\
                  064926.55+752124.97 & DAH & 9.0 (1.0)  \\
                  084516.87+611704.81 & DAH & 0.8 (0.2)  \\
                  151534.80+823028.99 & DZH & 3.1 (0.2) \\
                  160700.89$-$140423.88 & DAH & 0.6 (0.2) \\
                  215140.11+591734.85 & DAH & 0.7 (0.2) \\
                \hline
        \end{tabular}
        \\
\end{table}

\textbf{WD\,J0303+0607} is already extensively discussed in \citet{landstreet2020} where the authors detect a strong polarisation signal. It has large absorption bands of unknown nature in the optical, hence we use the spectral type DXP. \citet{landstreet2020} suggest that magnetic splitting of hydrogen lines from a huge (hundreds of MG) magnetic field is responsible for the observable features. The object is in a wide binary system and separated by only 11$^{\prime\prime}$ from its close and bright G0V companion (\citealt{landstreet2020} estimate a physical separation of 380 AU). As a result the \textit{Gaia} fluxes have large error bars, while the Pan-STARRS photometry is unreliable. Combined with the lack of spectroscopic parameters, this results in rather uncertain atmospheric parameters for this object, although with a strong hint at a large mass (1.18 $\pm$ 0.15 M$_{\odot}$), which is fully compatible with the mass determination in \citet{landstreet2020}.

We have found four DAH white dwarfs with obvious Zeeman splitting and average magnetic field strengths of $\sim$ 1~MG. \textbf{WD\,J2151+5917} is a cool white dwarf with a temperature of $\approx$ 5100~K where only a weak H$\alpha$ line is predicted. The line is tentatively split into three components separated by several angstroms, which has been confirmed in the meantime by \citet{landstreet2019}.  For all magnetic white dwarfs we estimate field strengths in Table~\ref{tab:Mag_st} from Zeeman splitting but do not derive spectroscopic atmospheric parameters, which is notoriously difficult \citep{kulebi09}.

\subsection{DB and DC white dwarfs}
                   
We have found two DBA white dwarfs shown in Fig.~\ref{fig:DB}. Both are at the very cool end of the DB range where spectroscopic fits are difficult \citep{koester15,rolland18}. Using our 3D model atmospheres we could derive spectroscopic parameters that are in reasonable agreement with \textit{Gaia} values for 
{\bf WD\,J0703+2534} with [H/He] = $-5.4$ $\pm$ 0.3. For {\bf WD\,J0552+1642} the helium lines are too weak for a meaningful spectroscopic fit but we find a hydrogen abundance of the order of [H/He] = $-4.5$.

The spectra of 76 DC white dwarfs are shown in Figs.~\ref{fig:DC1}-\ref{fig:DC4}. Only about a dozen have temperatures above $\approx$ 5100\,K where the helium-dominated nature of the atmosphere is unambiguous. A large number of these new white dwarfs have temperatures in the range 4800--5100\,K where H$\alpha$ is predicted to be marginal for pure-hydrogen composition. Higher S/N or higher resolution observations could be used to determine or confirm the atmospheric composition. The majority of new DC white dwarfs are cooler than 4800\,K where only detailed model atmosphere fits of the continuum fluxes could possibly suggest an atmospheric composition \citep{blouin19}.

The analysis of {\it Gaia} and Pan-STARRS photometric fits of objects cooler than about 5000\,K with either pure-H or pure-He model atmospheres, and with independent grids of models, has led to the finding that derived $\log g$ values are systematically lower by up to 0.1-0.2 dex compared to the average $\approx$ 8.0 value observed at higher temperatures \citep[][see also Paper II]{hollands18Gaia,blouin19}. This is also seen in Fig.~\ref{fig:distribution} for the sample of new white dwarfs observed in this work. Since white dwarfs are expected to cool at constant mass even for that low temperature regime \citep{tremblay16}, this is unlikely to be a real astrophysical effect. For a fixed mass-radius relation, apparent magnitude and parallax, the photometric surface gravity correlates with effective temperature given Stefan-Boltzmann law, which implies that the issue could either be caused by \textit{Gaia} temperatures that are too low or luminosities that are too large. In the former case, the amplitude of the colour correction necessary to obtain $\log g$ $\approx$ 8.0 values would be fairly large ($G_{\rm BP}-G_{\rm RP} \approx$ 0.10 mag) and vary strongly with temperature. It is unlikely to be a \textit{Gaia} calibration issue and it is therefore unclear if this is at all related to the colour offset observed for warmer DA white dwarfs and discussed in Section~\ref{DAsec}, which is much milder. The issue is marginally worse with \textit{Gaia} DR2 compared to Pan-STARRS (see Paper II) and as a consequence, we only use Pan-STARRS parameters for objects cooler than 4500\,K.

\textbf{WD\,J0639+2435} is in a wide binary with a bright F-star companion. Pan-STARRS photometry is contaminated as well as our red arm ISIS spectrum. Given the warm \textit{Gaia} temperature of 8500\,K and the lack of Balmer lines, we can nevertheless confirm that the atmosphere is helium dominated.

\textbf{WD\,J1305+7022} and \textbf{WD\,J2305+3922} are rare examples of ultra-massive (1.19 and 1.13 M$_{\odot}$, respectively) and cool (4800 and 6500\,K, respectively) DC white dwarfs. The former is cool enough that the atmosphere could be pure-hydrogen while WD\,J2305+3922 is helium-dominated with a relatively large upper limit on its hydrogen content ([H/He] $<$ $-2.5$). The progenitor of WD\,J2305+3922 could be a massive helium-rich DB white dwarf, which are rare \citep{tremblay19} but occasionally found \citep{richer20}. Alternatively it could also have experienced convective mixing earlier in its evolution \citep{cunningham20} or could be a cooled down example of warm DQ white dwarfs \citep{coutu19,koester20} where carbon has diffused downward, for instance from the shrinking of the convection zone or upwards diffusion of helium. 

\begin{figure*}
	\includegraphics[scale=0.55]{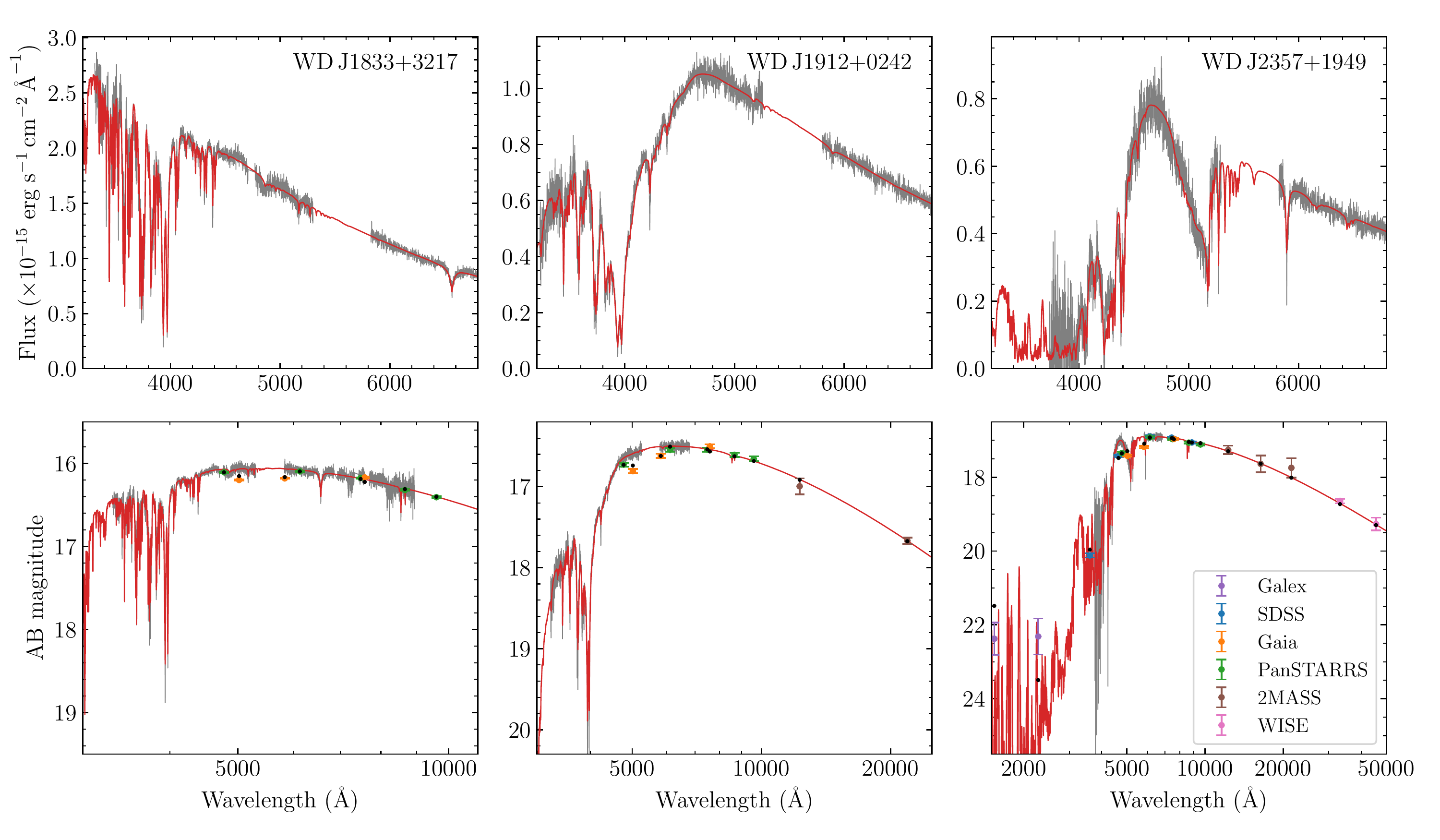}
	\caption{Simultaneous fits of spectroscopy and photometry for the DZA 183352.68+321757.25 (left panels), DZ 191246.12+024239.11 (middle panels) and DZ 235750.73+194905.90  (right panels). The top row of panels compare our best fit models to normalised spectroscopic observations. The spectroscopic observations are recalibrated onto the models to deal with flux-calibration quirks, but are still in physical flux units. The bottom panels compare our best fit models to catalogue photometry over a wider wavelength range. All three objects have helium-dominated atmospheres and fit parameters are given in Table~\ref{tab:marksfit_He}. For WD\,J1912$+$0242, the $K$-band is from UKIDDS rather than 2MASS.}
        \label{fig:DZMark}
\end{figure*}

\subsection{DZ white dwarfs}
\label{DZsec}

We show 10 new DZ and DZA white dwarfs in Fig.~\ref{fig:DZ}. We performed a combined spectroscopic and photometric analysis for the warmer subsample of these objects using the model atmospheres of \citet{koester10} as described in Section~\ref{sec:models}. \textit{Gaia} astrometry is used for all fits. In most cases we relied on multiple photometric data sets, including Pan-STARSS, SDSS, SkyMAPPER, UKIDSS, 2MASS and WISE, but generally neglected broadband \textit{Gaia} photometry. We describe the data sets for individual objects below. We have added a systematic error, by forcing a reduced $\chi^2$ of one, to account for possible systematic offsets between the various photometric surveys. Detailed fits are shown for the three most metal-rich objects in Fig.~\ref{fig:DZMark}. The resulting atmospheric parameters and metal abundances are presented in Tables~\ref{tab:marksfit_H} (hydrogen-dominated atmospheres) and \ref{tab:marksfit_He} (helium-dominated atmospheres).

\setlength{\tabcolsep}{2.2pt}

\begin{table*}
	\centering
	    \scriptsize
        \caption{Atmospheric parameters and chemical abundances of helium-atmosphere DZ and DZA white dwarfs.}
        \label{tab:marksfit_He}
        \begin{tabular}{llllcccccccc}
                \hline
                WD\,J name & SpT & $T_{\rm eff}$ [K] & $\log g$ & [H/He] & [Na/He] & [Mg/He] & [Ca/He] & [Ti/He] & [Cr/He] & [Fe/He] & [Ni/He] \\
                \hline
                \hline
                  020210.60+160203.31 & DZ & 5380 (60) & 8.40 (0.02) & $-$1.46 (0.03) & -- & -- & $-$9.43 (0.03) & -- & -- & -- & -- \\
                  055443.04$-$103521.34 & DZ & 6300 (40) & 8.06 (0.02) & -- & -- & -- & $-$11.82 (0.04) & -- & -- & -- & -- \\
                  183352.68+321757.25 & DZA & 7650 (60) & 8.05 (0.02) & $-$3.31 (0.02) & -- & $-$7.86 (0.04) & $-$8.72 (0.01) & $-$9.95 (0.05) & -- & $-$7.08 (0.01) & $-$8.42 (0.03) \\
                  191246.12+024239.11 & DZ & 6050 (80) & 8.15 (0.03) & $-$4.40 (0.17) & $-$9.70 (0.11) & $-$8.11 (0.01) & $-$9.11 (0.01) & --& --  & $-$8.53 (0.01) & -- \\
                  235750.73+194905.90 & DZ & 5700 (30) & 7.95 (0.02) & -- & $-$9.20 (0.02) & $-$6.79 (0.01) & $-$8.31 (0.01) & -- & $-$9.29 (0.04) & $-$7.31 (0.02) & -- \\
                \hline
        \end{tabular}
        \\
        Note: Atmospheric parameters are based on an iterative fit of photometry and spectroscopy. All quoted uncertainties represent the intrinsic fitting errors.
\end{table*}

\textbf{WD\,J0130+4414} is a cool $\approx$ 5000~K DZA  white dwarf with a weak H$\alpha$ line. Given the low temperature, the presence of hydrogen lines requires a pure-hydrogen atmosphere. The Ca H+K lines are also extremely narrow, confirming the atmospheric composition. The object has Pan-STARRS and SDSS photometry, both of which were used for the photometric fit. The spectrum shows a weak absorption feature from Na, which was not observed for the warmer DAZ stars discussed in Section~\ref{sec:DAZ}. Compared to Ca, the Na abundance is large \citep[see figure 6 of][]{blouin19DZ}.
We have verified that the Na detection is not from poor sky-subtraction. The line is broader than sky lines would be and the redshift is consistant with that of H$\alpha$. 

\textbf{WD\,J0202$+$1602} exhibits a cool $\approx$ 4800\,K DZ spectrum showing only lines from Ca\,\textsc{i/ii}. The star has photometry from SDSS, Pan-STARRS, UKIDSS, and WISE,
allowing model comparison across the full spectral energy distribution. Yet, in spite of the simple spectrum and ample photometric coverage, this object proved highly challenging to fit, and we consider our best attempts to constrain stellar parameters from a simultaneous fit of the photometric and spectroscopic observations to be unsatisfactory.

We found two solutions that were able to accurately reproduce either the spectrum or the photometry, but neither simultaneously. The first solution demonstrated a good fit to both the optical and infra-red photometry, and required a cool hydrogen-dominated atmosphere (parameters listed in Table~\ref{tab:marksfit_H}). However the low pressure in this H-rich atmosphere results in Ca lines that are all too narrow (FWHM $\approx 3$\,\AA\ in the model but $\approx 15$\,\AA\ wide in the data). A better fit to the line widths can be achieved at a lower temperature (close to $4200$\,K), though the Ca\,\textsc{i} line becomes too strong relative to the Ca\,\textsc{ii} doublet,
and collision-induced absorption (CIA) from H$_2$-H becomes apparent in the model, but is not observed in the data.

While we note that a helium-dominated atmosphere (with next to no hydrogen), can also reproduce most of the photometry with similar stellar parameters, the Ca lines were found to be so highly broadened that such a possibility can be disregarded entirely. Instead we find that a helium-dominated atmosphere with a moderate hydrogen component
($[\mathrm{H/He}] = -1.46$) can reproduce the spectrum well, though at a somewhat hotter $T_{\rm eff}$ (to maintain the Ca\,\textsc{i/ii} line ratio) and higher $\log g$
(parameters listed in Table~\ref{tab:marksfit_He}). For this He-rich solution, our photometric fit is extremely poor -- the model over-predicts the optical flux by about 0.5\,mag while under-predicting the infrared fluxes by a similar degree. Additionally, the model exhibits appreciable H$_2$-He CIA, while the observations do not. H$_2$-He CIA has been observed for other DZ stars, such as J0804$+$2239 which has a comparable $T_{\rm eff}$ and hydrogen abundance \citep{blouin19cia}. While \citet{blouin19cia} do find two photometric solutions for J0804$+$2239, these are either side of the H$_2$-He CIA maximum at $[\mathrm{H/He}] = -2.5$, with the higher hydrogen abundance solution also permitting an adequate spectroscopic fit. We advocate that the hydrogen-dominated solution is most likely to be correct, given its good agreement with all photometry and more typical surface gravity, though the reason for the overly narrow predicted Ca lines remains unexplained.

\textbf{WD\,J0554$-$1035} has a He-atmosphere as it is warm enough to show Balmer lines if it was
hydrogen-dominated. The Ca H+K lines are very weak suggesting there is very little opacity in this atmosphere.
We fitted the Pan-STARRS and SkyMapper photometry and spectroscopy fixing the hydrogen abundance to [H/He] = $-$4.0 throughout.

\textbf{WD\,J1515+8230} is a cool $\approx$ 4500\,K magnetic DZH for which abundance determinations are inherently more challenging. It likely has a hydrogen-rich atmosphere because the metal lines are much narrower than would be the case for a helium-dominated atmosphere at the same temperature. Because we are using non-magnetic models the fit is only to the $\pi$-components, hence we artificially increase the abundances by 0.48 dex to account for the other magnetic components. We find a magnetic field strength of 3.06 $\pm$ 0.14 MG.

\textbf{WD\,J1833+3217} (Fig.~\ref{fig:DZMark}) is a $\approx$ 7600\,K DZA with a He-rich atmosphere, strong metal contamination and obvious Balmer lines. The accreted material is found to be moderately Fe- and Ni-rich compared to known DZ stars \citep{hollands17,hollands18DZ} and the Balmer lines allow for a tight constraint on the hydrogen content of [H/He] = $-$3.31 $\pm$ 0.02. The line blanketing was important to include in the atmospheric structure when fitting the Pan-STARRS photometry to obtain a consistent solution with spectroscopy. Note that while 2MASS and WISE photometry are available for this object, they are obviously contaminated by another source located $<3^{\prime\prime}$ away, as shown by Pan-STARRS and \textit{Gaia} data. Ca, Mg and Fe abundances place the object firmly into the regime of polluted white dwarfs that are thought to have accreted material with core-Earth composition \citep[see figure 2 of][]{hollands18DZ}. 

For the He-rich atmosphere \textbf{WD\,J1912+0242} (Fig.~\ref{fig:DZMark}) we fitted against Pan-STARRS, 2MASS ($J$-band only) and UKIDSS ($K$-band only) photometry. The spectrum shows transitions from Na, Mg, Ca and Fe. While hydrogen lines are not seen, leaving it as a free parameter does find a preferential value of [H/He] = $-$4.40 $\pm$ 0.17.

For \textbf{WD\,J1922+0233} we initially derived $T_{\rm  eff} \approx$ 5800\,K and $\log g \approx 9.10$ using \textit{Gaia} photometry, which would imply the most massive polluted white dwarf known to date \citep{koester14,coutu19,veras2020}. However, a closer inspection of Pan-STARRS photometry reveals non blackbody-like optical fluxes sharply peaking in the $g$ and $r$ bands, with a drop-off in $i$, $z$ and $y$ that is sharper than a Rayleigh-Jeans tail. This leaves the strong possibility that the object exhibits collision-induced absorption (CIA) with a spectrum close to that of so-called ultra-cool white dwarfs. This would make it the first DZ to show strong optical CIA absorption, although a few DZ are known to show strong near-IR CIA absorption \citep{blouin19DZ}. Our GTC spectrum only covers the red part of the optical spectrum and as a consequence only the sodium D-line is detected. Given the lack of near-IR photometry we make no quantitative attempt to determine the atmospheric parameters. However, our preliminary analysis suggests that a hydrogen-dominated atmosphere and a very cool temperature is necessary to explain the sodium line. The lower temperature coupled with \textit{Gaia} absolute fluxes would also suggest a more moderate mass for this white dwarf.

\textbf{WD\,J2357+1949} (Fig.~\ref{fig:DZMark}) is a relatively warm DZ with a particularly large Mg \textsc{i} triplet. This object benefits from having photometry from Pan-STARRS, SDSS, 2MASS and WISE. We were able to fit five elements (Na, Mg, Ca, Cr and Fe), with Si scaled with Mg, Ti with Ca, and Ni with Fe. The hydrogen abundance is largely unconstrained but must be low ([H/He] $< -6$) to fit the lines and SDSS $u$ band. 

Finally, we postpone the detailed analysis of two of the coolest DZ white dwarfs, \textbf{WD\,J1824+1213} and \textbf{WD\,J2317+1830}, to a future work (Hollands et al., in preparation). These rare objects allow detailed microphysics study of cool and dense atmospheres in a way that is not possible with featureless DC objects \citep[see, e.g.,][]{blouin19DZ}. 

\subsection{DQ white dwarfs}

We show two DQ white dwarfs in Fig.~\ref{fig:DB}. We fitted both objects with the model atmosphere code of \citet{koester10} using an iterative procedure similar to that described for DZ stars. $T_{\rm eff}$ and $\log g$ rely mostly on the combined Pan-STARSS and \textit{Gaia} photometry while C/He abundances are determined from the spectra. Results for both objects are shown in Table~\ref{tab:marksfit_DQ}.

\begin{table}
	\centering
	    \scriptsize
        \caption{Atmospheric parameters and chemical abundances of DQ white dwarfs.}
        \label{tab:marksfit_DQ}
        \begin{tabular}{lllll}
                \hline
                WD\,J name & SpT & $T_{\rm eff}$ [K] & $\log g$ & [C/He] \\
                \hline
                \hline
                  052913.45+430025.89   & DQ & 8580	(20) & 7.94 (0.01) &  $-$4.73 (0.03) \\
                  171620.72$-$082118.60 & DQ & 5800 (10) & 7.73 (0.01) &  $-$7.25 (0.02) \\ 
                \hline
        \end{tabular}
        \\
        Note: Atmospheric parameters are based on an iterative fit of photometry and spectroscopy. All quoted uncertainties represent the intrinsic fitting errors.
\end{table}

\subsection{Non-white dwarfs}
\label{MS}

Figure~\ref{fig:STARa} and Table~\ref{tab:STAR} show 12 high-probability white dwarf candidates ($P_{\rm WD} > 0.75$) from \citet{gentile19} that turned out to be stellar objects. One additional source \textbf{WD\,J0456+6409} turned out to be a spurious \textit{Gaia} detection and there is no object of that magnitude at the predicted location in Pan-STARRS or other sky images. In all other cases the \textit{Gaia} source is real, isolated on the sky, and \textit{Gaia} colours are confirmed by Pan-STARRS data. Most observed stars have G/K spectral classes and therefore are orders of magnitude over-luminous compared to where \textit{Gaia} locates them in the H-R diagram. The most likely explanation is that the \textit{Gaia} parallaxes are greatly overestimated. The lower Balmer lines of WD\,J0727$-$0718 could be mistaken for a low-mass DAZ white dwarf but a fit of the spectrum is consistent with an A-type main-sequence star. Considering the full sample of 521 confirmed \textit{Gaia} white dwarfs in the northern 40\,pc hemisphere (Paper II), the contamination of the high $P_{\rm WD}$ sample is relatively small (1.3\%).

Nearly all the high-probability contaminants are flagged as {\it duplicated source}\footnote{See Chapter 10.2.2. of \url{https://gea.esac.esa.int/archive/documentation/GDR2/index.html} for information on the {\it duplicated source} flag.} in \textit{Gaia} DR2, which is not taken into account in our probability calculation. This flag signifies that the detection system on-board \textit{Gaia} generated more than one detection for these sources, but during on-ground processing those were identified as a single object and only one solution was kept. The parallax measurements for objects with the {\it duplicated source} seem to be inherently less reliable. However, $\approx$ 800 white dwarfs confirmed by SDSS spectroscopy can be correctly identified using their \textit{Gaia} parallax despite having the {\it duplicated source} flag, indicating that it can not be used to efficiently eliminate unreliable sources in larger volume samples.

Fig.~\ref{fig:STARb} and Table~\ref{tab:STAR} show 26 stellar objects for which \citet{gentile19} predict a low probability ($P_{\rm WD} <$ 0.75) of being a white dwarf. We emphasise that two more low probability candidates described in earlier sections turned out to be under-luminous white dwarfs (WD\,J1305+7022 and WD\,J2317+1830), both with significantly larger proper motions than the average for low probability white dwarf candidates. This confirms that low probability candidates can still reveal some surprises although at a large observational cost, with only a few per cent of these objects turning out to be white dwarfs in the present work. It is hoped that \textit{Gaia} DR3 will help in defining a cleaner distinction between peculiar (under- or over-luminous) white dwarfs and contaminants.

\begin{table}
	\centering
	    \scriptsize
        \caption{Spectroscopically confirmed main-sequence stars}
        \label{tab:STAR}
        \begin{tabular}{llllllll}
                \hline
                WD\,J name   & $P_{\rm WD}$ & Note \\
                \hline
                \hline
                002332.98+432029.26 		& 0.187 & --\\
                004940.60+033400.22 		& 0.043 & --\\
              $*$ 005645.62+551556.10 		& 0.997 & a,b\\
                010343.47+555941.53 		& 0.131 &--\\
                011519.50+573836.11 		& 0.420 &--\\
                011608.20+584642.12 		& 0.997 & a,b\\
                030433.12+361150.47 		& 0.001 &--\\
                044454.33+632408.23 		& 0.015 &--\\
                045620.38+640927.64 		& 0.997 & c \\
                052129.02+185236.19 		& 0.814 &b\\
                054615.88+380324.86 		& 0.292 &--\\
                061350.39+010424.07 		& 0.996 &a\\
                064643.38$-$090839.54 	    & 0.104 &--\\ 
                064711.68+243202.84 		& 0.128 &--\\
                072714.16$-$071837.09     	& 0.763 & b\\
                092138.08$-$014300.80 	    & 0.042 &--\\
                113726.34$-$112357.80 	    & 0.977 & b\\
               $*$ 125256.54$-$140607.88 	& 0.976 &a\\
                131843.05+381034.60 		& 0.990 &b\\
               $*$ 134252.41+003312.28 	    & 0.988 &a\\
                151421.37$-$110323.10 	    & 0.212 &--\\
                160430.57$-$192728.26 	    & 0.373 &--\\
                161105.78$-$045652.94 	    & 0.997& b\\
                170027.28$-$184958.45 	    & 0.002 &--\\
                181747.59+191218.39 		& 0.089 &--\\
                182431.36+193723.82 		& 0.051 &--\\
                185136.02+221307.15 		& 0.988 &b\\
                194333.65+222513.78 		& 0.065 &--\\
                194530.78+164339.17 		& 0.092 &--\\
                194843.46$-$073635.55    	& 0.113 &--\\
                195513.90+222458.79 		& 0.001 &--\\
                200748.71$-$040717.02    	& 0.265 &--\\
                201437.22+231607.23         & 0.054 &--\\
                205009.26+291929.59 		& 0.919 & a,b\\
                205241.82+294828.65 		& 0.058 &--\\
                213132.74+332302.32 		& 0.644 &--\\
                213723.27+224811.81 		& 0.178 &--\\
                214756.33+225203.56 		& 0.128 &--\\
                223544.75+391451.39 		& 0.112 &--\\
                    \hline
        \end{tabular}\\
        Notes: (a) high \textit{Gaia} DR2 astrometric excess noise, (b) marked as duplicate source, (c) spurious \textit{Gaia} DR2 source.  Low probability white dwarf candidates ($P_{\rm WD} < 0.75$) are intrinsically less reliable \citep{gentile19}. Objects with an asterisk symbol have a parallax value outside of 40\,pc but may still be within that volume at 1$\sigma$.
\end{table}

\section{Summary}
\label{conclusions}
The volume-limited 20\,pc white dwarf sample has long been a benchmark to study white dwarf evolution, stellar formation history, Galactic kinematics, the local binary population and stellar magnetism \citep{giammichele12,tremblay14,holberg16,toonen17,landstreet2019}. Thanks to \textit{Gaia} DR2 the sample is now relatively well defined with a large spectroscopic completeness \citep{hollands18Gaia}. Assembling a spectroscopically complete sample for the eight-times larger, 40\,pc volume is a far greater challenge. \citet{limoges13,limoges15} initiated this important work by securing $\approx$ 300 spectra of new white dwarfs likely within 40\,pc and mostly within the northern hemisphere. We have pursued the goal of enhancing the spectroscopic completeness of the 40\,pc sample by following-up 230 white dwarf candidates from \textit{Gaia} DR2. We have described spectral types for 191 white dwarfs within that volume, in the vast majority confirmed as stellar remnants for the first time. We have reported on several examples of rare classes of white dwarfs, including a handful of ultra-massive remnants, a DA star with peculiar so-far unexplained Balmer line emission, one of the closest DZ white dwarfs accreting from a disrupted planetesimal with core-Earth composition and possibly the first ultra-cool DZ star.

We have now reached high-level spectroscopic completeness in the northern 40\,pc hemisphere. From cross-matching \textit{Gaia} DR2 white dwarf candidates from \citet{gentile19} with past spectroscopic catalogues \citep[e.g.,][]{limoges15,subasavage17} and present observations, we identify 521 spectroscopically confirmed white dwarfs in the companion Paper II. Only three high-probability white dwarf candidates from \citet{gentile19} do not have a spectral type in the northern 40\,pc sample. The completeness of the \textit{Gaia} DR2 catalogue itself and selection of \citet{gentile19} are expected to be very high. \citet{hollands18Gaia} estimate that \textit{Gaia} DR2 has found at least 96\% of all white dwarfs within 20\,pc and the completeness is expected to be similar within 40\,pc, especially since the \textit{Gaia} DR2 detection rate is poorer at very close distances ($<$10\,pc) and high proper-motions \citep{hollands18Gaia}. The northern 40\,pc spectroscopic sample, an increase by a factor of four in size  compared to the previous 20\,pc sample, provides the observational constraints required to study, as outlined in Paper II, the mass, temperature and age white dwarf distributions, spectral evolution, properties of magnetic and metal-rich subtypes, the binary fraction and the crystallisation of white dwarfs. 

We have also contributed to enhance the spectroscopic completeness of the southern hemisphere 40\,pc sample, where observations are still underway. Including the 64 white dwarfs confirmed in this work, at least 320 white dwarfs in the southern 40\,pc sample have a known spectral type, but there remains $\approx$ 200 high-probability candidates without spectroscopy, a sharp contrast with the northern hemisphere. With upcoming multi-object spectroscopic (MOS) surveys on 4-meter class telescopes such as WEAVE, 4MOST and DESI, there is hope for a major increase in the size of volume-limited white dwarf samples. These surveys may take decades to cover the full sky, hence the relevance of continued dedicated studies. However, magnitude-limited spectroscopic surveys in portions of the sky such as the SDSS \citep{kepler19} can rival with volume-limited samples if biases and completeness are well understood.

\section*{Acknowledgements}
The research leading to these results has received funding from the European Research Council under the European Union's Horizon 2020 research and innovation programme n. 677706 (WD3D). This article is based on observations made in the Observatorios de Canarias del IAC with the WHT operated on the island of La Palma by the Isaac Newton Group of Telescopes in the Observatorio del Roque de los Muchachos. Based on observations made with the Gran Telescopio Canarias (GTC), installed at the Spanish Observatorio del Roque de los Muchachos of the Instituto de Astrof{\'i}sica de Canarias, in the island of La Palma. This work presents results from the European Space Agency (ESA) space mission Gaia. Gaia data are being processed by the Gaia Data Processing and Analysis Consortium (DPAC). Funding for the DPAC is provided by national institutions, in particular the institutions participating in the Gaia MultiLateral Agreement (MLA). B.T.G. was supported by the UK STFC grants ST/P000495 and  ST/T000406/1. ARM acknowledges  support  from  the  MINECO under  the  Ram\'on  y  Cajal programme (RYC-2016-20254), the AYA2017-86274-P grant and the AGAUR grant SGR-661/2017. MRS thanks for support from Fondecyt (grant 1181404). DdM acknowledges financial support from ASI-INAF I/037/12/0 and ASI-INAF n.2017-14-H.0 and from INAF ``INAF main streams'', Presidential Decree 43/2018. RR has received funding from the postdoctoral fellowship programme Beatriu de Pin\'os, funded by the Secretary of Universities and Research (Government of Catalonia) and by the Horizon 2020 programme of research and innovation of the European Union under the Maria Sk\l{}odowska-Curie grant agreement No 801370.

\section*{Data Availability Statement}
The raw data underlying this article are available in the ING Data Archives and The Gran Telescopio CANARIAS Public Archive. The reduced spectra will be shared on reasonable request to the corresponding author.

\bibliographystyle{mnras}
\bibliography{mybib} 

\clearpage
\onecolumn
{\bf \large \noindent Affiliations}\\
    \begin{description}
    \item $^{1}$ Department of Physics, University of Warwick, Coventry, CV4 7AL, UK
    \item$^{2}$ European Southern Observatory, Karl Schwarzschild Stra{\ss}e 2, Garching, 85748, Germany
    \item$^{3}$ Instituto de Astrof\'isica de Canarias, 38205 La Laguna, Tenerife, Spain
    \item$^{4}$ Departamento de Astrof\'isica, Universidad de La Laguna, 38206 La Laguna,
       Tenerife, Spain
    \item$^{5}$ Institut f\"ur Theoretische Physik und Astrophysik, University of Kiel,
       24098 Kiel, Germany
 \item$^{6}$Smithsonian Astrophysical Observatory, 60 Garden Street,Cambridge, MA 02138 USA
    \item$^{7}$ Institut de Recherche en Astrophysique et Plan\'etologie, CNRS, Universit\'e de Toulouse, CNES,
14 Avenue Edouard Belin, 31400, Toulouse, France)
    \item$^{8}$ Department of Physics and Astronomy, University College London, London WC1E 6BT, UK
    \item$^{9}$ Space sciences, Technologies and Astrophysics Research (STAR) Institute, Universit\'e de Li\`ege,
19C All\'ee du 6 Ao\^ut, B-4000 Li\`ege, Belgium
    \item$^{10}$ Department of Astronomy, Boston University, 725 Commonwealth Ave., Boston, MA 02215, USA
    \item$^{11}$ Astronomisches Rechen-Institut, Zentrum f\"ur Astronomie der Universit\"at Heidelberg, D-69120 Heidelberg, Germany
    \item$^{12}$ Instituto de F\'isica da Universidade Federal do Rio Grande do Sul, Porto Alegre, RS Brazil
    \item$^{13}$ Gemini Observatory, 670 N'Aohoku Pl., Hilo, Hawaii, 96720, U.S.A.
    \item$^{14}$ INAF - Osservatorio Astronomico di Capodimonte, Salita Moiariello 16, 80131 Naples, Italy
    \item$^{15}$ Department of Physics and Astronomy, University of Sheffield, Sheffield, S3 7RH, UK
    \item$^{16}$ Institut f\"ur Physik und Astronomie, Universit\"at Potsdam, Haus 28, Karl-Liebknecht-Str. 24/25,
D-14476 Potsdam-Golm, Germany
    \item$^{17}$ Departament de F\'{\i}sica, Universitat Polit\`{e}cnica de Catalunya, c/Esteve Terrades 5, 08860 Castelldefels, Spain
    \item$^{18}$ Institut d'Estudis Espacials de Catalunya, Ed. Nexus-201, c/Gran Capit\`a 2-4, 08034 Barcelona, Spain
    \item$^{19}$ Key Laboratory of Space Astronomy and Technology, National Astronomical Observatories,
Chinese Academy of Sciences, Beijing 100101, P. R. China
    \item$^{20}$ Instituto de F\'{\i}sica y Astronom\'{\i}a de la Universidad de Valpara\'{\i}so, Av. Gran Breta\~na 1111, Valpara\'{\i}so, Chile.
    \item$^{21}$ Millennium Nucleus for Planet Formation, NPF, Universidad de Valpara{\'i}so, Valpara{\'i}so, Chile
     \item$^{22}$ NAF-Osservatorio Astrofisico di Torino, Strada dell’Osservatorio 20, 10025, Pino Torinese, Italy
    \item$^{23}$ Institute for Gravitational Wave Astronomy, School of Physics and Astronomy, University of Birmingham,
Birmingham, B15 2TT, UK
    \end{description}
\appendix

\section{Online Figures}

\begin{figure*}
	\includegraphics[viewport= 1 20 700 750,scale=0.85]{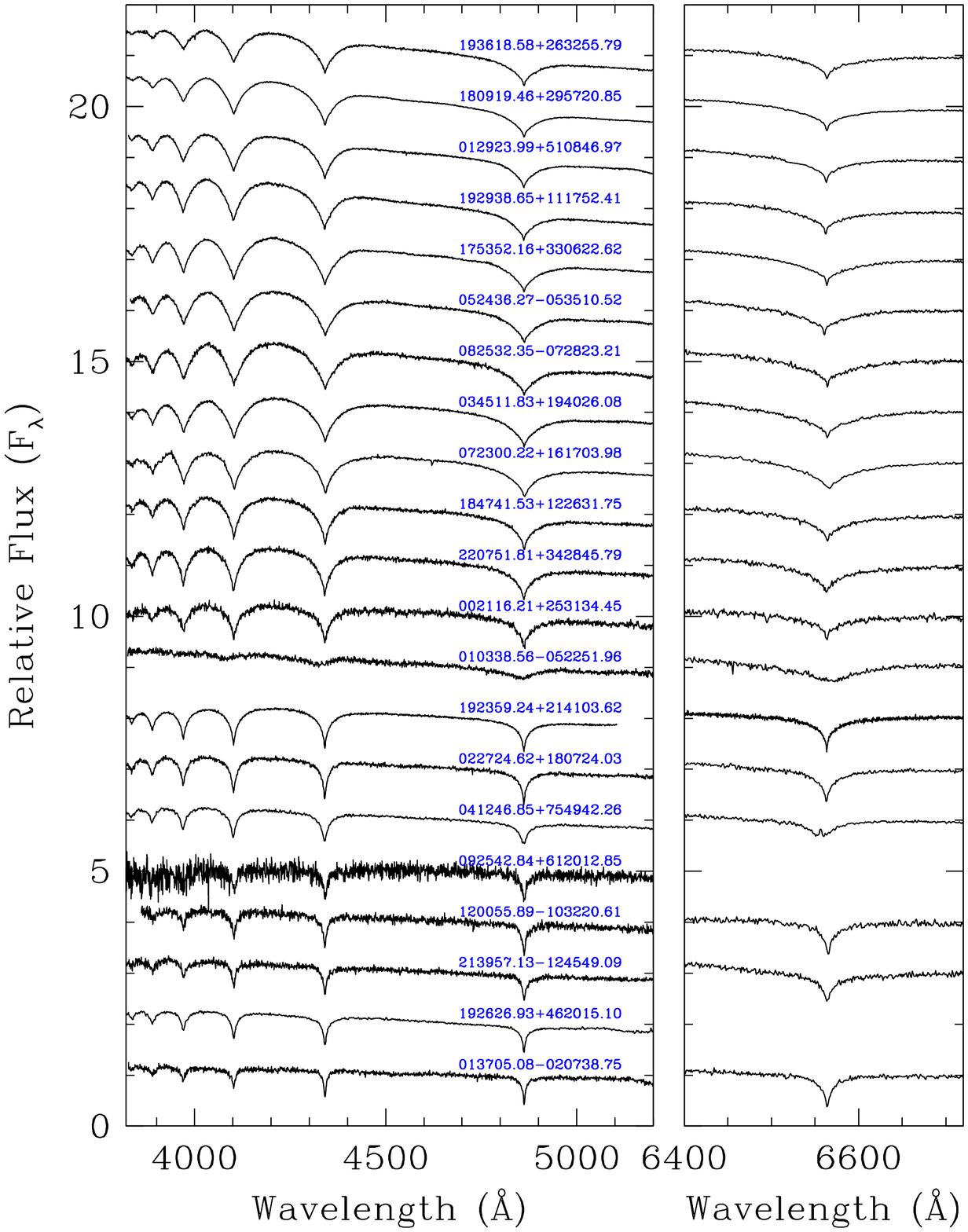}
	\caption{Spectroscopic observations of 89 DA white dwarfs ordered with decreasing photometric temperature (continued on Figs.~ \ref{fig:DA2}-\ref{fig:DA4}).}
        \label{fig:DA1}
\end{figure*}

\begin{figure*}
	\includegraphics[viewport= 1 20 700 750,scale=0.85]{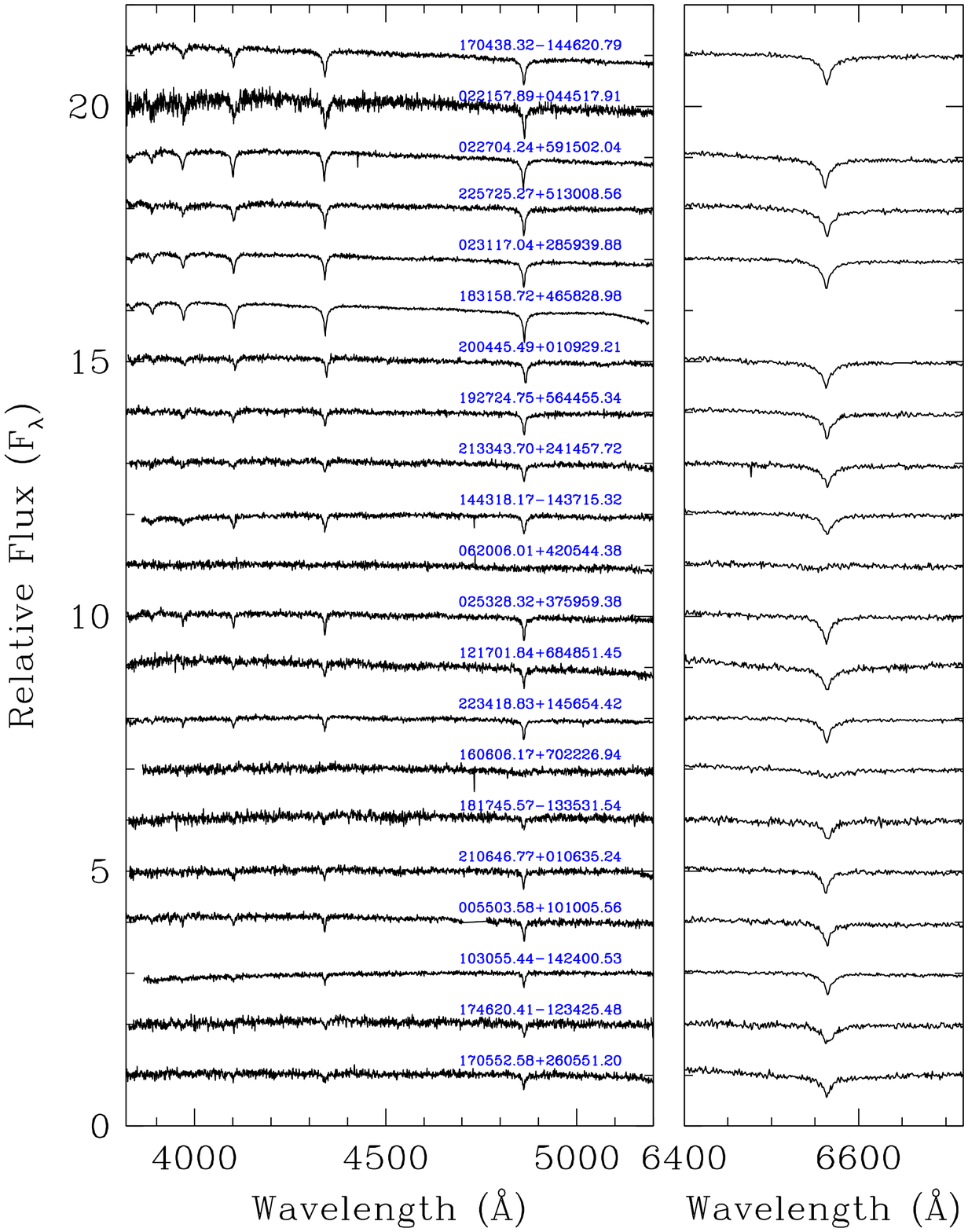}
	\caption{Spectroscopic observations of 89 DA white dwarfs ordered with decreasing photometric temperature (continued 2/4).}
        \label{fig:DA2}
\end{figure*}

\begin{figure*}
	\includegraphics[viewport= 1 20 700 750,scale=0.85]{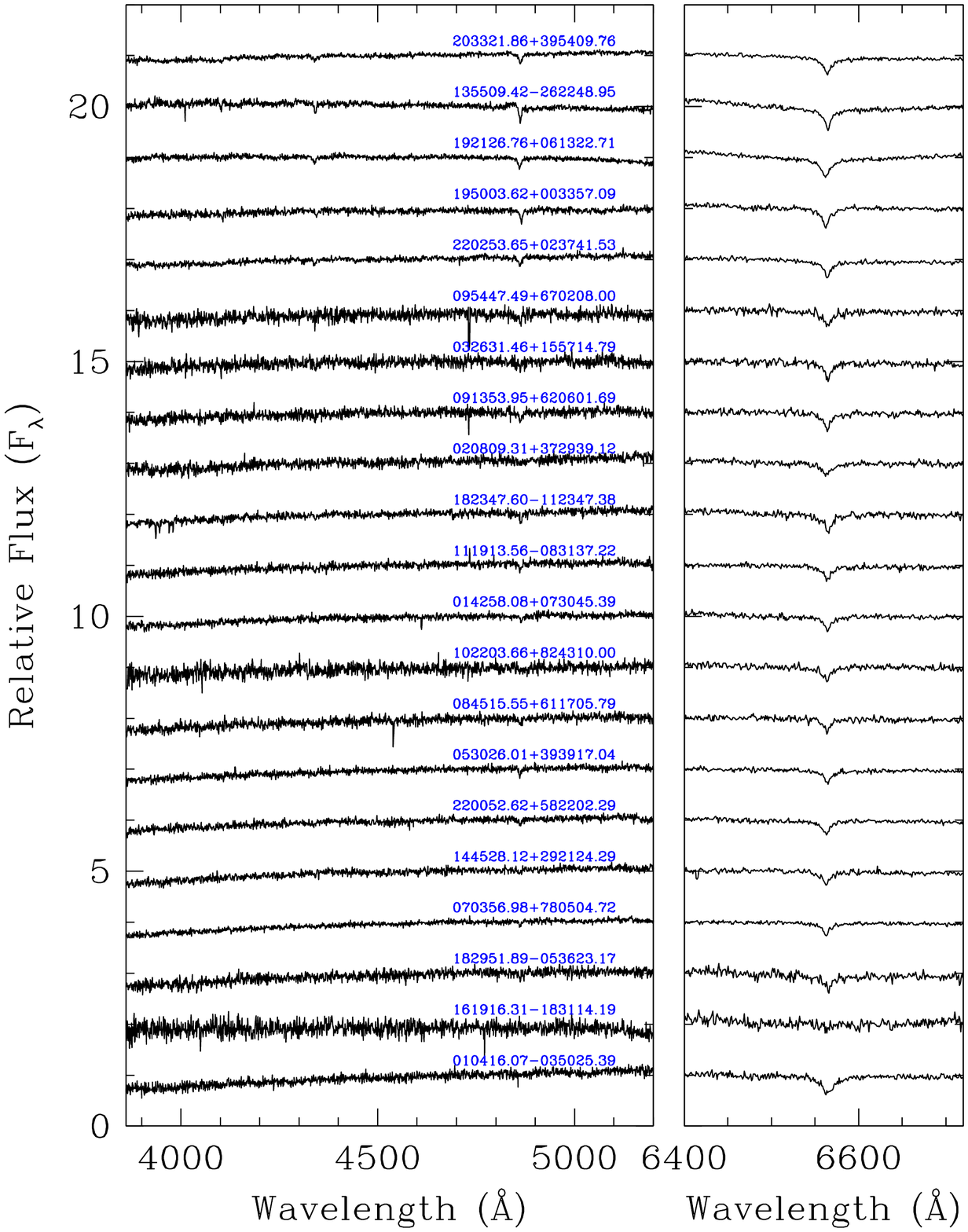}
	\caption{Spectroscopic observations of 89 DA white dwarfs ordered with decreasing photometric temperature (continued 3/4).}
        \label{fig:DA3}
\end{figure*}

\begin{figure*}
	\includegraphics[viewport= 1 20 700 750,scale=0.85]{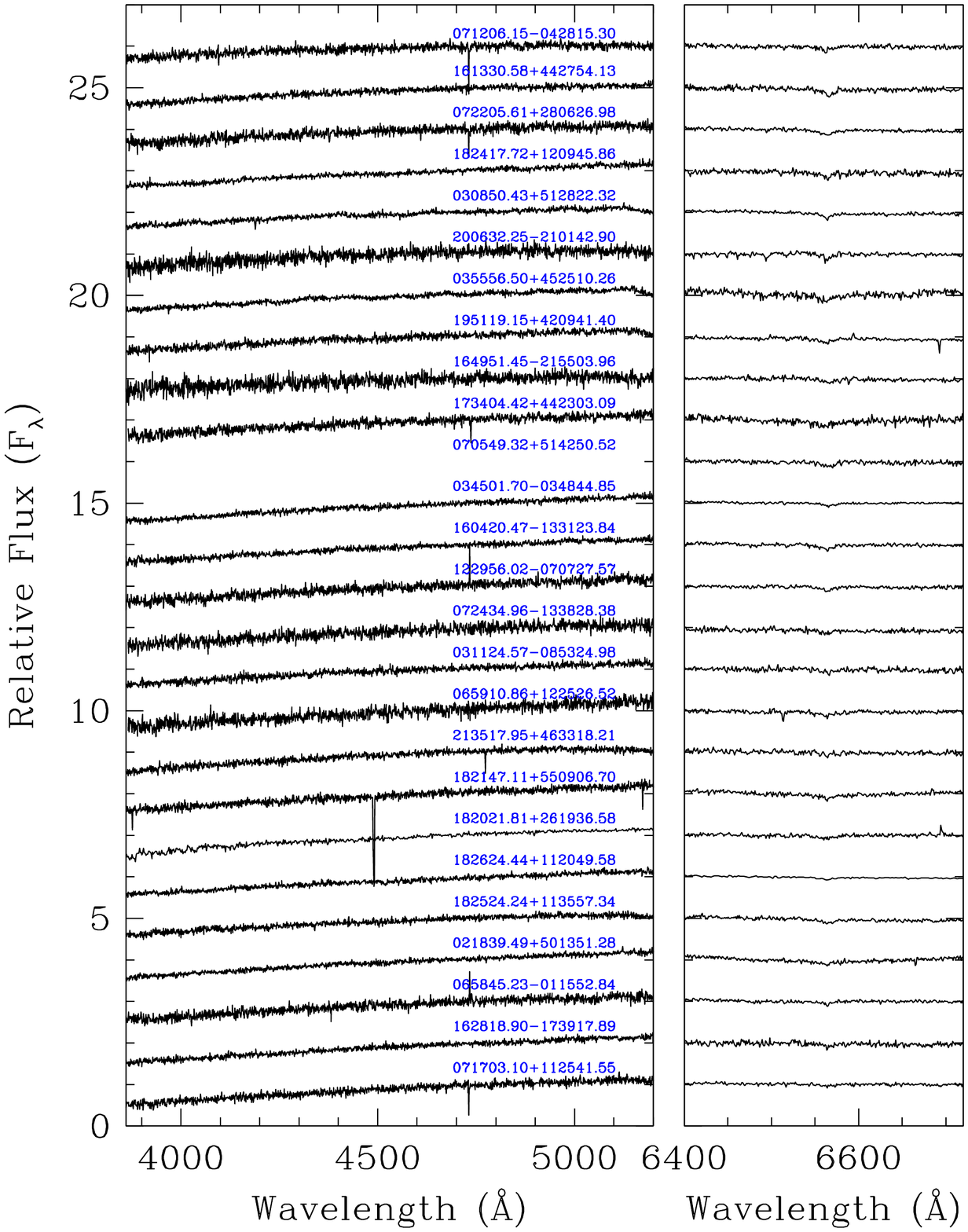}
	\caption{Spectroscopic observations of 89 DA white dwarfs ordered with decreasing photometric temperature (continued 4/4).}
        \label{fig:DA4}
\end{figure*}

\begin{figure*}
	\includegraphics[viewport= 1 20 700 750,scale=0.85]{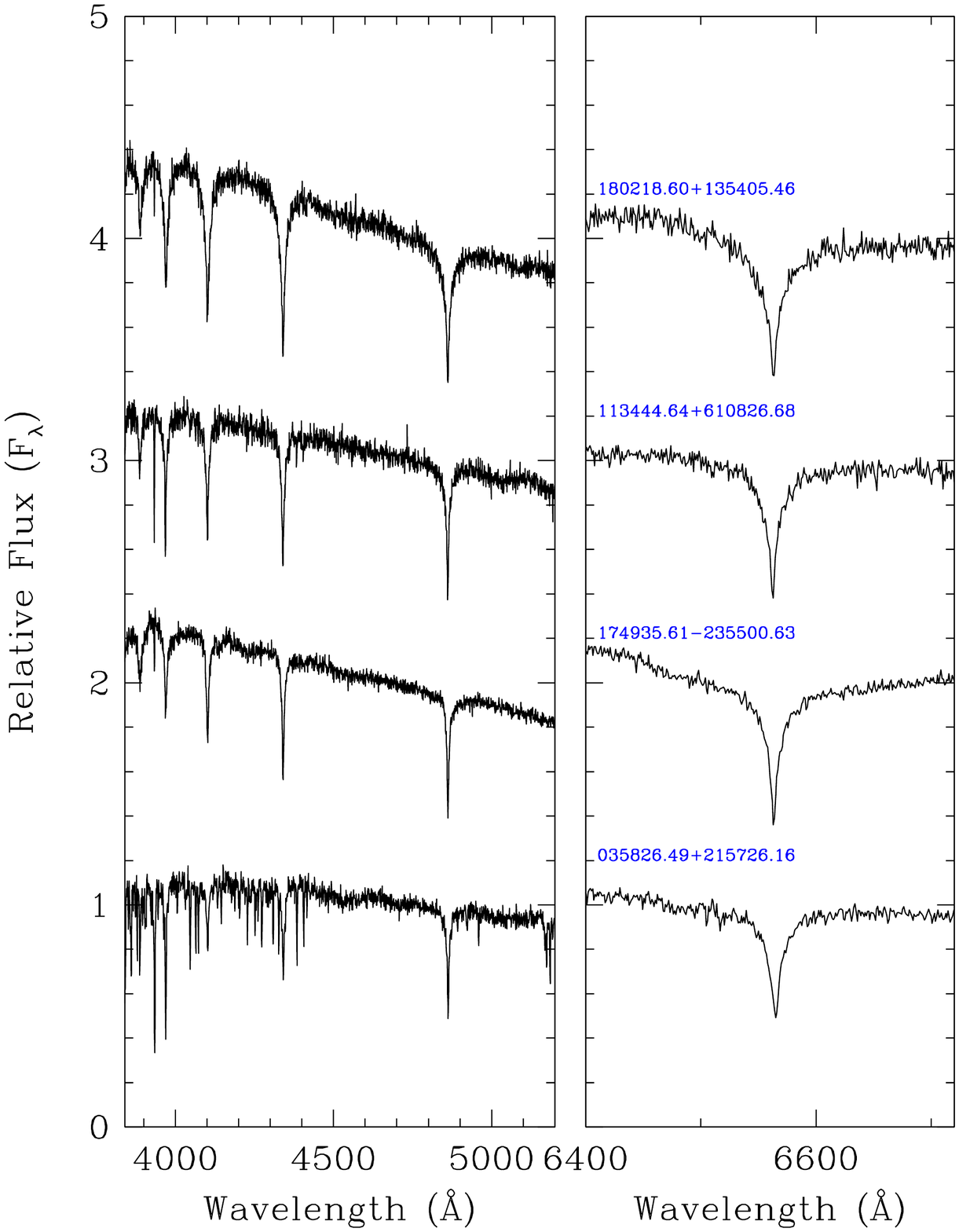}
	\caption{Spectroscopic observations of 4 DAZ white dwarfs ordered with decreasing photometric temperature.}
        \label{fig:DAZ}
\end{figure*}

\begin{figure*}
	\includegraphics[viewport= 1 20 700 750,scale=0.85]{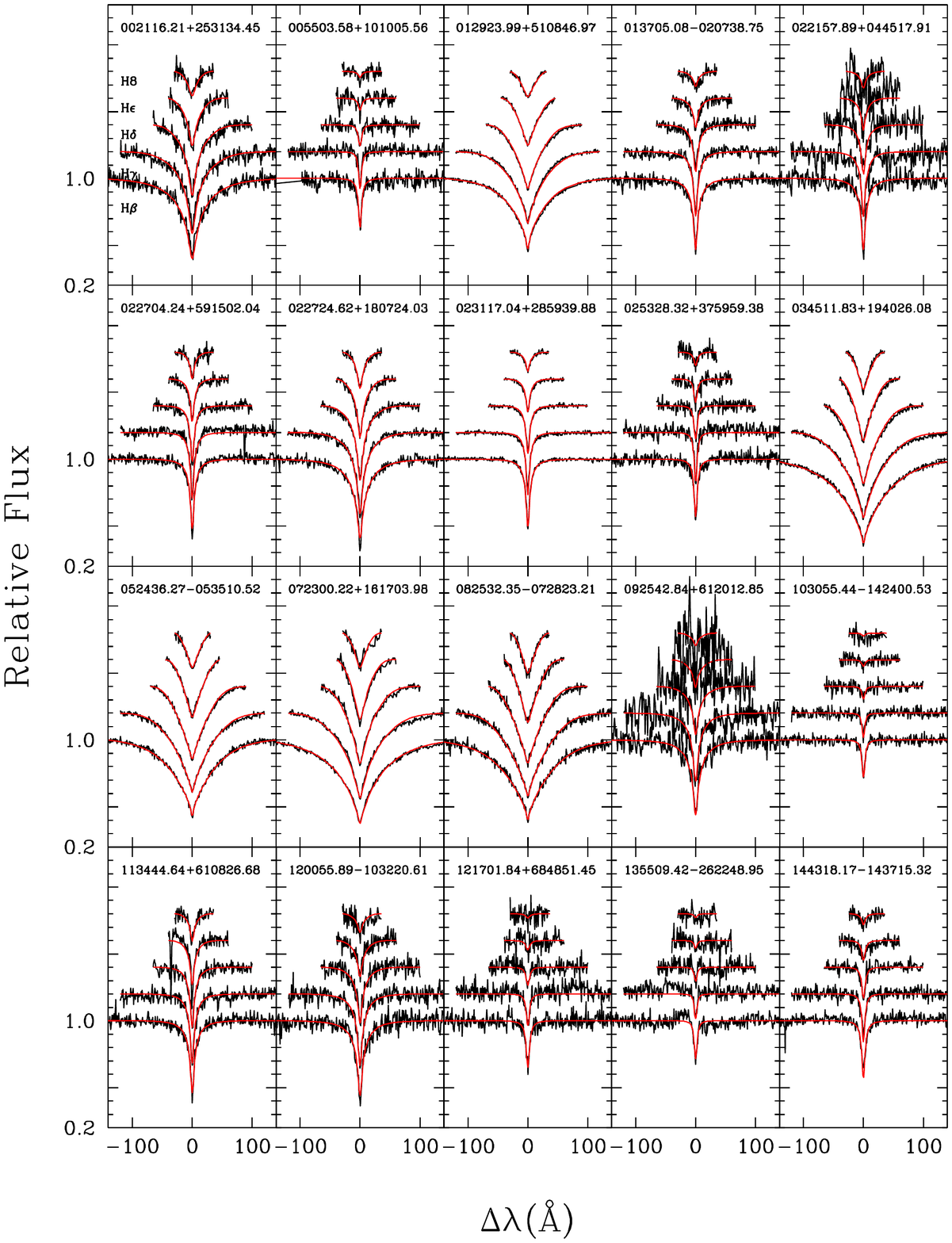}
	\caption{Fits to the normalised Balmer lines for 40 DA and DAZ white dwarfs (continued on Fig.~\ref{fig:fitsDA2}). Atmospheric parameters are given in Table~\ref{tab:final_all}. Fits to the peculiar DA WD\,J0103$-$05225 and WD\,J0412+7549 are shown instead in Figs.~\ref{fig:WDJ0103} and \ref{fig:WDJ0412}, respectively. The metal-rich DAZ WD\,J0358+2157 is also excluded. Two He-rich DA white dwarfs and all DA with \textit{Gaia} temperatures below 6000\,K are shown instead in Figs.~\ref{fig:HalphaDA1}-\ref{fig:HalphaDA4}.}
        \label{fig:fitsDA1}
\end{figure*}

\begin{figure*}
	\includegraphics[viewport= 1 20 700 750,scale=0.85]{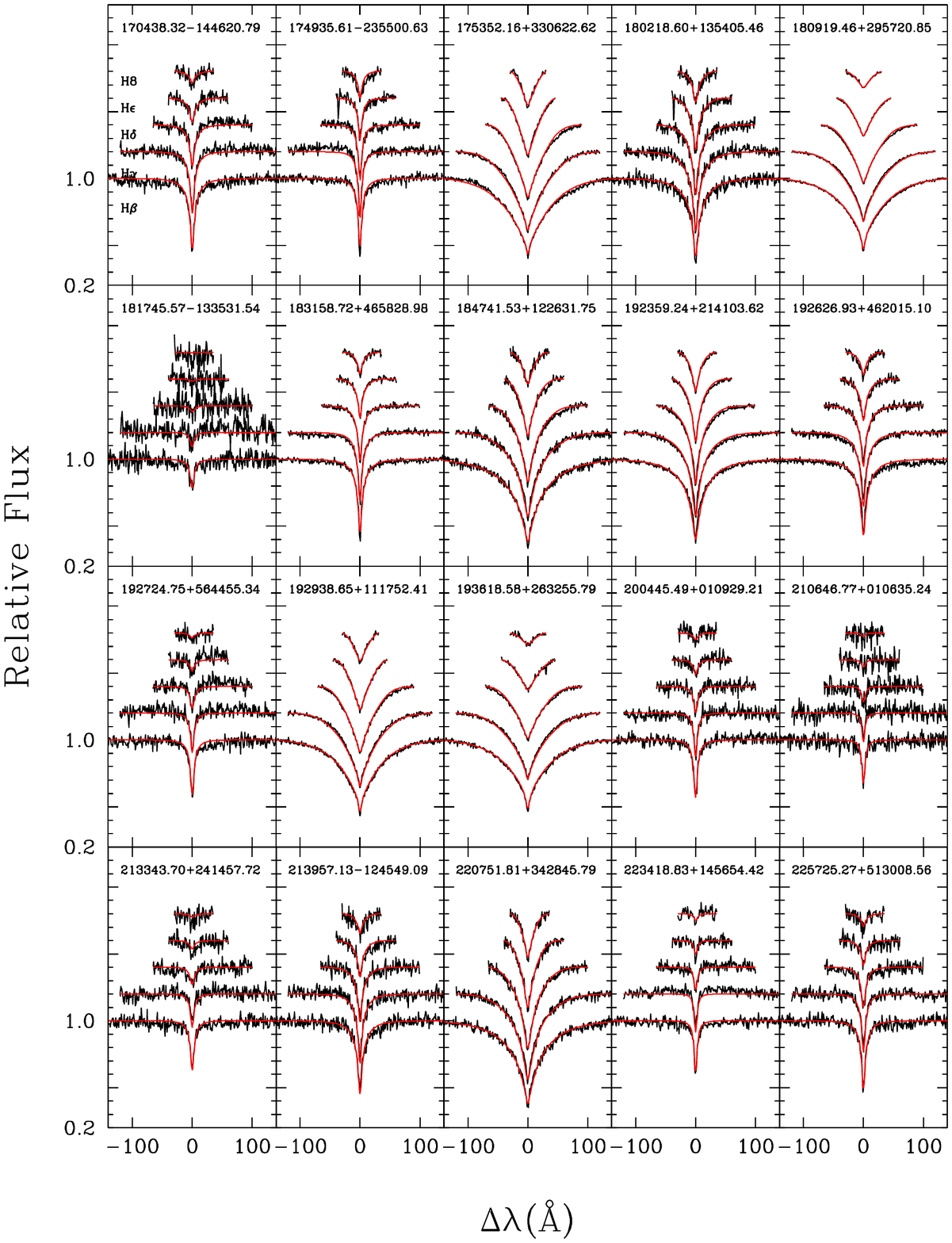}
	\caption{Fits to the normalised Balmer lines for 40 DA and DAZ white dwarfs (continued 2/2).}
        \label{fig:fitsDA2}
\end{figure*}

\begin{figure*}
	\includegraphics[viewport= 1 20 700 750,scale=0.85]{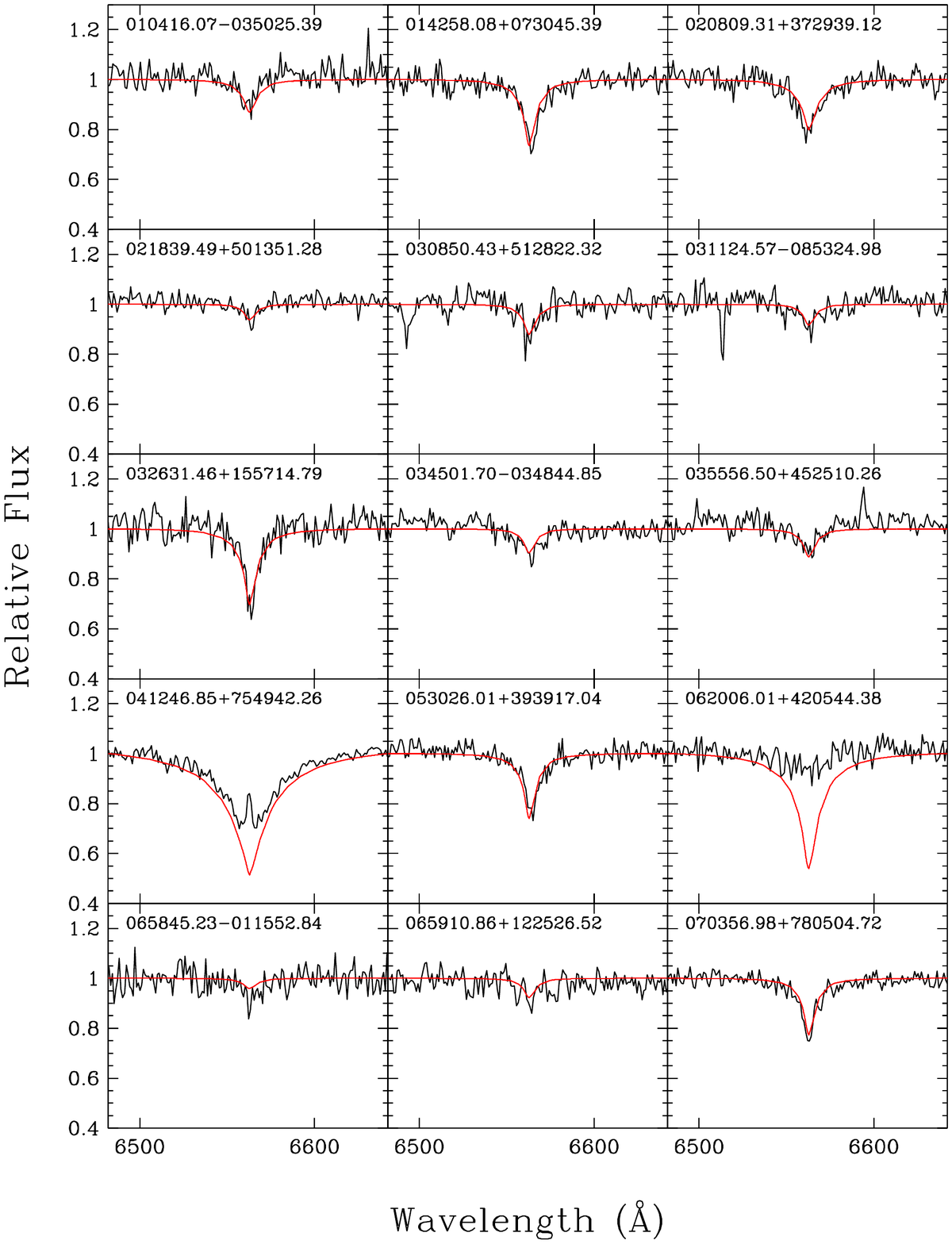}
	\caption{Comparison of observed and predicted H$\alpha$ line profiles from \textit{Gaia} photometric parameters. 51 DA stars are shown, including all of those with \textit{Gaia} parameters below 6000\,K in addition to the DAe WD\,J0412+75494 and He-rich DA WD\,J0620+4205 and WD\,J1606+7022. A correction of +2.7\% to the Gaia temperatures was applied. Continued on Figs.~\ref{fig:HalphaDA2}-\ref{fig:HalphaDA4}.}
        \label{fig:HalphaDA1}
\end{figure*}

\begin{figure*}
	\includegraphics[viewport= 1 20 700 750,scale=0.85]{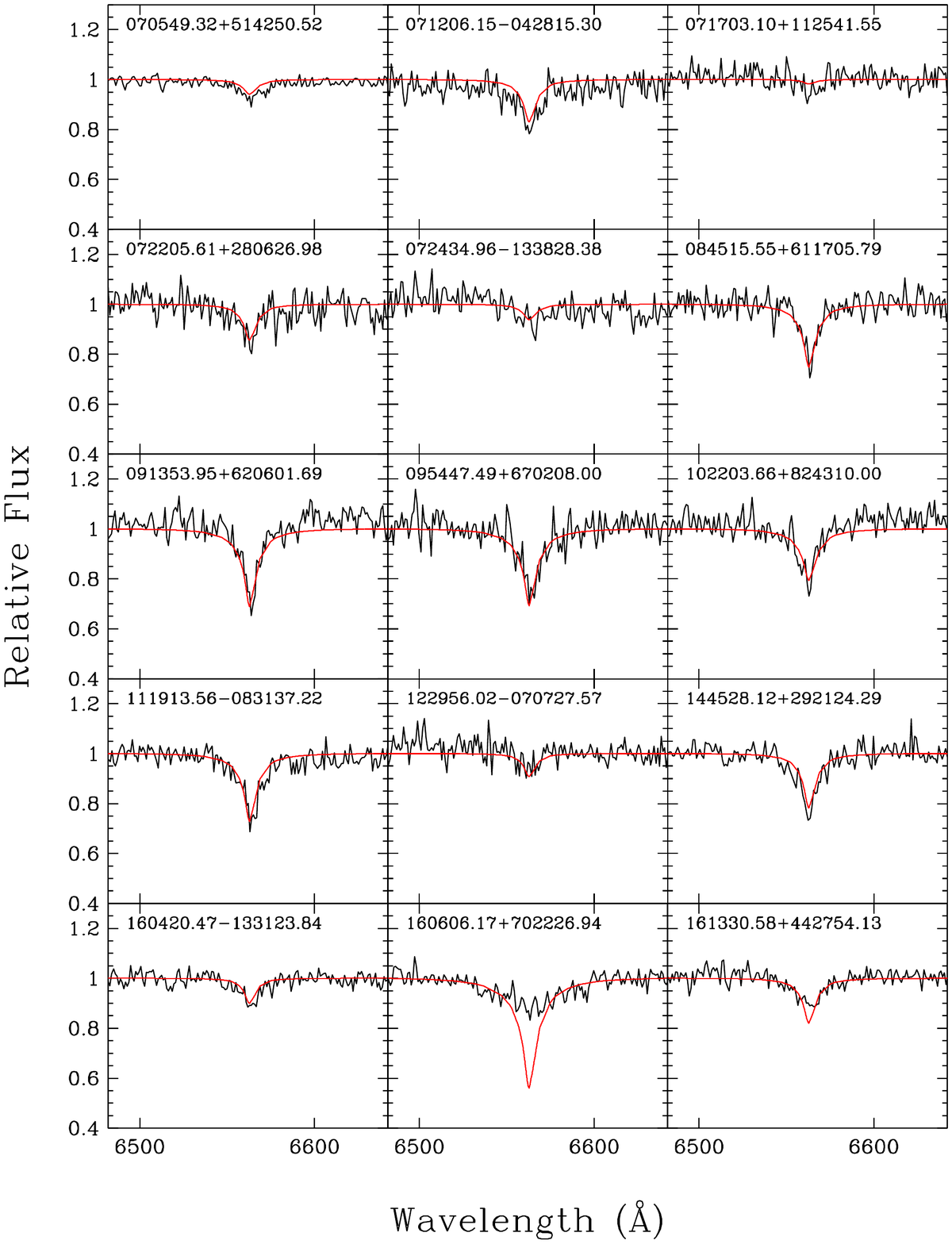}
	\caption{Comparison of observed and predicted H$\alpha$ line profiles from \textit{Gaia} photometric parameters (continued 2/4).}
        \label{fig:HalphaDA2}
\end{figure*}

\begin{figure*}
	\includegraphics[viewport= 1 20 700 750,scale=0.85]{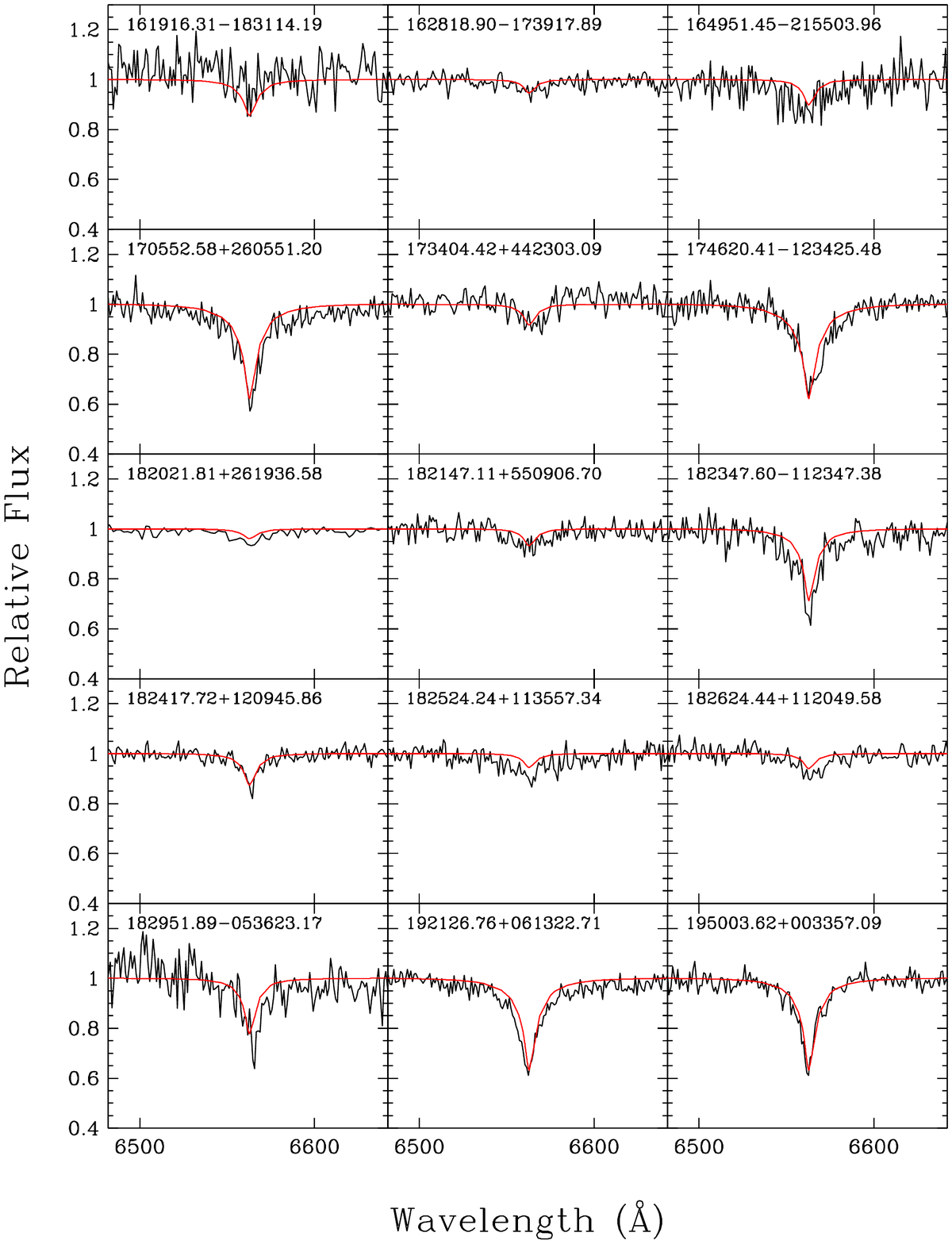}
	\caption{Comparison of observed and predicted H$\alpha$ line profiles from \textit{Gaia} photometric parameters (continued 3/4).}
        \label{fig:HalphaDA3}
\end{figure*}

\begin{figure*}
	\includegraphics[viewport= 1 20 700 750,scale=0.85]{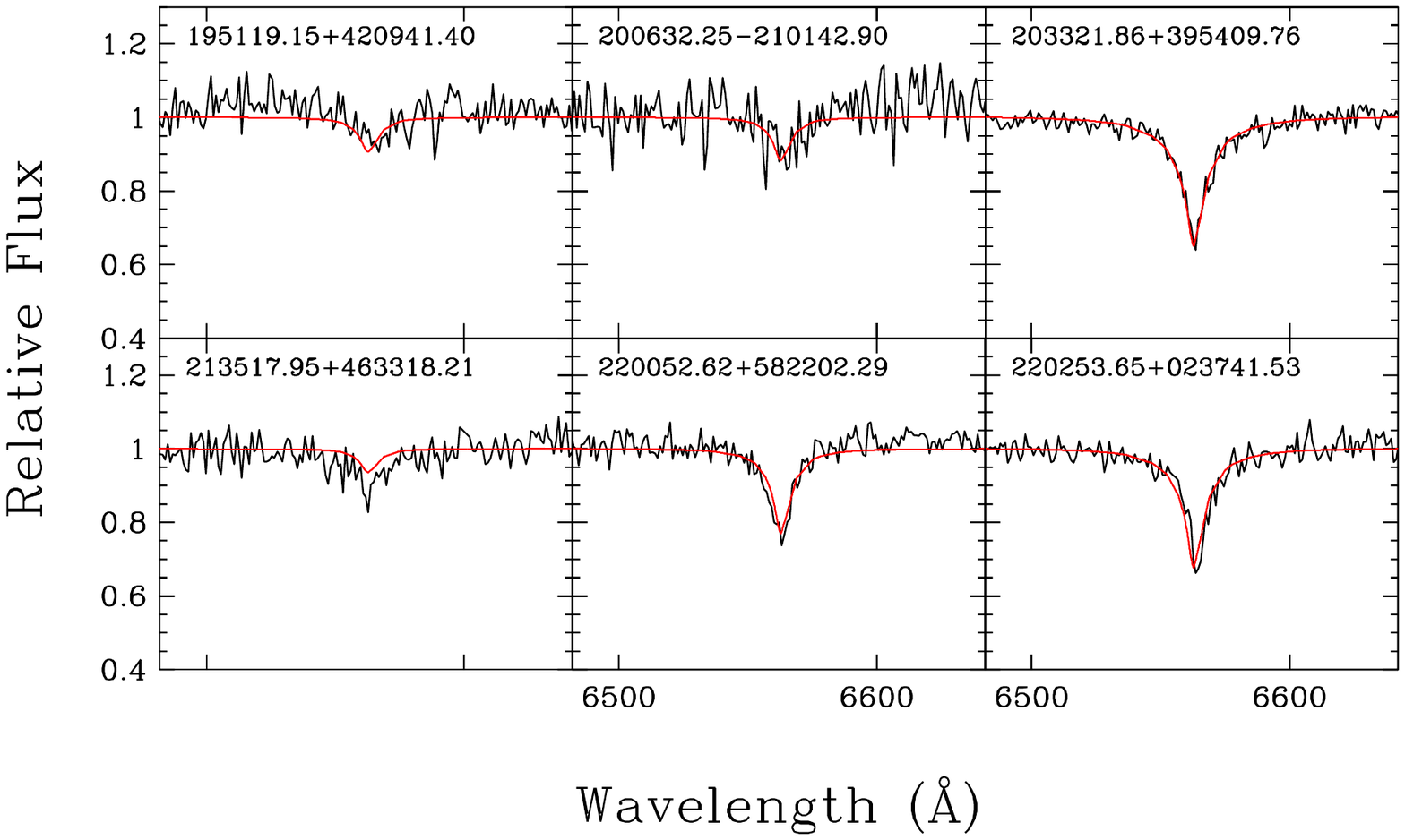}
	\caption{Comparison of observed and predicted H$\alpha$ line profiles from \textit{Gaia} photometric parameters (continued 4/4).}
        \label{fig:HalphaDA4}
\end{figure*}

\begin{figure*}
	\includegraphics[viewport= 1 20 700 750,scale=0.85]{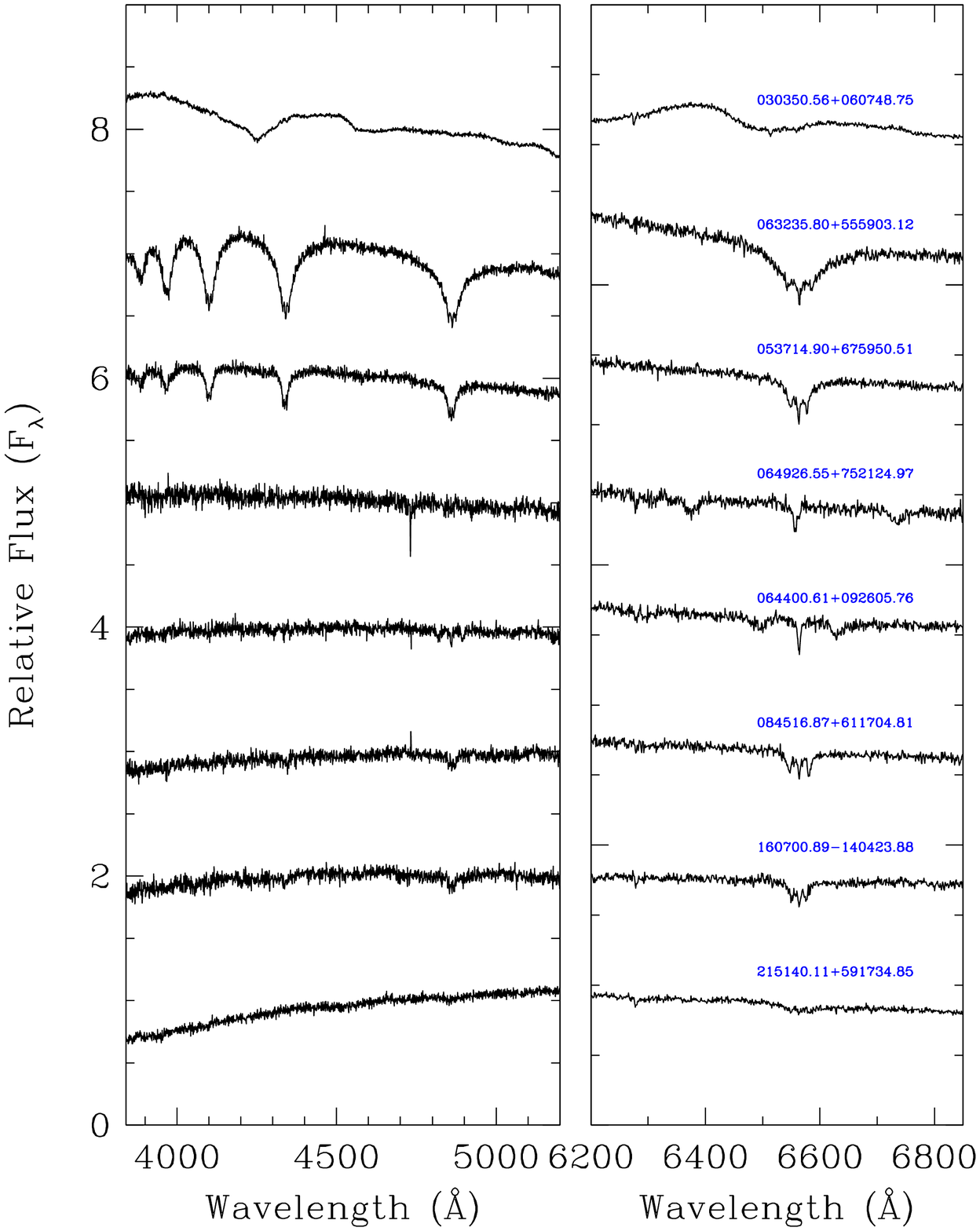}
	\caption{Spectroscopic observations of eight magnetic white dwarfs ordered with decreasing temperature and excluding the DZH WD\,J1515+8230, shown instead on Fig.~\ref{fig:DZ}.}
        \label{fig:Mag}
\end{figure*}

\begin{figure*}
	\includegraphics[viewport= 1 20 700 750,scale=0.85]{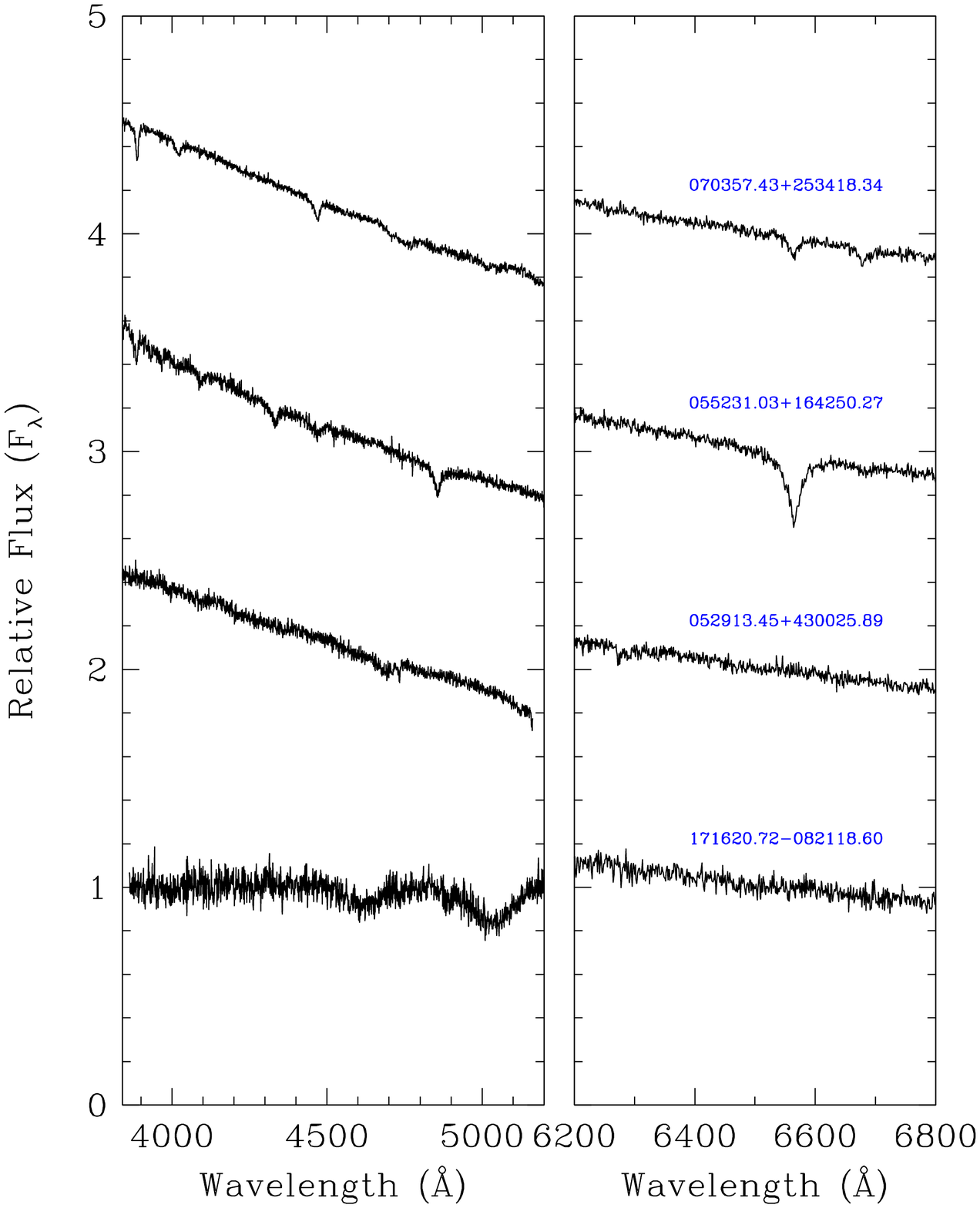}
	\caption{Spectroscopic observations of two DBA (top) and two DQ white dwarfs (bottom).}
        \label{fig:DB}
\end{figure*}

\begin{figure*}
	\includegraphics[viewport= 1 20 700 750,scale=0.85]{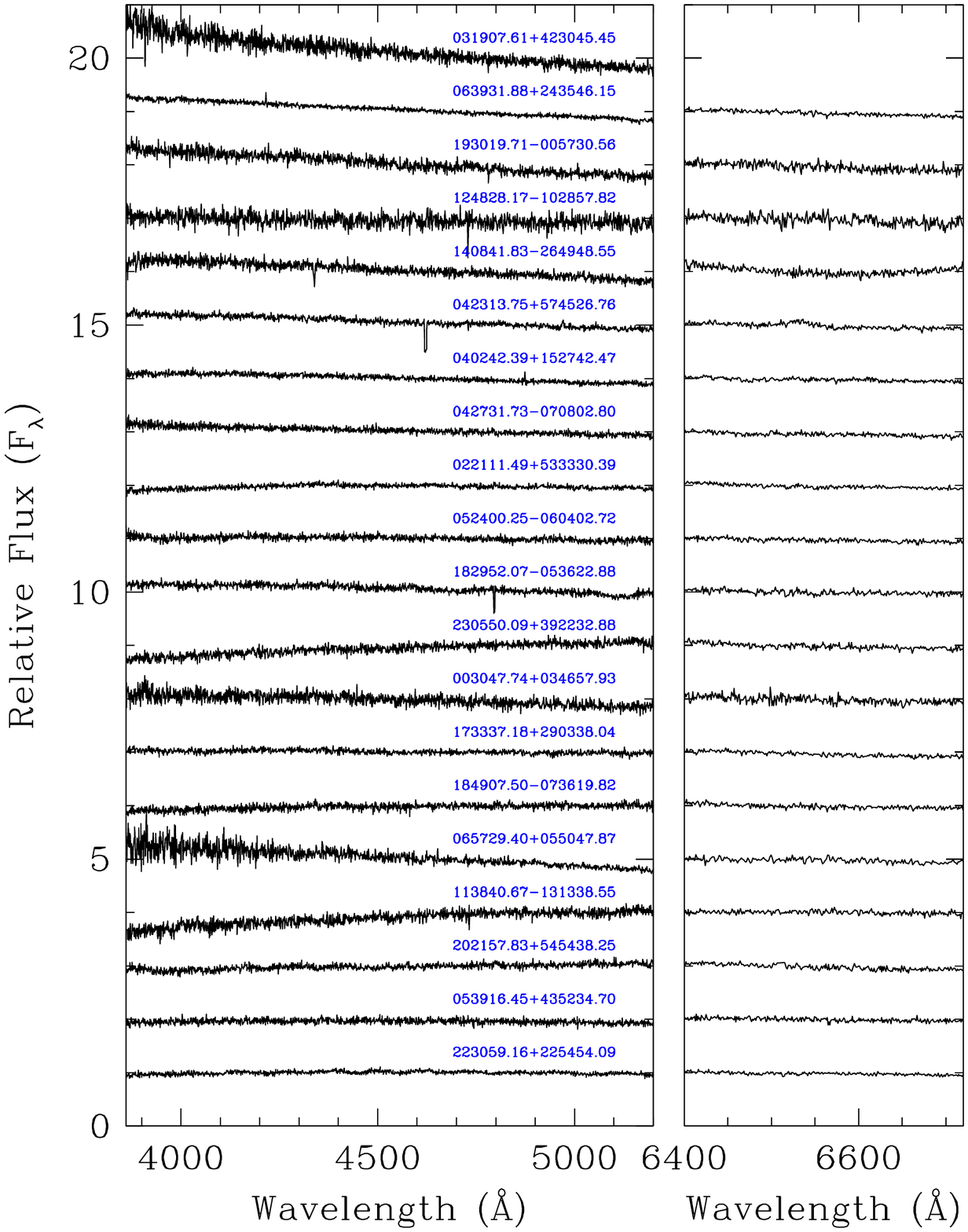}
	\caption{Spectroscopic observations of 76 DC white dwarfs ordered with decreasing photometric temperature (continued on Figs.~\ref{fig:DC2}-\ref{fig:DC4}).}
        \label{fig:DC1}
\end{figure*}

\begin{figure*}
	\includegraphics[viewport= 1 20 700 750,scale=0.85]{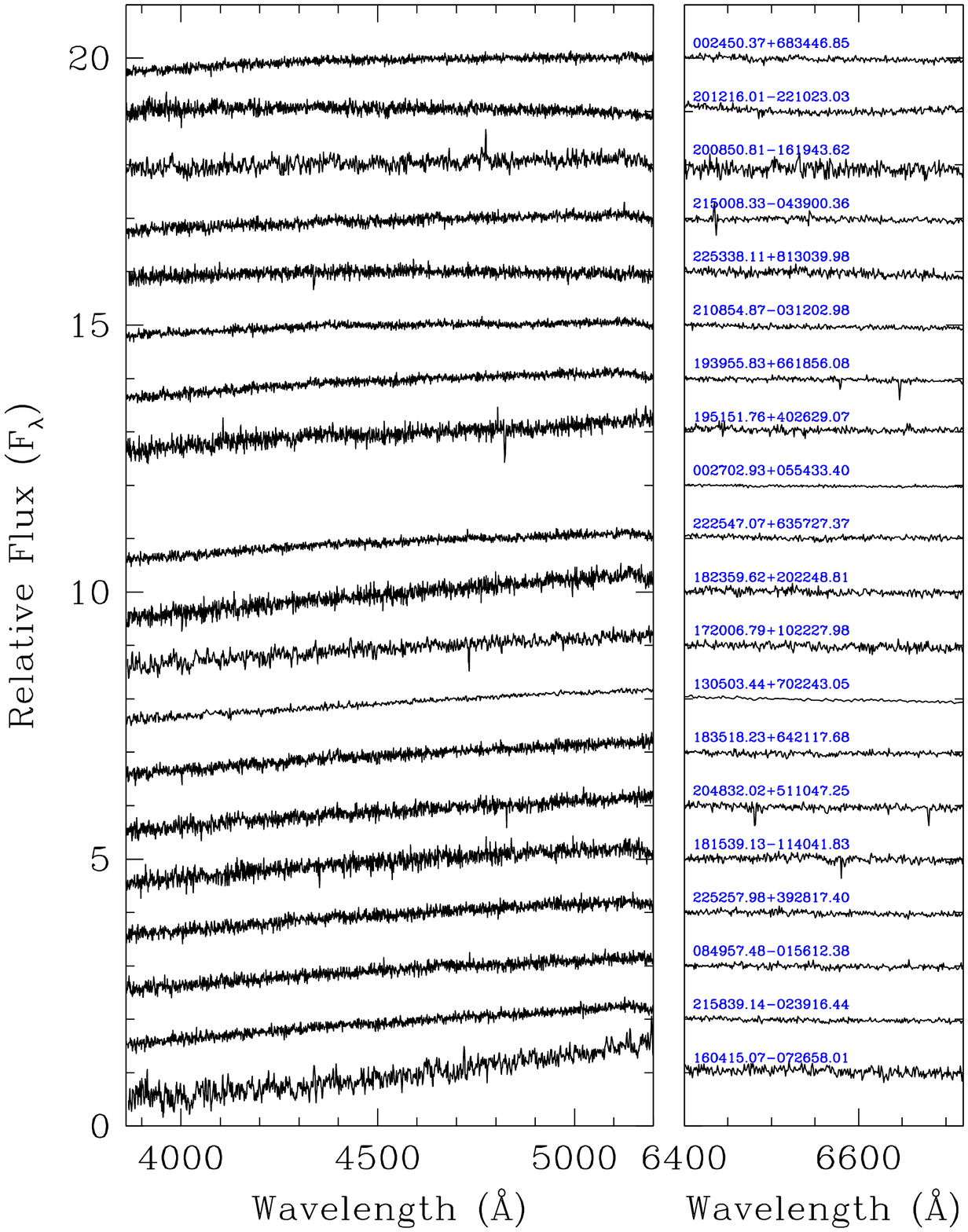}
	\caption{Spectroscopic observations of 76 DC white dwarfs ordered with decreasing photometric temperature (continued 2/4).}
        \label{fig:DC2}
\end{figure*}

\begin{figure*}
	\includegraphics[viewport= 1 20 700 750,scale=0.85]{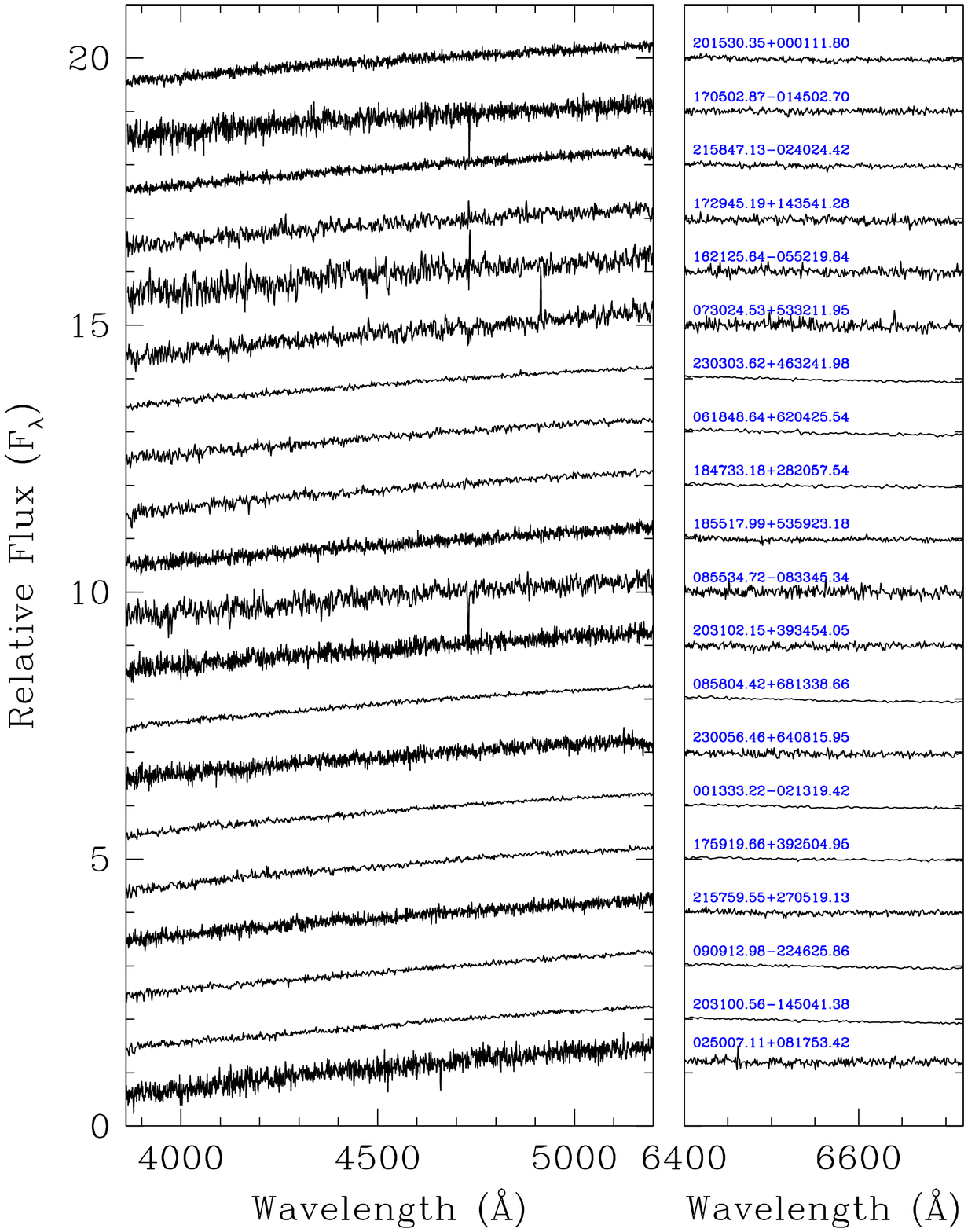}
	\caption{Spectroscopic observations of 76 DC white dwarfs ordered with decreasing photometric temperature (continued 3/4).}
        \label{fig:DC3}
\end{figure*}

\begin{figure*}
	\includegraphics[viewport= 1 20 700 750,scale=0.85]{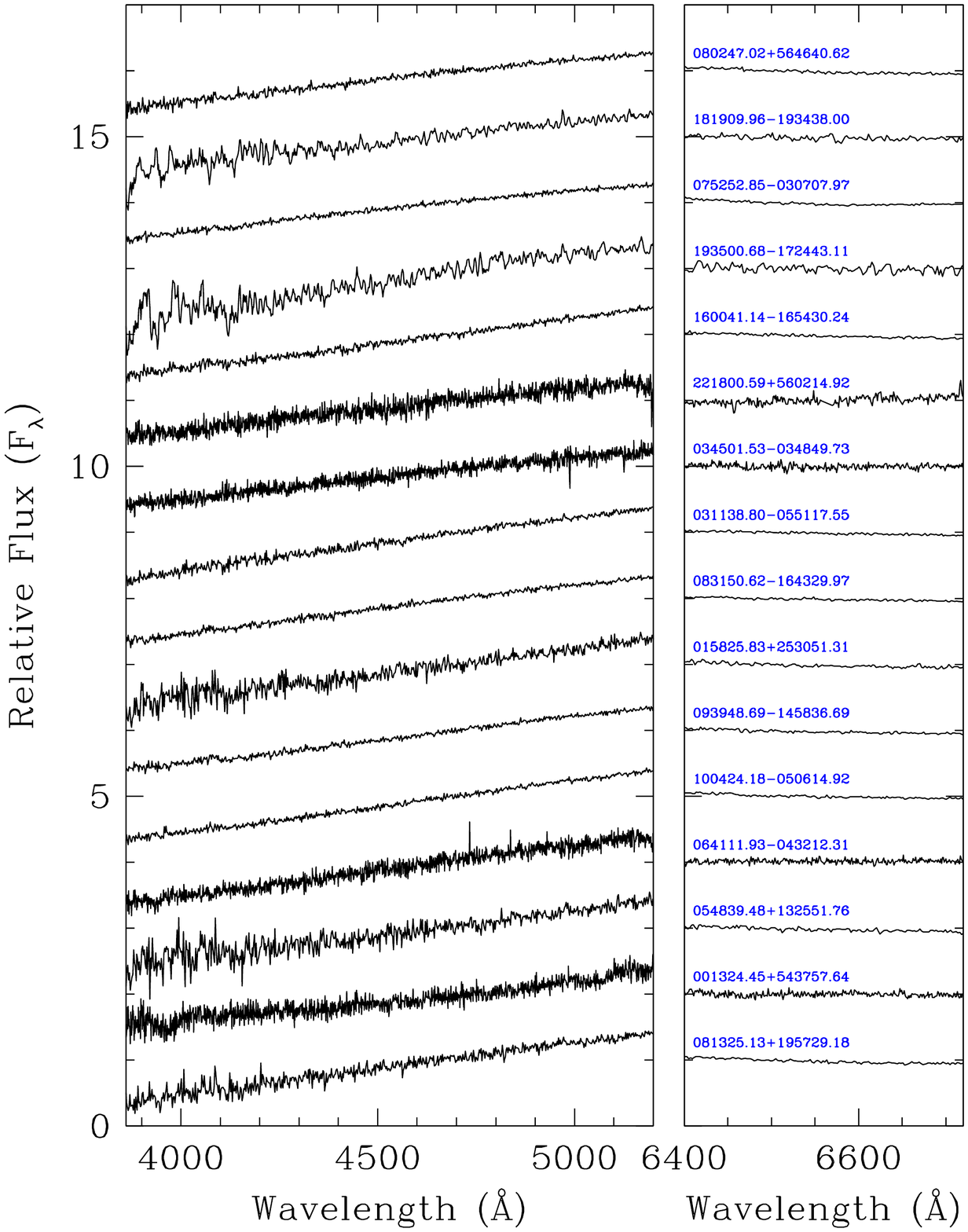}
	\caption{Spectroscopic observations of 76 DC white dwarfs ordered with decreasing photometric temperature (continued 4/4).}
        \label{fig:DC4}
\end{figure*}

\begin{figure*}
	\includegraphics[viewport= 1 20 700 750,scale=0.85]{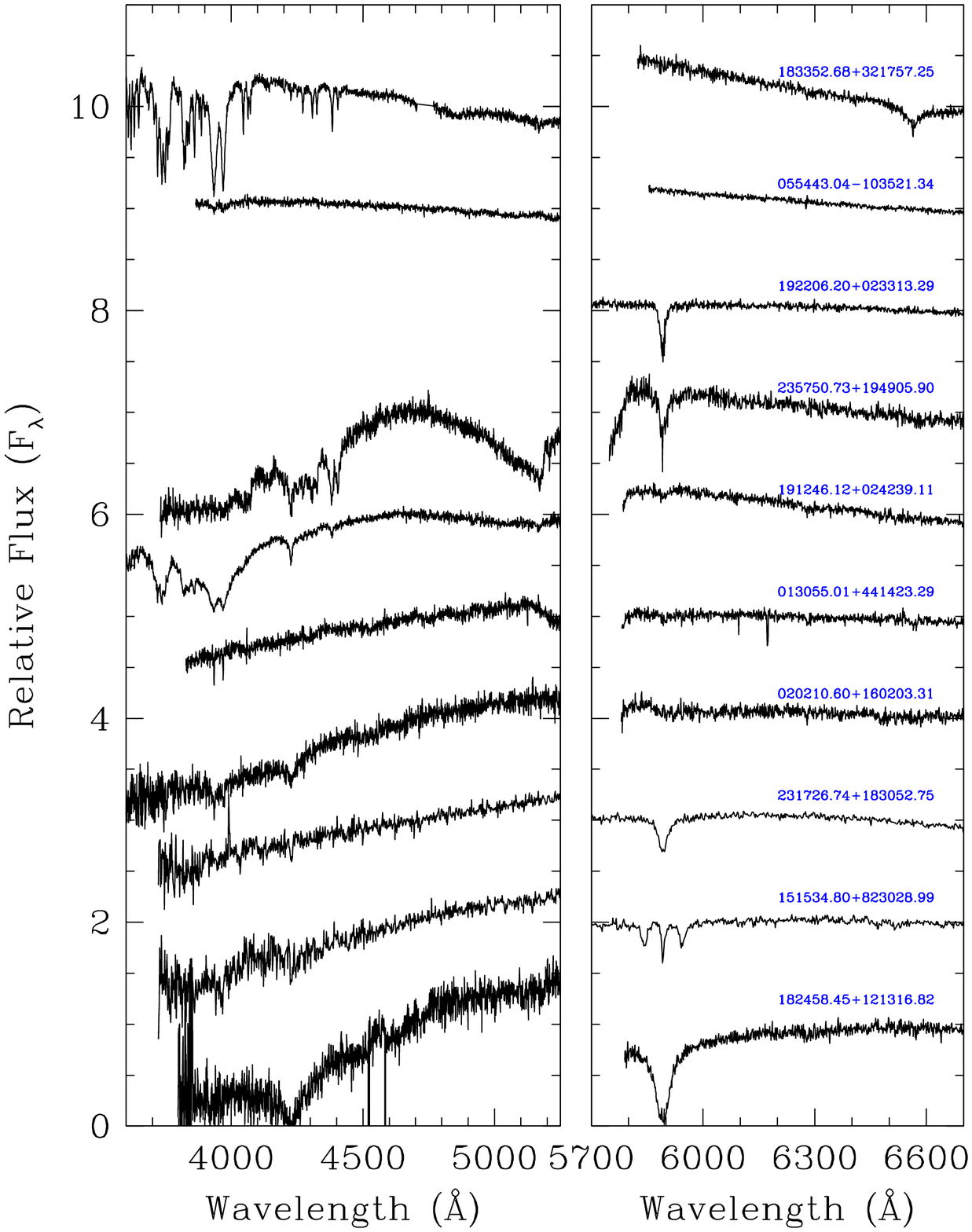}
	\caption{Spectroscopic observations of 10 DZ, DZA and DZH white dwarfs.}
        \label{fig:DZ}
\end{figure*}

\begin{figure*}
	\includegraphics[viewport= 1 20 700 750,scale=0.85]{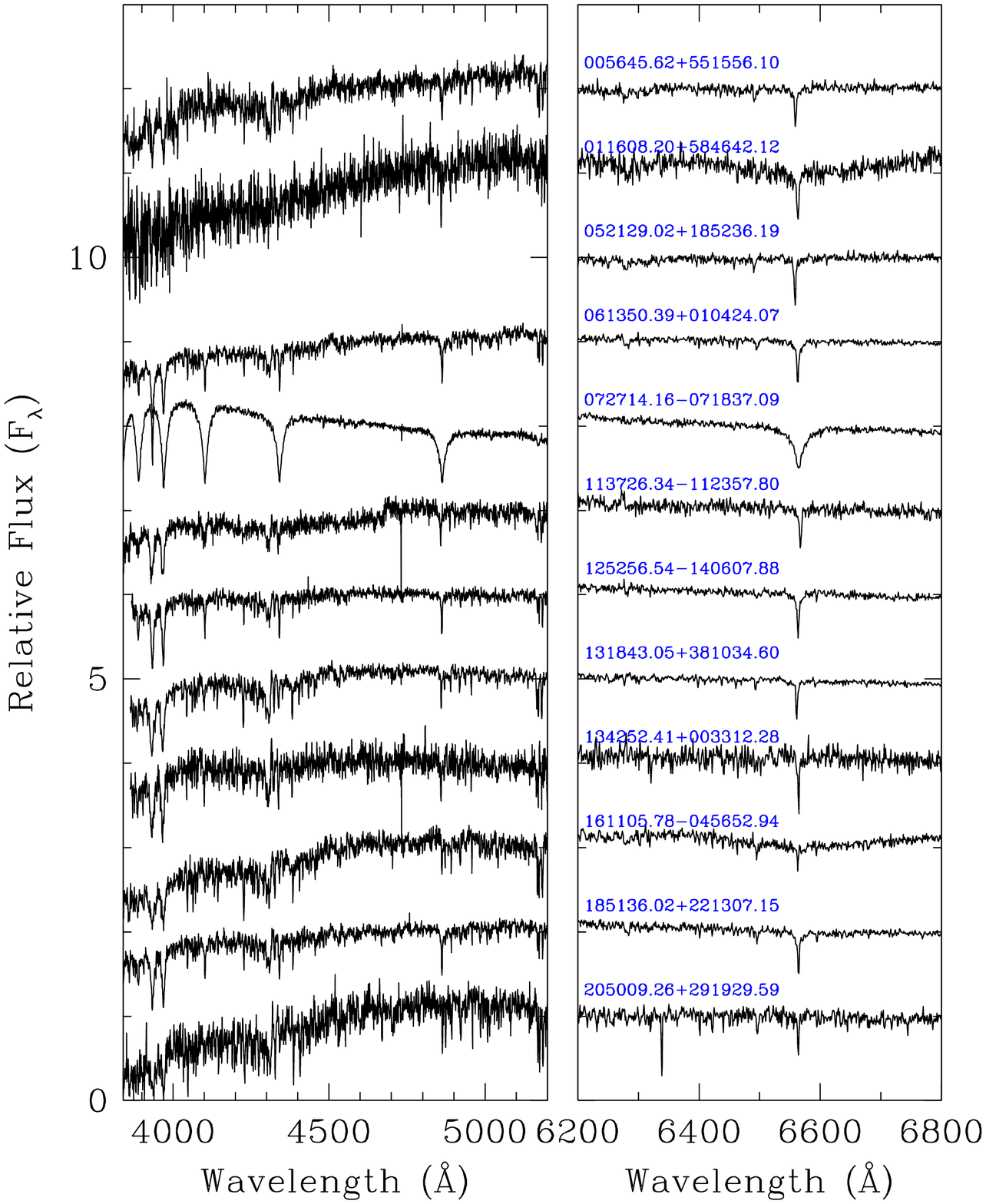}
	\caption{Spectroscopic observations of main-sequence stars that are high-probability white dwarf candidates ($P_{\rm WD}>0.75$) in the catalogue of \citet{gentile19}.}
        \label{fig:STARa}
\end{figure*}

\begin{figure*}
	\includegraphics[viewport= 1 20 700 750,scale=0.85]{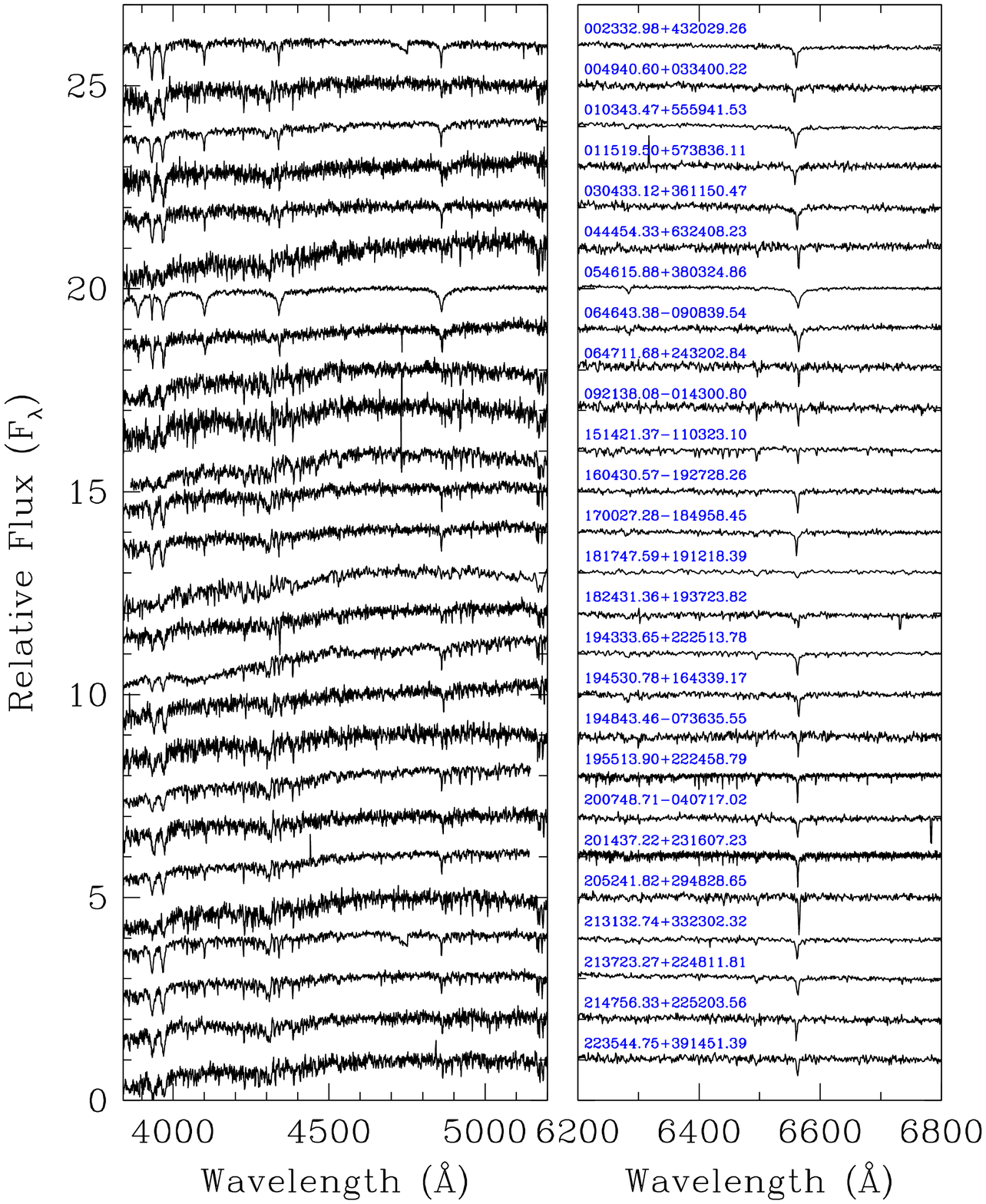}	\caption{Spectroscopic observations of main-sequence stars that are low-probability white dwarf candidates ($P_{\rm WD}<0.75$) in the catalogue of \citet{gentile19}.}
        \label{fig:STARb}
\end{figure*}











\bsp	
\label{lastpage}
\end{document}